\documentclass[fleqn,12pt]{wlscirep_mod}

\pdfoutput=1

\usepackage{scalerel}
\usepackage{tikz}
\usetikzlibrary{svg.path}

\definecolor{orcidlogocol}{HTML}{A6CE39}
\tikzset{
  orcidlogo/.pic={
    \fill[orcidlogocol] svg{M256,128c0,70.7-57.3,128-128,128C57.3,256,0,198.7,0,128C0,57.3,57.3,0,128,0C198.7,0,256,57.3,256,128z};
    \fill[white] svg{M86.3,186.2H70.9V79.1h15.4v48.4V186.2z}
                 svg{M108.9,79.1h41.6c39.6,0,57,28.3,57,53.6c0,27.5-21.5,53.6-56.8,53.6h-41.8V79.1z M124.3,172.4h24.5c34.9,0,42.9-26.5,42.9-39.7c0-21.5-13.7-39.7-43.7-39.7h-23.7V172.4z}
                 svg{M88.7,56.8c0,5.5-4.5,10.1-10.1,10.1c-5.6,0-10.1-4.6-10.1-10.1c0-5.6,4.5-10.1,10.1-10.1C84.2,46.7,88.7,51.3,88.7,56.8z};
  }
}

\newcommand\orcidicon[1]{\href{https://orcid.org/#1}{\hspace{-0.1cm}\mbox{
\begin{tikzpicture}[yscale=-1,transform shape]
\includegraphics[width=0.017\textwidth]{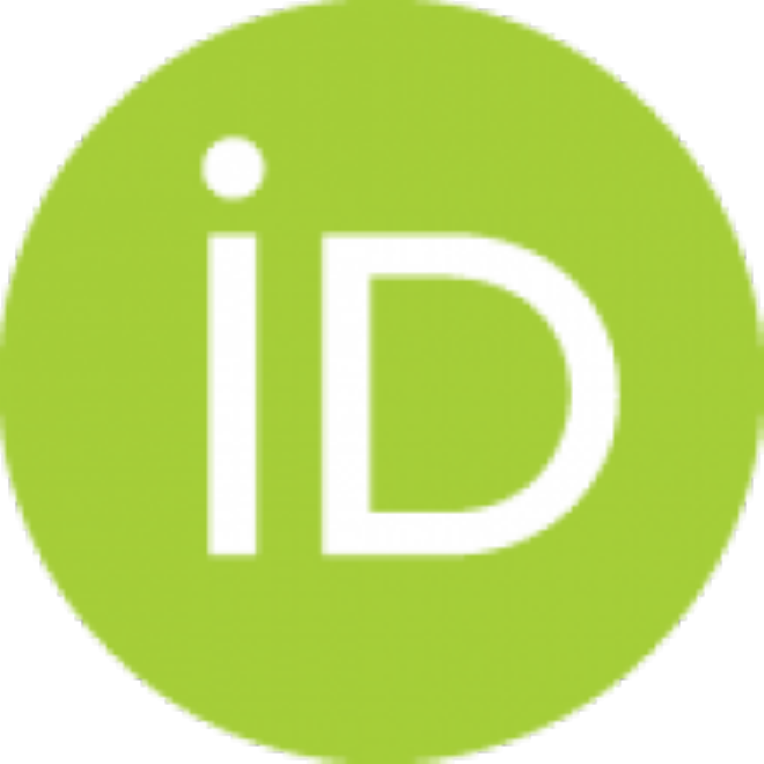} 
\end{tikzpicture}
}\hspace{2mm}}}

\usepackage{array}
\usepackage[ justification=justified,format=plain]{caption}
\usepackage{multirow}

\newcounter{partextdummy}
\newcommand*{\CreateLink}[2]{%
  \begingroup
    \renewcommand*{\thepartextdummy}{#2}%
    \ifhmode
      \raisebox{2ex}[0pt][0pt]{%
        \refstepcounter{partextdummy}%
        \label{#1}%
      }%
    \else
      \refstepcounter{partextdummy}%
      \label{#1}%
    \fi
  \endgroup
  \ignorespaces
}
\newcommand*{\LinkTo}{\ref}

\usepackage[utf8]{inputenc}
\usepackage[T1]{fontenc}
\usepackage{graphicx}
\usepackage{amsmath}
\usepackage{amsfonts}
\usepackage{amssymb}
\providecommand{\U}[1]{\protect\rule{.1in}{.1in}}

\usepackage[most]{tcolorbox}

\usepackage{yfonts}
\usepackage[normalem]{ulem}

\usepackage{array}
\usepackage{hhline}

\usepackage{hyperref}

\usepackage{ragged2e}
\usepackage[ justification=justified,
   format=plain]{caption}

\newcommand{\Hz}{\rm{\, Hz }}

\newcommand{\beq}{\begin{equation}}
\newcommand{\eeq}{\end{equation}}
\newcommand{\ba}{\begin{array}}
\newcommand{\ea}{\end{array}}

\newcommand{\E}{E_{53}}

\newcommand{\ee}{\epsilon_{e,-1}}
\newcommand{\eB}{\epsilon_{B,-2}}

\renewcommand{\O}{{\mathcal {O}}}

\newcounter{arxiv}
\newcounter{tr}
\usepackage[normalem]{ulem}

\ifnum \value{tr}>5

\newcommand{\deleted}[1]{{\color{red} Husne - Deleted: } \sout{\color{red} {#1}}}

\newcommand{\authorcomment}[1]{{\color{purple} Husne - Comment :} {\color{cyan} #1}}

\else

\newcommand{\deleted}[1]{}

\newcommand{\authorcomment}[1]{}

\fi

\ifnum \value{tr}>5

\newcommand{\deletedD}[1]{{\color{red} Damien - Deleted: } \sout{\color{red} {#1}}}

\newcommand{\authorcommentD}[1]{{\color{purple} Damien - Comment :} {\color{cyan} #1}}

\else

\newcommand{\deletedD}[1]{}

\newcommand{\authorcommentD}[1]{}

\fi

\begin{document}


\title{ A wind environment and Lorentz factors of tens explain gamma-ray bursts X-ray plateau }


\author[1,*, \orcidicon{0000-0002-8852-7530}  ]{H\"usne Dereli-B\'egu\'e}
\author[1, \orcidicon{0000-0001-8667-0889}]{Asaf Pe'er}
\author[2, \orcidicon{0000-0002-9769-8016}]{Felix Ryde}
\author[3, \orcidicon{0000-0001-9309-7873}]{Samantha R. Oates}
\author[4,5, \orcidicon{0000-0002-9725-2524}]{Bing Zhang}
\author[6, \orcidicon{0000-0003-4442-8546}]{Maria G. Dainotti}
\affil[1]{Department of Physics, Bar-Ilan University, Ramat-Gan 52900, Israel}
\affil[2]{Department of Physics, KTH Royal Institute of Technology and The Oskar Klein Centre, SE-106 91 Stockholm, Sweden}
\affil[3]{School of Physics and Astronomy \& Institute for Gravitational Wave Astronomy, University of Birmingham, Birmingham B15 2TT, UK}
\affil[4]{Nevada Center for Astrophysics, University of Nevada, Las Vegas, Nevada, NV 89154}
\affil[5]{Department of Physics and Astronomy, University of Nevada, Las Vegas, Nevada, NV 89154}
\affil[6]{National Astronomical Observatory of Japan, 2-21-1 Osawa, Mitaka, Tokyo 181-8588}
\affil[*]{husnedereli@gmail.com}




\begin{abstract}
Gamma-ray bursts (GRBs) are known to have the most relativistic jets, with initial Lorentz factors in the order of a few hundreds. Many GRBs display an early X-ray light-curve plateau, which was not theoretically expected and therefore puzzled the community for many years. Here, we show that this observed signal is naturally obtained within the classical GRB “fireball” model, provided that the initial Lorentz factor is rather a few tens, and the expansion occurs into a medium-low density “wind”. The range of Lorentz factors in GRB jets is thus much wider than previously thought and bridges an observational gap between mildly relativistic jets inferred in active galactic nuclei, to highly relativistic jets deduced in few extreme GRBs. Furthermore, long GRB progenitors are either not Wolf-Rayet stars, or the wind properties during the final stellar evolution phase are different than at earlier times. We discuss several testable predictions of this model.
\end{abstract}

\flushbottom
\maketitle

\thispagestyle{empty}


Gamma-ray bursts (GRBs) are the most energetic explosions known
in the Universe and are also known to have the most relativistic jets, with
initial expansion Lorentz factors of $100< \Gamma_i <1000$
\cite{KP91, Fenimore93, Racusin11}. One of the most puzzling results in the
study of GRBs is the existence of a long “plateau” in
the early X-ray light curve (up to thousands of seconds)
\cite{Nousek06, OBrien06, Zhang06, Srinivasaragavan20} of a significant
fraction of GRBs (43\% until 2009 \cite{Evans09} and
56\% until 2019 \cite{Srinivasaragavan20}). This plateau,
not predicted theoretically \cite{MR93}, was first detected by the Neil
Gehrels \textit{Swift} Observatory \cite{Gehrels04} in 2005, and despite
the long time passed since its discovery, its origin is still highly debate
in the literature, with many authors suggesting various extensions to the
classical “fireball” model in order to explain it.



The first and most commonly used idea is the continuous energy injection from a central compact object which can be a newly formed black hole \cite{Zhang06, GK06, Nousek06} or a millisecond magnetar \cite{Metzger11}. Other notable ideas include two components \cite{Berger03, Huang04, Granot06} or multi component \cite{Toma06} jet models; forward shock emission in homogeneous media \cite{Toma06}; scattering by dust/modification of ambient density by gamma-ray trigger \cite{Ioka06, SD07}; dominant reverse shock emission \cite{UB07, Genet07}; evolving micro-physical parameters \cite{Ioka06}; and viewing angle effects in which jets are viewed off-axis \cite{EG06, Toma06, Eichler14}. While each of these ideas is capable of explaining the observed plateau under certain conditions, they all require an external addition to the basic "fireball" model scenario, and in some cases they cannot address the full set of properties of the plateau phase (such as the flux, slope or duration). A thorough discussion on the advantages and weaknesses of each of these ideas appear in Ref.\cite{KZ15}. We provide a short comparison with some of the recent proposed ideas in Supplementary Discussion,~\LinkTo{sec:SuppDisc}.

Plateaus are seen in the X-ray light curves of both short ($\lesssim 2$ s) and long ($\gtrsim 2$ s) GRBs \cite{Kouveliotou1993}, and may be associated with the properties of the progenitors. While it is widely believed that short GRBs originate from compact binary merger \cite{Eichler1989}, the progenitors of long GRBs are thought to be the explosion of (very) massive stars ($\gtrsim$10 $M_\odot$) emitting strong stellar winds \cite{Woosley1993, MacFadyen1999}. This idea is strongly supported by GRB-SN associations \cite{Galama1998, Hjorth2003}, suggesting that Wolf-Rayet stars are the most likely progenitors of long-duration GRBs \cite{Woosley2006}. Additional supporting evidence are host galaxy studies \cite{Fruchter2006} and relatively low metalicity \cite{Modjaz2008}. The low metalicity implies the expected mass-loss rates to be smaller than those in typical Wolf-Rayet stars in our galaxy \cite{Vink2005}, $\dot M = 10^{-5} \, M_\odot \, {\rm yr}^{-1}$. This implies wind velocity of $v_w = 10^{8} \,{\rm cm\, s^{-1}}$ as a characteristics of a GRB progenitor. Indeed, multiple spectral components from the GRBs and SNe at the optical band are seen with speeds of $5\times 10^7 \,{\rm cm\, s^{-1}}$ and $3\times 10^8 \,{\rm cm\, s^{-1}}$. \cite{Woosley2006} These properties characterize the wind strength of the progenitor, therefore, affect the observed properties of the GRBs. As we show in this work, they may strongly affect the afterglow emission, and specifically the plateau emission.

Here we study a sample of GRBs with plateau phase seen in both X-ray and optical bands. We consider a much simpler idea than previously discussed, which does not require any modification of the classical GRB "fireball" model. Rather, we simply look at a different region of the parameter space: a flow having an initial Lorentz factor of the order of a few tens, propagating into a "wind" (decaying density) ambient medium, with a typical density of up to two orders of magnitude below the expectation from a wind produced by a Wolf-Rayet star. We compute the physical parameters assuming synchrotron emission from a power-law distribution of electrons accelerated at the forward shock. As we show, this model provides a natural explanation to the observed signals. We discuss the implication of the results on the properties of GRB progenitors and the resulting jets, and show how they provide a novel tool to infer the physical properties inside the jet.


\section*{Results} 

\subsection*{Sample selection and data analysis}

We selected 13 GRBs based on the criteria defined in methods subsection
\LinkTo{sec:SampleSelection} below. In analyzing the optical and X-ray light
curves (see Supplementary Method \LinkTo{sec:SuppMethod1}), we identified two achromatic temporal breaks \cite{Zhang06}: one during the
transition from the "plateau" to the decaying lightcurve, which we interpret as transition from the coasting to the self-similar expansion (this break marks the
end of the "plateau" and denoted by "$T_{\rm a}$"); and a second, later break, which is identified as a jet-break, marked as "$T_{\rm b}$". In some
bursts a second break could not be detected due to poor data
sampling at very late times. The temporal slopes during
the plateau phase, self-similar phase and after the jet break are marked
as $\alpha_{\rm p}$, $\alpha_{\rm A1}$ and $\alpha_{\rm A2}$ respectively
and are given in Supplementary 
Tables~\ref{tab:ClassI}~---~\ref{tab:ClassIII}. 


\subsection*{Theoretical regions}

The optical and X-ray light curves do not necessarily follow the
same power-law decay in either of the dynamical
phases (plateau or self-similar).  This can easily be understood
in the framework of synchrotron emission from forward-shock accelerated
electrons. The injected electrons assume a power-law distribution
with power-law index $p$, namely $N_{el}(\gamma) d \gamma \propto \gamma^{-p}$
above a minimum value $\gamma_m$; below this value, one can assume the electrons to have a Maxwellian (or quasi-Maxwellian)
energy distribution \cite{MR93, SPN98}. This assumption leads to a broken
power-law spectra and light curves, whose shapes, in the relevant observed
bands, depend on whether the peak frequency, $\nu_m$ (corresponding Lorentz factor
$\gamma_m$) is above or below the cooling frequency, $\nu_c$ (corresponding
 Lorentz factor $\gamma_c$, for which the rate
of energy lost by synchrotron emission is equal to the rate of energy lost
by adiabatic cooling). For a given observed frequency (optical [typically
the U-band]: $\nu_{\rm U}$ or X-rays: $\nu_{\rm X}$), different possibilities
of the expected light curve and spectra exist.
These possibilities are listed in Supplementary Table \ref{tab:summary}
and are displayed in Supplementary Fig. \ref{fig:summary_regions_plateau}.
The parameter
space defined by the fast cooling regime ($\nu_m>\nu_c$) is split in three
regions marked as A, B, C; while the parameter space for slow
cooling regime ($\nu_m<\nu_c$) contains the regions marked
as D, E, F, respectively. At low frequencies, one needs to consider synchrotron
self-absorption, which can safely be neglected being below the optical
frequency at all observed times (see details, methods subsection \LinkTo{sec:Theoreticalmodel}).

\subsection*{Sample classification}

After analyzing the data, we split the sample into 3 classes based on the X-ray and optical light curves during the plateau phase (corresponding to Supplementary Tables \ref{tab:ClassI} --- \ref{tab:ClassIII}). These classes match very well the theoretical predictions.  Class~I is characterized by a flat X-ray light curve ($F_\nu \propto t_{\rm obs.}^{0.0 .. -0.2}$), corresponding to regions C and F in Supplementary Table \ref{tab:summary} and a decaying optical light curve ($F_\nu \propto t_{\rm obs.}^{-0.5 .. -0.7}$), region E in Supplementary Table \ref{tab:summary}. Theoretically, this class corresponds to GRBs having their cooling frequency between the optical and the X-ray bands, namely $\nu_{\rm U} < \nu_c < \nu_{\rm X}$. In class~II, both the X-ray and the optical light curves are flat. Theoretically, this is expected for a low cooling frequency, $\nu_c < \nu_{\rm U} < \nu_{\rm X}$. In class~III, both X-ray and optical light curves are decaying. This is expected when the cooling frequency is high, $\nu_{\rm U} < \nu_{\rm X} < \nu_c$. Interestingly, after excluding faint flares, all GRBs in our sample fall into one of these three classes. Furthermore, even tough the selection criteria for the different classes is based only on X-ray and optical light curves, there seems to be a connection between the energy of a burst, the duration of its plateau, and the its class. This is shown in Fig. \ref{fig:EisoandTa}.


\subsection*{Closure relations and determination of the electron power-law indices}

The theoretical model used herein imposes a relation between the spectral
and the temporal slopes. These so-called 'closure relations' are unique to each class and each observational band (optical and X-rays). Therefore, they can be used to assess the validity of our model.
For each GRB, the X-ray spectral slopes ($\beta_{\rm p}$ for the plateau phase, $\beta_{\rm A1}$ for the self-similar phase, $\beta_{\rm A2}$ for the spectral slope after the jet break) were obtained from the online \textit{Swift} repository for the same time range as the temporal slopes, and optical spectral slopes were retrieved from the literature (the relevant references are given in Supplementary Table \ref{tab:ClassI} --- \ref{tab:ClassIII}).
We used the spectral and temporal slopes in each
phase (plateau and self-similar) independently to further confront our theory to the data.
Specifically, we checked that the closure relations relevant for regimes (C, E, F)
for each band, each phase
and each class are satisfied independently. The results are presented in
Supplementary Figs. \ref{fig:CR-plateau} and \ref{fig:CR-self-similar} for
the plateau phase and the self-similar phase respectively. From these figures,
it is clear that all the data - spectral and temporal, both in X-ray and optical
- are consistent with the theoretical closure relations of the model.


Furthermore, from the closure relation, we deduced the power-law index $p$ of
the accelerated electrons by using 8 independent measurements, namely the temporal and spectral indices
of both the optical and X-ray data during both the plateau and self-similar phases. This was done for both regions E and F. 
As we demonstrate in Fig. \ref{fig:ElectronPowerlawindex-091029}, 
we find that the power-law index of the accelerated electrons does not change in between the dynamical phases. We find that
for all GRBs in our sample, the power-law index, $p$ is in the range
$1.8 \lesssim p \lesssim 2.5$. For those bursts whose X-ray light curve is
identified as being in region E, the electron power-law index is narrowly
clustered around $p~\simeq~2$, while a larger spread is found for other GRBs.
In 3 cases, namely GRBs 060614, 060729 and 110213A, we find that the values derived
from the optical spectral slopes are inconsistent with the other measurements
(see caption of Table \ref{tab:ElectronPowerlawindex}), in which case we use 6
out of 8 independent measurements in determining the region and the power-law index.
The results are summarized in Table \ref{tab:ElectronPowerlawindex}.

We find that the optical data of 6 GRBs out of 13 in our sample
(listed in Supplementary Table~\ref{tab:ClassII}) are compatible with being
in region F, implying that the cooling frequency at the end of the plateau
phase is below the observed optical band, $\nu_c<\nu_{\rm U}$. For another
4 GRBs out of 13 (listed in Supplementary Table \ref{tab:ClassI}),
the optical light curve decays (corresponding to region E), while the X-ray
light curve is flat, namely for these bursts $\nu_{\rm U} < \nu_c < \nu_{\rm X}$.
For the remaining 3 GRBs out of 13 (listed in Supplementary Table
\ref{tab:ClassIII}) both the optical and X-ray light curves decay, and are
therefore compatible with the cooling frequency being above the observed
X-ray band $\nu_{\rm X}<\nu_c$.

\subsection*{The derived values of the physical parameters}

We use here a simple theoretical model for the afterglow: 
the emission is produced by synchrotron
radiation from electrons accelerated to a power law distribution at the forward shock, generated by the
propagation of the ejecta 
into a "wind" medium, characterized by a decaying density:
$\rho (r) = 5  \times 10^{11}  A_\star r^{-2} {\rm~ g~cm^{-3}}$. Here, the
normalisation is obtained by assuming a wind mass-loss rate of
$\dot M = 10^{-5} \, M_\odot \, {\rm yr}^{-1}$ and a wind velocity of
$v_w = 10^{8} \,{\rm cm~ s^{-1}}$ characteristics of a GRB progenitor
(see, Supplementary Method \LinkTo{sec:SuppMethod2}, Equation \eqref{eq:2.9}).
The excellent agreement between the data and the theory enables us
to determine or constrain the parameters of the outflow and the wind,
in particular the proportionality constant $A_\star$ of the wind density,
the initial jet Lorentz factor, $\Gamma_i$, the fraction of energy in the
electrons, $\epsilon_e$ and the magnetization, $\epsilon_B$. All relevant
parameters used in the analysis are given in Table \ref{tab:key_parameters}
(see, Supplementary Method \LinkTo{sec:SuppMethod1c} for details).
For the 10 GRBs in Supplementary Tables \ref{tab:ClassI} and \ref{tab:ClassII},
the X-ray flux and the transition time that marks the end of the plateau phase
enable a direct deduction of $\epsilon_e$, while for the 3 GRBs in Supplementary
Table \ref{tab:ClassIII}, only a lower limit is available. For the 4 GRBs
in Supplementary Table \ref{tab:ClassI} we can directly infer the combined
value of $A_\star \Gamma_i^4$. For a given value of $A_\star$, the value
of $\epsilon_B$ is solely determined. Thus, an independent
estimate of $\epsilon_B$ enables to break the degeneracy.

In Fig. \ref{fig:Gamma-A_star} we use known limits
($10^{-5} \lesssim \epsilon_B \leq 0.1$) \cite{Santana2014, Duran2014} to
constrain the values of $A_\star$ and $\Gamma_i$. The lower values of
$\epsilon_B$ are obtained from fitting the optical-to-X-ray light curves
of bursts within the framework of the
decaying "afterglow" model \cite{Santana2014} as well as the analysis of
bright LAT GRBs \cite{Wang2013, Nava2017}. Such a low value of
$\epsilon_B= 10^{-5}$ would not be obtained from
the faint GRBs in class II due to the physical limitation on the
Lorentz factor ($\Gamma>1$). Therefore, we find that for GRBs in class II,
$\epsilon_B$ cannot be smaller than $10^{-3}$. However, for the GRBs in
classes I and III, such a restriction is not necessary, and $\epsilon_B$
can be as small as $10^{-5}$. Correspondingly, if indeed $\epsilon_B$
obtain such a low value, the density would be large. Assuming the value
$\epsilon_B = 10^{-5}$, we show the micro physical parameters of GRBs in
classes I and III in Fig. \ref{fig:Gamma-A_star} in orange.

A tighter constraint on the minimum value of the coasting Lorentz factor
$\Gamma_i$, thereby on the value of the magnetization of GRBs in class II
can be put using the requirement that the prompt emission radius,
$R_{\rm E} = 2 \Gamma_i^2 c \Delta t_{\rm min}$ be above the photospheric radius,
$R_{\rm ph} = L_{\rm iso} \sigma_T / (8 \pi m_p c^3 \Gamma_i^3)$ \cite{Piran1999}.
Here, $\Delta \rm t_{min}$ is the minimum observed variability timescale
during the prompt phase, $L_{\rm iso}$ is the isotropic luminosity,
$\sigma_T$ is the Thomson cross section, $m_p$ is proton mass and $c$
is the speed of light. For typical GRB parameters (including sub-second
variability,  $\Delta t_{\rm min}=0.1$~s and isotropic luminosity,
$L_{\rm iso} = 10^{50.5}$~erg/s), the requirement $R_{\rm E} \geq R_{\rm ph}$
results in a minimum Lorentz factor $\Gamma_i \gtrsim 30$
\cite{Piran1999, Daigne2002}. A few GRBs in our sample, mainly in class II
have an estimated Lorentz factor lower than this
limiting value. However, all these GRBs are low luminosity GRBs, implying
relatively low signal-to-noise ratio ($S/N \leq 10$) during the
prompt phase  \cite{Golkhou2014}. Only in one case (GRB 060729) a reliable
variability is measured, giving $\Delta t_{\min} = 4.99$~s \cite{Golkhou2014},
although the general trend of low luminosity GRBs having
$\Delta t_{\min} > 1.0$~s is clearly observed \cite{Golkhou2014, Sonbas2015}. 
Using the {\it Swift}-BAT light curves of all other low $\Gamma_i$ GRBs
in our sample we estimate that the observed variability
of these GRBs is much longer, and we choose as a conservative estimate
$\Delta t \geq \Delta t_{\min} = 5$~s. Most importantly, given the degeneracy
between $\Gamma_i - A_\star$, for the GRBs with the lowest Lorentz factor,
namely GRBs 060614, 060729, 100418A and  GRB 171205A the required criteria is
achieved for value of $\epsilon_B \geq 0.1$, which enforces (relatively)
high $\Gamma_i$ and low $A_\star$ (the left hand side in Fig.
\ref{fig:Gamma-A_star}). We therefore mark the value $\epsilon_B = 0.1$
by blue arrows in Fig. \ref{fig:Gamma-A_star}. The derived Lorentz factors
for these bursts are $\approx 10$, while the density is characterized by
$A_\star = 10^{-2}$. The ratio of prompt emission radius to photospheric
radius of all GRBs in our sample are plotted in
Fig.~\ref{fig:ratio_of_radii_Eiso}, demonstrating that despite the low
values of the Lorentz factors we obtain, the prompt emission radius is
always above the photospheric radius.

For the 6 GRBs in Supplementary Table \ref{tab:ClassII}, only a lower limit on the value of $\epsilon_B$ can be deduced. This is still highly valuable, as physically $\epsilon_B < 1.0$.  For the 3 GRBs in Supplementary Table \ref{tab:ClassIII}, only an upper limit on $\epsilon_B$ is obtained. We use these limits to constrain the combination of $A_\star \Gamma_i^{4}$. They are presented in Table \ref{tab:outflow_parameters} and Fig. \ref{fig:Gamma-A_star}. In the figure, the values of $\Gamma_i$ and $A_{\star}$ are marked by lines, each correspond to a different GRB. These are, from top to bottom: GRBs 080607, 110213A, 060714, 080310, 061121, 060605, 130831A, 091029, 050319, 060729, 060614, 100418A and 171205A. While directly deduced values are marked by squares, upper and lower values are marked by arrows.

\section*{Discussion} 
\label{sec:Discussion}

A histogram of the initial Lorentz factor $\Gamma_i$ for the 13
GRBs in our sample is shown in Fig. \ref{fig:hist-Gamma}. The average
value of the Lorentz factor deduced is $\langle \Gamma_i \rangle \approx 51$
(the median is 32), although the range span is between
$~ 2 \lesssim \Gamma_i \le 218$. These values may initially seem at odds
with the typical values discussed in the literature, of $\Gamma_i \gtrsim 100$
of GRB jets. However, a closer look reveals that in fact there is no
contradiction. 


There are several ways of inferring the value of the Lorentz factor in GRB
jets. The most widely used method is the opacity argument
\cite{KP91, Fenimore93}, which is commonly used in deducing that
observed GeV photons by \textit{Fermi}-Large Area Telescope (LAT)) must originate from a
region expanding with a Lorentz factor $\Gamma_i \gtrsim 100's$.
This argument, though,
is only valid when $>$~MeV photons are observed. 
A previous search \cite{Dainotti21} of GRBs observed from 2008 until May
2016 by \textit{Fermi}-LAT that appeared in the $2^{nd}$ catalogue
\cite{Ajello19} and are (1) fitted with a broken power-law, (2) have the Test
Statistic, $TS>64$ and (3) have known redshift (in total 13 GRBs) demonstrates that
although 3 GRBs out of the thirteen show evidence for a shallow decay phase in the
LAT data, only one (the hard-short GRB 090510) show any evidence for a
decaying plateau in the \textit{Swift}-XRT data. For this specific burst,
the shallowest decay segment in its X-ray afterglow light-curve can
only marginally be associated to a plateau, having an X-ray slope of
$-0.69_{-0.06}^{+0.05}$, to be compared to the $-0.7$ limit used in
this study. These arguments are consistent with earlier findings by
Ref.\cite{Yamazaki20} for 23 GRBs triggered by \textit{Swift}-BAT
and subsequently detected by \textit{Fermi}-LAT \cite{Ajello19}.

The second method relies on identifying an early optical flash and
interpreting it as originating from the reverse shock \cite{MR97, ZKM03, Racusin11}.
Since the reverse shock exists during the transition from the coasting to the
decaying (self-similar) phase, identifying its emission constrains the transition
time, from which, assuming the energy and ambient density are known, the
initial Lorentz factor can be deduced. However, a clear signature of a
reverse shock emission is nearly never identified \cite{Molinari07, Liang10}
as opposed to flares common to both the X-ray and optical data
\cite{Burrows05, Chincarini07, Falcone07, Chincarini10} and does not
exist in any of the bursts in our sample. 

When a strong thermal component exists during the prompt phase, it is possible
to use it to infer the Lorentz factor at the initial phase of the expansion
\cite{Peer07}. We therefore searched (i) all bursts with known strong
thermal component as appeared in Refs. \cite{YDR19, Acuner20, DPR20}. None
of those showed any evidence for an X-ray plateau. (ii) Similarly,
none of the bursts in our sample show any evidence for a thermal emission.

To conclude, we find that in all cases where there is any evidence for an 
initial Lorentz factor $\Gamma_i \gtrsim$ a few hundreds, no X-ray plateau
exists, and vice versa: for all bursts that show a plateau, no significant
indication for high Lorentz factor exist, neither at high energy, thermal
or optical photons. We further emphasis that GRBs with plateau phase which
have very low Lorentz factor (namely, GRBs in class II) lack any evidence
of MeV emission, implying that the opacity argument cannot be used at all
in these bursts.


One may argue that such a difference in the Lorentz factor cannot
only manifest itself in the afterglow phase, but should be manifested
in the prompt emission spectra as well. To further test this hypothesis,
we therefore compared the peak energy of the 13 bursts in our sample
to a reference sample of selected 12 GRB without plateau phase, as presented
in table 6 of Ref. \cite{Liang10}. In that work, the authors
estimated the Lorentz factor of these bursts using X-ray onset bump
or early peak in the optical data and found that the Lorentz factor
is of the order of few hundreds in all cases. We
therefore concluded that this is a good reference
for comparing the distribution of observed peak energies, $E_{\rm pk}$. In order to
ensure consistency, we did not use the values of $E_{\rm pk}$ as given in
Ref. \cite{Liang10} (measured from different instruments e.g. Konus-WIND).
Rather, we calculated $E_{\rm pk}$ directly from the \textit{Swift}-BAT data.
This ensures that there are no biases between the samples.
In the calculation, we used the correlation between the peak energy
and spectral index derived from fitting a single power-law to a large
\textit{Fermi}-GBM and \textit{Swift}-BAT data as parameterized by Ref. \cite{Virgili2012} as such
a method is commonly used in the literature \cite{Zhang07, Racusin09}. The method
gives consistent results ($E_{\rm pk}$) with the deduced value from the
\textit{Fermi}-GBM data.
In Fig. \ref{fig:logEpeak_13GRBs_with_plateau_12GRBs_without_plateau}, we
compare the peak energy distribution of these two samples. A clear separation
is found: those GRBs which have a higher Lorentz factor indeed have a
consistently higher $E_{\rm pk}$ than the GRBs in our sample. In addition to
the clear differences between the peak energy distribution of “plateau” and
without plateau, we also found clear differences between the high energy
spectral indices (presented in the \textit{Fermi}-GBM GRB catalog
\cite{vonKienlin2020}) of bursts in these samples. 


Furthermore, it is known that short GRBs have, on the average, higher
$E_{\rm pk}$ than long GRBs \cite{DPR20}, while it was argued by Ref.
\cite{Srinivasaragavan20} that 43/222 short GRBs do show a plateau.
However, we point out that the criteria used by Ref. \cite{Srinivasaragavan20}
for a plateau, namely a break in the X-ray light curve, is much less
restrictive than the one used here (namely, X-ray temporal
index $>-0.7$). Using the criteria in this work, none of the short GRB light
curves considered by Ref. \cite{Srinivasaragavan20} would be classified
as having a flat X-ray light curve (considered as class II, with low
Lorentz factor of $\gtrsim$~few), or was observed in the MeV range by \textit{Fermi}-GBM
\cite{vonKienlin2020}. Therefore, the compactness argument does not
apply for these bursts. 


The results we have here, therefore, complement and extend the known range
of Lorentz factors in GRBs. The values of $A_{\star}$
we find are typically up
to 2 orders of magnitude lower than the fiducial value of $A_\star = 1$
(pending on the exact value of the magnetization parameter, $\epsilon_B$;
see Fig. \ref{fig:Gamma-A_star}). We therefore conclude that the expansion
occurs into a low-density “wind”, having density which may be somewhat lower
than the expectation from a Wolf-Rayet star ($A_\star \simeq 0.5-1.0$)
\cite{Chevalier04, Eldridge06}. This result therefore implies that either
Wolf-Rayet stars are not the progenitors of GRBs with ”plateau”, or that the
properties of the wind ejected by the star prior to its final collapse are
different than in earlier stages of its life.

Indeed, we cannot consider this as an evidence against Wolf-Rayet progenitor
stars, as very little is known about the final stages of the evolution of
the most massive stars (luminous blue variables and Wolf-Rayet stars), of
which some lead to an evolutionary channel which end up as GRBs. Rapid
evolutionary stages of such stars are expected during the last 10's of centuries
of their life, which will have profound affect on the circumstellar wind
profiles. Instabilities will cause elevation of the outer envelope potentially leading to occasional giant eruption events, with major mass ejections in
several consecutive periods. These mass ejections lead to circumstellar
nebulae and wind blown bubbles~\cite{Crowther2007, Toal2013}. Observations
of galactic Wolf-Rayet stars indicate shell structures and nebulae at 1-10
pc scales, and in some cases, reveals the existence of low density cavities within these nebulae
\cite{Toal2013}. We thus view one of the merits of this work as providing
further information that could potentially help understanding the nature
of these objects.

Clearly, the fact that a substantial fraction of GRBs have a Lorentz factor
of tens rather than hundreds bridges an important observational gap. Other
astronomical objects known to have jets such as X-ray binaries or
active-galactic nuclei (AGNs) have mildly relativistic jets, with
$\Gamma_i\lesssim 20$, while earlier estimates of the initial Lorentz
factor in GRB jets are in the hundreds. Our result, therefore implies
that the range of initial jet velocities that exist in nature does not
have a ‘gap’ in the range $\Gamma_i$ of tens, but
is rather continuous from mildly relativistic
to $\lesssim$~1000. This is shown by the histogram presented
in Fig. \ref{fig:hist-Gamma_BLac}.


Here, we consider a simple model in explaining the  X-ray plateau in GRB afterglows, which does not require any
modification of the classical GRB "fireball" model. Rather, we simply look
at a different region of the parameter space: we consider an outflow
having an initial Lorentz factor of the order of few tens, propagating into
a "wind" environment, with a typical density of up to two orders of magnitude
below the expectation from a wind produced by a Wolf-Rayet star. We follow a
similar idea that was proposed by Ref. \cite{ShM12}, but did not gain popularity,
as (i) the deduced values of the Lorentz factor are lower than the `fiducial'
values, $\Gamma_i \gtrsim 100$; and (ii) it was mistakenly claimed that this
model can only account for achromatic afterglow, and can therefore explain only
a sub-sample of the GRB population \cite{KZ15}. As we showed here, (i) there is
no contradiction in the deduced value of the Lorentz factor, and (ii) the claim
for an achromatic afterglow break is incorrect, as the optical and X-ray bands
are not necessarily in the same regimes. 

In our work, we considerably extended
this simple idea theoretically and thoroughly confronted it to observations. We show that whenever there
is enough data to perform a fit in both X-ray and optical bands, the break time
in between these two bands is compatible and data in both bands can be interpreted
within the single theoretical model presented here. We
further carried out a more careful analysis on a much larger data set, allowing
for a larger freedom (with more than a single break) and removal
of flares on both the X-ray and optical data. We extended the theory to include
all possible regimes. We showed that all observed light-curves can be explained by at least one of these regimes.
We then showed how the confrontation of our model to the data can
be used to infer the values of the density, Lorentz factor, magnetization and fraction
of energy carried by the electrons. {\bf Moreover, our
model provides several testable predictions about bursts with long "plateau".
Such bursts (i) are not expected to show high energy ($\gtrsim$~GeV) emission;
(ii) are not expected to show strong thermal component; and (iii) the typical
variability time during the prompt phase is expected to be long, of the order
of few seconds. Exact constraints can be put on a case-by-case basis, using
the equations we provide below (see, methods subsection \LinkTo{sec:Theoreticalmodel}).}
 
While the idea presented in this work is very simple, clearly it has
very far reaching consequences:
[a] On the nature of long GRB progenitors, which can either (i) not be a Wolf-Rayet star, or (ii) imply that the properties of the wind ejected by these stars prior to their final explosion is very different than the properties of the wind ejected at earlier times.
[b] On our understanding of the nature of the explosion itself, which produce a much wider range of initial jet Lorentz factor, which in many cases are in the range of tens and in others are in the hundreds.


\section*{Methods} 
\label{sec:Methods}

\subsection*{Sample Selection}
\CreateLink{sec:SampleSelection}{`Sample Selection'}

To study the properties of the plateau, we use the sample of 222 GRBs with known
redshifts and plateau phases defined in Ref.\cite{Srinivasaragavan20}.
These GRBs were detected by the Neil Gehrels \textit{Swift} Observatory from January
2005 until August 2019. They represent 56\% of all GRBs with known redshifts
observed by the \textit{Swift} satellite in this period. 


In order to make our analysis as reliable as possible, we limit the bursts used
to the ones having the best quality observations. Therefore, we added three criteria
to the ones used by Ref.\cite{Srinivasaragavan20}.
We require (i) a long lasting (from $10^2$ to $10^5$ s) plateau phase with a temporal
X-ray slope larger than $-0.7$, followed by a power-law decay phase at later times 
(interpreted here as the self-similar phase). (ii) Sufficient number of
data points ($\gtrsim$~5)
during the plateau and self-similar phases to enable the fits to give well constrained 
parameter (see, Supplementary Method \LinkTo{sec:SuppMethod1a}). For the analysis to be valid, we
excluded all X-ray flares from the analyzed data. We found 130 GRBs
matching these two criteria. (iii) Finally, we require an optical counterpart
at around the same time as the X-ray data. We searched the optical catalogue of
Ref.\cite{Dainotti20} and found that 24 GRBs in our sample 
have an optical counterpart. Out of these, we had full access to the
optical data of 13 GRBs, which are listed in Supplementary Tables \ref{tab:ClassI},
\ref{tab:ClassII}, \ref{tab:ClassIII}. The analysis of X-ray and optical
light curves of these 13 GRBs is detailed in
Supplementary Method~\LinkTo{sec:SuppMethod1}. 


\subsection*{Theoretical model} 
\CreateLink{sec:Theoreticalmodel}{`Theoretical model'}
The key to understanding the observations in the framework of our model is
the realization that the end of the plateau corresponds to the transition from a
coasting phase (steady state in which all the energy has been converted to kinetic
energy) to a self-similar expansion phase (decaying phase in which the kinetic
energy is being converted back to radiation by the shocks)
of the expanding plasma. In this model, the emission originates entirely from ambient electrons
collected and heated by the forward shock wave, propagating at relativistic speeds
inside a "wind" (decaying density) ambient medium. As we show
(see, Supplementary Method \LinkTo{sec:SuppMethod2}) this assumption
about the decay of the ambient density is crucial in explaining the observations.
During the transition from the coasting phase to the decaying
phase a reverse shock crosses the expanding plasma. However,
the contribution from electrons heated by the reverse shock is suppressed due to
(i) the declining ambient density, which implies that the ratio of plasma density
to ambient density remains constant (under the assumption of a
conical expansion), and (ii) its slower speed, which translates into less
energetic electrons that emit at much lower frequencies than forward shock heated
electrons, implying that the contribution to the optical and X-ray bands is negligible. 
A detailed derivation of the theoretical model is provided in
Supplementary Method~\LinkTo{sec:SuppMethod2}.

\section*{Data Availability}
The data used in this paper are publicly available. The processed
data that support the findings of this study are available from the
corresponding author upon reasonable request. The X-ray
and the optical light curves of 13 GRBs with the overlaid fit parameters
are presented in the Supplementary Information File as Supplementary Figures.
The source data for all figures are provided with this paper. The authors declare
that all other data supporting the findings of this study are available
within the paper and its supplementary information files.


\section*{Additional information}
Correspondence and requests for materials should be addressed to H\"usne Dereli-B\'egu\'e \\ (email: husnedereli@gmail.com) or Asaf Pe'er (email: asaf.peer@biu.ac.il).



\bibliography{Myreferences}




\section*{Acknowledgements}
We wish to thank Dr. Damien B\'egu\'e for enlightening conversations throughout the project as well as Dr. Mukesh Vyas and Dr. Filip Samuelsson for the discussion on the prompt properties of our sample. This work made use of data supplied by the UK Swift Science Data Centre at the University of Leicester. H.D.-B. and A.P. is supported by the European Research Council via ERC consolidating grant 773062 (acronym O.M.J.). F.R. is supported by the G\"oran Gustafsson Foundation for Research in Natural Sciences and Medicine. We acknowledge support from the Swedish National Space Agency (196/16), the Swedish Research Council (Vetenskapsr\aa det, 2018-03513), and the Swedish Foundation for international Cooperation in Research and Higher Education (STINT, IB2019-8160).


\section*{Author contributions statement}
H.D.-B., A.P. and F.R. wrote the manuscript. H.D.-B. has performed sample selection and temporal analysis of the data, theoretical calculations and interpretation. A.P. provided theoretical calculations, interpretation and insight. F.R. assists in the sample selection, interpretation and insight of the results. S.-R.O. provides the \textit{Swift}-UVOT count rate light curves and assists in the correction and conversion processes of the data. S.-R.O. also enlighten us with her questions and comments. B.Z. assists in the theoretical calculations, discussion and representation of the results. M.-G.D. provides the X-ray sample of 222 GRBs and the optical sample of 102 GRBs with all required parameters and assists in the discussion of those two samples. The parameters of those samples are used for initial discussion. M.-G.D. assists in the general discussion of the paper as well as the discussion of the Fermi-LAT paper with flat phase. All authors reviewed the manuscript.


\subsection*{Competing interests}
The authors declare no competing financial interests.


\begin{table}[ht!]
\centering
\small
\begin{tabular}{c|cc|cc|cc|cc}
\hline
GRB name & \multicolumn{2}{|c|}{\begin{tabular}{c}Regions \\ in  X-ray band\end{tabular}} &  \multicolumn{2}{|c|}{$p_{\rm X}$} &  \multicolumn{2}{|c|}{\begin{tabular}{c} Regions \\ in optical band\end{tabular}} & \multicolumn{2}{|c}{$p_{\rm U}$} \\
\hline
& Temporal& Spectral & Temporal& Spectral & Temporal & Spectral & Temporal & Spectral \\
\hline
 Listed in class I & & & & & & & & \\
\hline
080607 & F & F & $\sim$ 2.5 & 2.0 & E & E & 2.0 & $\sim$2.5\\
091029 &  F & F & 2.3 & 2.0 & F(E) & E & 2.4(1.8) & 1.9 \\
110213A & F & F & $\sim$ 2.0 & 2.0 & E & E & 2.1 & 3.2\\
130831A & F & F/E & 2.0 & 1.8/$\sim$2.4 & E & --- & 2.1 & ---\\
\hline
 Listed in class II & & & & & & & & \\
\hline
060605 & F & F & $\sim$ 2.1 & 2.3 & F & F &  $\sim$ 2.2 & $\sim$2.3\\ 
060614 & F & F & $\sim$ 2.3 & 1.8 & F & F &  2.0 &  0.7\\
060729 & F & F &  2.0 & 2.0 & F & F & 2.3 & $\sim$1.2\\
080310 & F & F & $\sim$ 2.5 & 2.0 & F & F &  2.0 & 1.9\\
100418A & F & F & 2.2 & $\sim$1.9 & F & F & $\sim$ 1.8 & $\sim$2.3\\
171205A & F & F/E & 2.0 & $\sim$1.8/2.4 & F & F & $\sim$ 1.8 & $\sim$1.8\\
\hline
 Listed in class III & & & & & & & & \\
\hline
050319 & E & F/E & 2.1 & 2.1/3.2 & E & E & $\sim$ 1.8 - 2.1 & 2.0\\
060714 & F/E & F/E & $\sim$ 2.4/2.0 & 2.0/2.8 & F/E & E & 2.0/ $\sim$ 1.8 & $\sim$ 2.4\\
061121 & F/E & F/E & 2.5/$\sim$ 2.1 & 2.0/$\sim$3.2 &  E & E &  2.0 & 2.3\\
\hline
\end{tabular}
\caption{\label{tab:ElectronPowerlawindex} \textbf{Region and electron power-law
index ($p$) in both X-ray and optical bands using both temporal and spectral
indices.} In Column 1, GRB names are ordered by classes (I, II, III listed
in Supplementary Tables \ref{tab:ClassI}, \ref{tab:ClassII},
\ref{tab:ClassIII} respectively).
In columns 2-5, we use the temporal and spectral X-ray data to determine both
the electron power law index ($p_X$) and the region (E or F) characteristics of the emission, see Supplementary Fig. \ref{fig:summary_regions_plateau}. The
optical data is used in a similar way in columns 6-9. For GRBs 060614, 060729
and 110213A, the power-law indices obtained using X-ray and optical temporal
indices as well as the X-ray spectral index are all consistent with each other,
while the values derived using the optical spectral data deviate. Since no errors
are given in the literature for GRBs 060614 and 060729 (see Supplementary Table \ref{tab:ClassII}),
we cannot estimate the reliability of the optical spectral indices for these bursts,
and we therefore accept the power law obtained using 6 out of 8 independent
measurements. For GRB 110213A, there is no available spectral index in the
optical band during the plateau phase, and the second peak in the  optical
band after the plateau phase (see, Supplementary Method \LinkTo{sec:SuppMethod1b}) might be affecting the spectral index
during the afterglow phase (see Supplementary Table \ref{tab:ClassI}). In all three bursts
which we categorize as being in class III, there seem to be a discrepancy
between the X-ray spectral and temporal data: while the temporal data
clearly indicates the X-ray to be in region E the spectral data favours
region F. However, this is because, in all three bursts there is a break
in the X-ray light curve during the plateau phase, or an early flare.
These may indicate a change in region, and may affect the spectral measurement. }
\end{table}
~

\begin{table}[ht!]
\centering
\tiny
\begin{tabular}{ccccccccccccc}
\hline
GRB name & z & ${d_L}$ & S (15-150~keV) & $\alpha_{\gamma}$ & $\log({E_{\rm iso}/\rm erg})$ &$T_{\rm 90}$ & ${T_{\rm a,X}}$ & $\nu F_{\nu} ({\rm X})$ & $\nu F_{\nu}({\rm U})$ & ${T_{\rm ref.,U}}$ &  $\nu F_{\nu}({\rm X})$ & $\nu F_{\nu}({\rm U})$\\
 &  & (Mpc) &  (${10^{-7}~\rm erg~cm^{-2}}$) & &  & ($s$) & ($10^3$ s) & ($ 10^{-12}{\rm~erg}$ & (${ 10^{-12}~\rm erg~}$ & (s) &  (${10^{-12}~\rm erg~}$ & ($ 10^{-12}{\rm~erg}$\\
&  & & & & & & & ${\rm cm}^{-2}{\rm~s}^{-1}$) & ${\rm cm}^{-2}{\rm~s}^{-1}$) & &  ${\rm cm}^{-2}{\rm~s}^{-1}$) & ${\rm ~cm}^{-2}{\rm~s}^{-1}$)\\
\hline
Listed in class I & & & & & & & & & & & & \\
\hline
080607 & 3.036 & 26150 &	240$\pm$0.0 & 1.31$\pm$0.04&  53.27$\pm$0.02 &79.0 & $2.23^{+0.32}_{-0.27}$ & 56$\pm 12.6$ & 0.13$\pm 0.01 $ & 1010 & 82.7$\pm 18.5$ & 0.18$\pm 0.01$ \\ 
091029 & 2.752 & 23222 &	24$\pm$1.0 & 1.46$\pm$0.27&  52.31$\pm$0.16 & 39.2 & $14.1^{+1.59}_{-3.37}$ & 1.34$\pm 0.23$ & 0.117$\pm0.029$ & 1170 & 2.8$\pm 0.6$ & 0.3 $\pm 0.1$	\\	
110213A & 1.46 & 10662 & 59$\pm$4.0 & 1.83$\pm$0.12&  52.45$\pm$0.06 & 48.0 & $1.35^{+0.14}_{-0.18}$ & 350$\pm 77$ & 8.45$\pm 0.33$ & 1130 & 218$\pm 48$ & 7.7$\pm0.4 $ \\
130831A & 0.4791 & 2704 &	65$\pm$0.2 & 1.93$\pm$0.05 &  51.57 $\pm$0.01  & 32.5 & $0.75^{+0.08}_{- 0.07}$ & 259$\pm 56$ & 46.2$\pm 2.8$ & 732 & 259$\pm 56$ & 46.2$\pm 2.8 $	\\	
\hline
Listed in class II & & & & & & & & & & & & \\
\hline
060605 & 3.78 & 34014 & 7.0$\pm$0.9 & 1.55$\pm$0.20 &  52.00$\pm$0.15  & 79.1 &  $4.88^{+1.24}_{-1.85}$ & 2.0$\pm 0.45$ & 1.12$\pm 0.114$ & 534 & 16.3$\pm 3.3$ & 9.51$\pm 0.42$	\\
060614 & 0.125 & 586 & 204$\pm$3.6	& 2.02$\pm$0.04 &  50.87$\pm$0.01 & 108.7& $34.1^{+2.28}_{-2.64}$ & 3.28$\pm 0.63$ & 0.626$\pm 0.143$ & 4838 & 2.1$\pm 0.5$ & 0.50$\pm 0.12$ \\
060729 & 0.54 & 3124 & 26$\pm$2.1 & 1.75$\pm$0.14&  51.25$\pm$0.04 & 115.3 & $38.1^{+3.35}_{-2.11}$ & 6.35$\pm 1.43$ & 3.34$\pm 0.75$ & 1160 & 10.6$\pm2.4$ & 4.13$\pm 0.94$ \\	
080310 & 2.42 & 19862 & 23$\pm$2.0 & 2.32$\pm$0.16&  52.67$\pm$0.09 & 365.0 & $10.9^{+ 0.94}_{-0.88}$ & 3.41$\pm 0.76$ & 0.258$\pm 0.059$ & 1505 & 4.9$\pm 0.9$ & 1.41$\pm 0.32$ \\	
100418A & 0.6235 & 3721 & 3.4$\pm$0.5 & 2.16$\pm$0.25&  50.57$\pm$0.08 & 7.0 &  $79.3^{+20.6}_{-12.8}$ & 0.86$\pm 0.22$ & 0.258$\pm 0.005$ & 1000 & 0.15$\pm 0.04$ & 0.06$\pm 0.01$ \\	
171205A & 0.0368 & 162 &	36$\pm$3.0 &	1.41$\pm$0.14&  49.03$\pm$0.04 & 189.4 & $91.0^{+8.60}_{-8.45}$ & 1.02$\pm 0.24$ & 0.568$\pm 0.068$ & 10834 & 0.61$\pm 0.16$ & 1.44$\pm $ 0.09 \\
\hline
Listed in class III & & & & & & & & & & & & \\
\hline
050319 & 3.24 & 28280 &13.1$\pm$1.5 & 2.02$\pm$0.19 & 52.48$\pm$0.13 & 152.5 & $32.0^{+4.36}_{-4.27}$ & 1.24$\pm 0.28$ & 0.144$\pm 0.026$ & 1120 & 5.02$\pm 1.13$ & 1.20$\pm 0.17$ \\
060714 & 2.711 & 22803 & 28.3$\pm$1.7 &	1.93$\pm$0.11 &  52.64$\pm$0.10  & 115.0 &  $4.54^{+1.38}_{-1.01}$ & 13.3$\pm 2.92$ & 0.180$\pm 0.053$ & 1069 & 13.6$\pm 3.07$ & 0.31$\pm 0.08$ \\
061121 & 1.314 & 9355 & 137$\pm$2.0 & 1.41$\pm$0.03&  52.58$\pm$0.01  & 81.3 & $9.11^{+0.53}_{-0.59}$ & 61.5$\pm 12.3$ & 0.916$\pm 0.250$ & 1173 & 67$\pm 15$ & 1.66$\pm 0.40$ \\	
\hline
\end{tabular}
\caption{\label{tab:key_parameters} \textbf{Some key parameters of the 13 GRBs in our sample.} Columns 1 --- 13 are the GRB name, redshift, luminosity distance, fluence, photon spectral index, isotropic equivalent energy in log scale, burst duration, time at the end of the X-ray plateau phase, $\nu F_{\nu}$ X-ray flux and $\nu F_{\nu}$ optical flux at ${T_{\rm a,X}}$, a reference time in optical band at around 1000 s, $\nu F_{\nu}$ X-ray flux and $\nu F_{\nu}$ optical flux at ${T_{\rm ref.,\rm U}}$ respectively. The errors correspond to a significance of one sigma. Note that the fluence of GRB 080607 has no error in the online \textit{Swift} GRB table. See, Supplementary Method \LinkTo{sec:SuppMethod1c} for the definition of each parameters.}
\end{table}
~

\begin{table}[ht!]
\centering
\tiny
\begin{tabular}{|c|c|cc|c|cc|c|cc|c|}
\hline
 & & \multicolumn{2}{c|}{$\epsilon_{B} = 10^{-2}$} & &\multicolumn{2}{c|}{$\epsilon_{B} = 10^{-3}$} & & \multicolumn{2}{c|}{$\epsilon_{B} = 10^{-5}$ }  &\\
 \hline
 GRB name & $\epsilon_{e}$ & $\Gamma_i$ & A$_{\star}$  & & $\Gamma_i$ & A$_{\star}$  & & $\Gamma_i$ & A$_{\star}$ &  \begin{tabular}{c}Ratio \\ $\nu F_{\nu}({\rm X})/\nu F_{\nu}({\rm U})$ \end{tabular}\\ 
\hline 
 Listed in class I &direct value & direct value &direct value & & direct value &direct value & & direct value &direct value  & \\
\hline
080607 & 1.2$\times 10^{-2}$ & 387 & $10^{-4}$ & & 218 & $ 10^{-3}$ & & 122 & $ 10^{-2}$  & 314$\pm$ 74  \\
091029 & 1.4$\times 10^{-2}$ & 52  & 5$\times 10^{-3}$ & & 44 &  $ 10^{-2}$ & & 14 & 1.0  & 4.5$\pm$1.4 \\
110213A & 8.3$\times 10^{-2}$  & 136 & $10^{-3}$ & & 91 &  5$\times 10^{-3}$ & & 43 & 0.1  & 45.4$\pm$2.2  \\
130831A & 2.7$\times 10^{-2}$ & 56 & 5$\times 10^{-3}$ &  & 32 & 5$\times 10^{-2}$ & & 15 & 1.0  & 5.6$\pm$1.3  \\
\hhline{|=|=|==|=|==|=|==|=|}  
Listed in class II & direct value & upper limit & lower limit & & upper limit & lower limit &  &  &  &  \\
\hline
060605 & 2.5$\times 10^{-2}$ & 51 & $ 10^{-2}$ & & 28 & 0.1 &   &  & &  0.21$\pm$0.05  \\
060614 & 4.9$\times 10^{-3}$ & 6 & 0.1 & & 4 & 1.0 & &  &  &  6.6$\pm$2.0 \\
060729 & 9.2$\times 10^{-2}$ & 8 & 0.1 & & 5 & 1.0 & & &  &  1.5$\pm$0.5  \\
080310 & $10^{-2}$ & 37 & 5$\times 10^{-2}$ & & 32 & 0.1 & &  & &  2.4$\pm$0.8  \\
100418A & 1.7$\times 10^{-1}$ & 6 & 5$\times 10^{-2}$ & & 5 & 0.1 & & & &  13.7$\pm$4.6  \\
171205A & 2.3$\times 10^{-2}$ & 2 & 5$\times 10^{-2}$ & &1.7 & 0.1 & & & &  0.7$\pm$0.2  \\
\hhline{|=|=|==|=|==|=|==|=|}  
  & \multicolumn{3}{c|}{ $\epsilon_{B} = 10^{-2}$} & \multicolumn{3}{c|}{$\epsilon_{B} = 10^{-3}$} & \multicolumn{3}{c|}{$\epsilon_{B} = 10^{-5}$ }  & \\
\hline
& $\epsilon_{e}$ & $\Gamma_i$ & A$_{\star}$  & $\epsilon_{e}$ &$\Gamma_i$ & A$_{\star}$   & $\epsilon_{e}$ & $\Gamma_i$ & A$_{\star}$ & \begin{tabular}{c}Ratio \\ $\nu F_{\nu}({\rm X})/\nu F_{\nu}({\rm U})$ \end{tabular}\\ 
\hline
 Listed in class III & lower limit & lower limit & upper limit&  & lower limit & upper limit & lower limit & lower limit & upper limit &   \\
050319 & 3.1 & 97 & 3$\times 10^{-4}$ & 1.6$\times 10^{-2}$& 61 & 2$\times 10^{-3}$ & 2.1$\times 10^{-6}$  & 27 & 5$\times 10^{-2}$ &  1.03$\pm$0.27  \\
060714 & 8.0$\times 10^{-2}$ & 147 & 5$\times 10^{-4}$ & 5.5$\times 10^{-4}$ & 94 & 3$\times 10^{-3}$& 3.3$\times 10^{-8}$  & 39 & 0.1 &  43.6$\pm$14.8  \\
061121 & 2.8$\times 10^{-1}$ & 106 & 5$\times 10^{-4}$ & 1.9$\times 10^{-3}$ & 70 & 3$\times 10^{-3}$& 1.1$\times 10^{-7}$  & 28 & 0.1 &  37.1$\pm$11.6 \\
\hline
\end{tabular}
\caption{\label{tab:outflow_parameters}  \textbf{
Model parameters of the outflow and wind.} $\epsilon_e$ is
the fraction of energy in the electrons, $A_\star$ is the wind density, $\Gamma_i$
is the initial jet Lorentz factor. Direct value of $\epsilon_e$ is computed by
using the information in the X-ray data for the GRBs listed in class I and II
respectively. The values obtained (using the end of plateau time and X-ray flux
in Supplementary Equation (\ref{eq:ee_final})) are
surprisingly close to the fiducial values, $\epsilon_e \simeq 10^{-1}$
often obtained by fitting late time afterglow data.
In addition, direct values of $A_\star$ and $\Gamma_i$ are obtained by assuming
that the fraction of energy in the magnetic field is
$\epsilon_{B} = 10^{-2}$, $10^{-3}$, $10^{-5}$ for the GRBs listed in class I.
Moreover, an (external) upper limit on $\epsilon_B$, (e.g.,  $\epsilon_{B} \leq 10^{-2}$)
is used to compute an upper limit on $\Gamma_i$ and lower limit on $A_\star$ for the
GRBs listed in class II. Vice-versa, an external knowledge on lower limit
(e.g.,  $\epsilon_{B} \geq 10^{-5}$) can be used to compute a lower limit on the
value of $\Gamma_i$ and an upper limit on $A_\star$ as well as a lower
limit on $\epsilon_e$ for the GRBs listed in class III (see,
Supplementary Method \LinkTo{sec:SuppMethod2}). 
The ratio [$1/(\nu F_{\nu} (\rm U)/\nu F_{\nu} (\rm X))$] 
is such that $\nu F_{\nu} (\rm U)$ is calculated at
$T_{\rm ref.,U}$ and $\nu F_{\nu} (\rm X)$ is calculated at $T_{\rm a,X}$ (see Table
\ref{tab:key_parameters}). The errors correspond to a significance of one sigma. The ratios are consistent with the theoretical
predictions in all three different classes. For the low luminous GRBs with
the lowest Lorentz factor (GRBs 060614, 060729, 100418A and 171205A)
in class II, when considering $\epsilon_B = 0.1$, the obtained values are
$\Gamma_i = 11, 15, 9, 4$ and $A_{\star} = 10^{-2}, 10^{-2}, 10^{-2}, 5 \times 10^{-3}$, respectively.}
\end{table}

\begin{figure}[ht!]
\centering
\includegraphics[width=\linewidth]{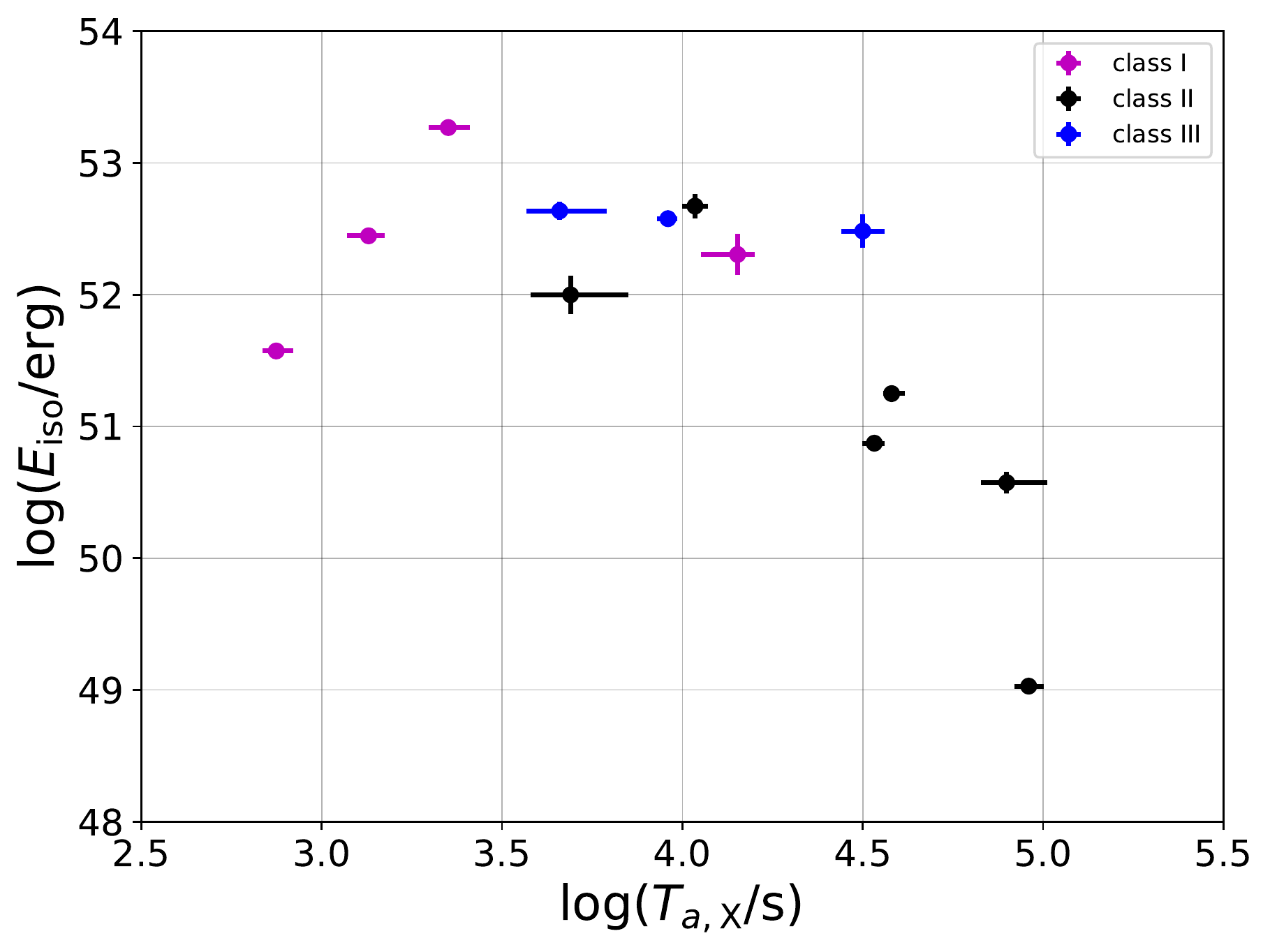}
\caption{  \textbf{ Total Isotropic Energy, ${E_{\rm iso}}$ as a function
of the time at the end of the plateau phase, ${T_{\rm a,X}}$ in the X-ray band.} Purple, black and blue points
represent GRBs in the three different classes I, II, III respectively (see
Supplementary Tables \ref{tab:ClassI}, \ref{tab:ClassII}, \ref{tab:ClassIII}). The errors correspond to a significance of one sigma. Note that
these three classes, selected only based on the temporal indices of their
X-ray and optical afterglow light curves, are correlated with the prompt
phase energy as well as the break time: each class occupies a different region in the ${E_{\rm iso}}$ – ${T_{\rm a,X}}$ parameter space.
This fact provides a further, independent tool that increases our confidence
in the selection criteria we use for these classes. The source data to reproduce this figure are
provided as a Source Data file.} 
\label{fig:EisoandTa}
\end{figure}
~

\begin{figure}[ht!]
\centering
\begin{tabular}{cc}
\includegraphics[width=\linewidth]{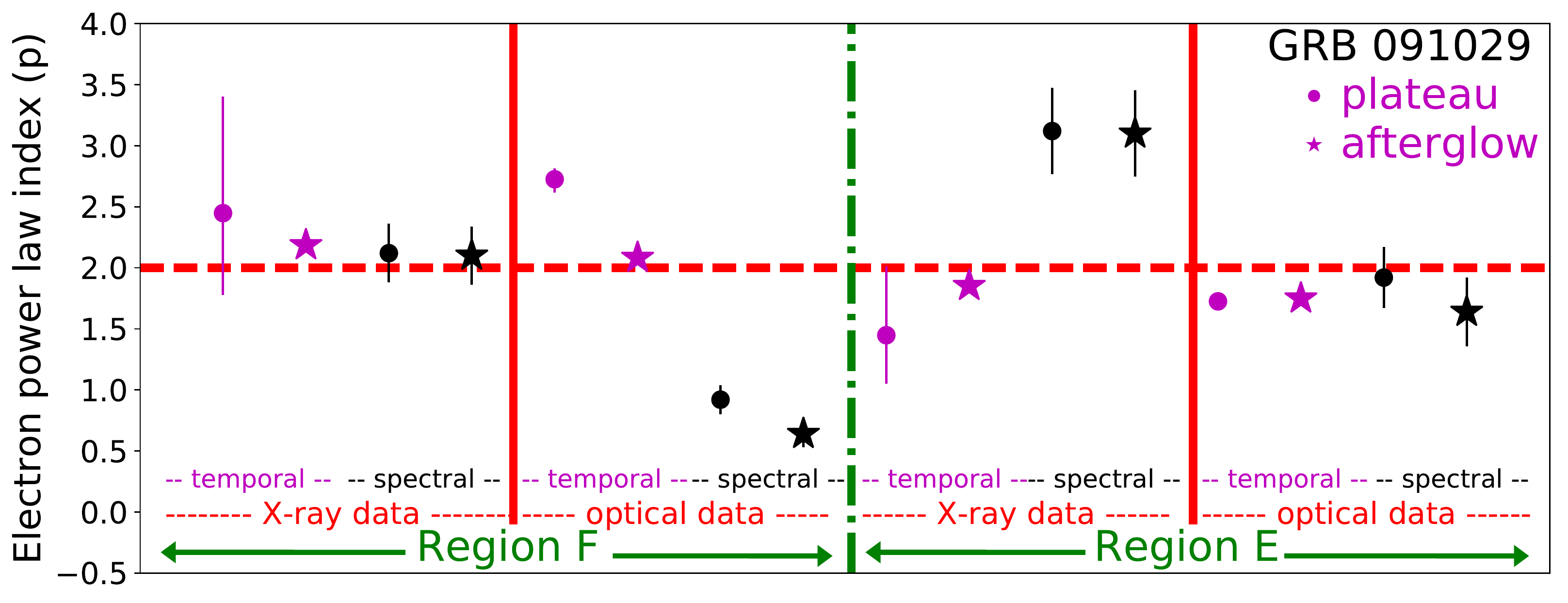}
\end{tabular}
\caption{\textbf{ An example demonstrating how we determine the relevant region for a given source as well as the electron power-law index ($p$) shown in Table \ref{tab:ElectronPowerlawindex}.}
For each burst (in the cases presented here, GRB 091029), we consider
eight independent indicators, namely the temporal (purple) and spectral
(black) slopes of both the X-ray and optical data during both the
plateau (dot) and self-similar (star) phases. These eight independent
measurements are inserted into the theoretical predictions given in
Supplementary Table \ref{tab:summary} to calculate the values of the electron
power-law index, under the assumption that the data is in region F
(left side) or E (right side). The obtained eight independent measurements
(for each region, F and E), are then displayed side
by side. The vertical green dash-dot line separates the analysis carried
under the assumption that the emission is in region F (left) and E (right),
the vertical red lines separates X-ray and optical data and the dashed
horizontal line marks $p = 2$. The errors correspond to a significance of one sigma. For X-ray
data, the independent calculations converge in region F to a single value
of $p \simeq 2.15$ while the assumption that the outflow is in region E
leads to a diverging result. Similarly, for optical data, the assumption
that the emission is in region F leads to a divergence, while the assumption
that it is in region E provides a
consistent value of $p\simeq 1.85$. We therefore conclude
that the X-ray emission of GRB 091029 is in region F, and the optical emission
is in region E. Therefore, this burst is classified as being in class I.
The source data necessary to reproduce this Figure are provided as a
Source Data file.
}\label{fig:ElectronPowerlawindex-091029}
\end{figure}

~

\begin{figure}[ht!]
\centering
\includegraphics[width=\linewidth]{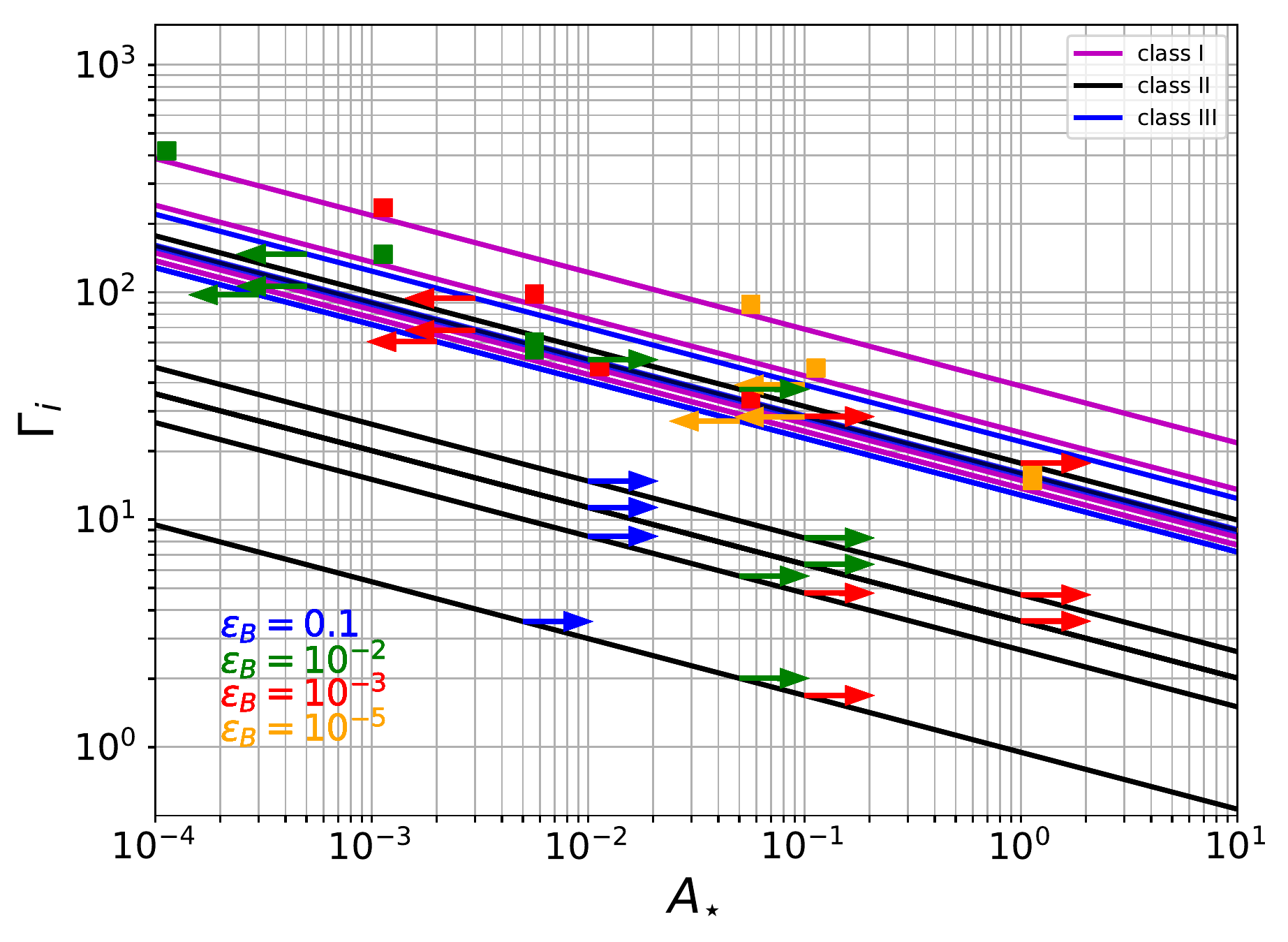}
\caption{\textbf{ The initial GRB jet Lorentz factor, $\Gamma_i$ and the
ambient density, $A_{\star}$.} They are marked by lines, each corresponds to a
different GRB. From top to bottom: GRBs 080607, 110213A, 060714, 080310,
061121, 060605, 130831A, 091029, 050319, 060729, 060614, 100418A, and
171205A. Purple, black and blue lines represent the GRBs in classes I, II
and III respectively. For a given $A_{\star}$, we determine the value of
$\Gamma_i$ by using Supplementary Equation (\ref{eq:6.4})
and knowing the burst energy and transition time $T_{\rm a,X}$.
This gives the lines. In order to further constrain the values of these parameters,
we assume knowledge of magnetization, $\epsilon_B$, and use Supplementary Equation
(\ref{eq:eB_classI}) to deduce direct values of $A_{\star}$ and $\Gamma_i$
(squares) for class I. For classes II and III, we instead use Supplementary
Equations (\ref{eq:6.10}) and (\ref{eq:6.13}) to compute the lower (upper) and
upper (lower) limits of $A_{\star}$ ($\Gamma_i$) respectively.
These limites are represented by arrows. In all classes, the
constraint put by the magnetization ($\epsilon_B$) is inversely proportional
to the ambient density. We mark on the plot the values obtained
for $\epsilon_{B} = 0.1, 10^{-2}$, $10^{-3}$ and $10^{-5}$ which are
associated to the blue, green, red and orange colors,
respectively. The lowest value of Lorentz factor (4, blue arrow) is obtained
for GRB 171205A. Due to its low luminosity, $L_{\rm iso} = 5.6
\times 10^{46}$~erg/s, the prompt emission radius is above the photospheric
radius, as shown in Fig. \ref{fig:ratio_of_radii_Eiso}. In addition,
this GRB is found to exhibit a black-body emission with a low temperature
in the X-ray spectra, later on this component cooling into the UV and
optical range over time \protect\cite{Valan2021, Izzo2019}. The source data
to reproduce this figure are provided as a Source Data file. }
\label{fig:Gamma-A_star}
\end{figure}
~

\begin{figure}[ht!]
\centering
\includegraphics[width=\linewidth]{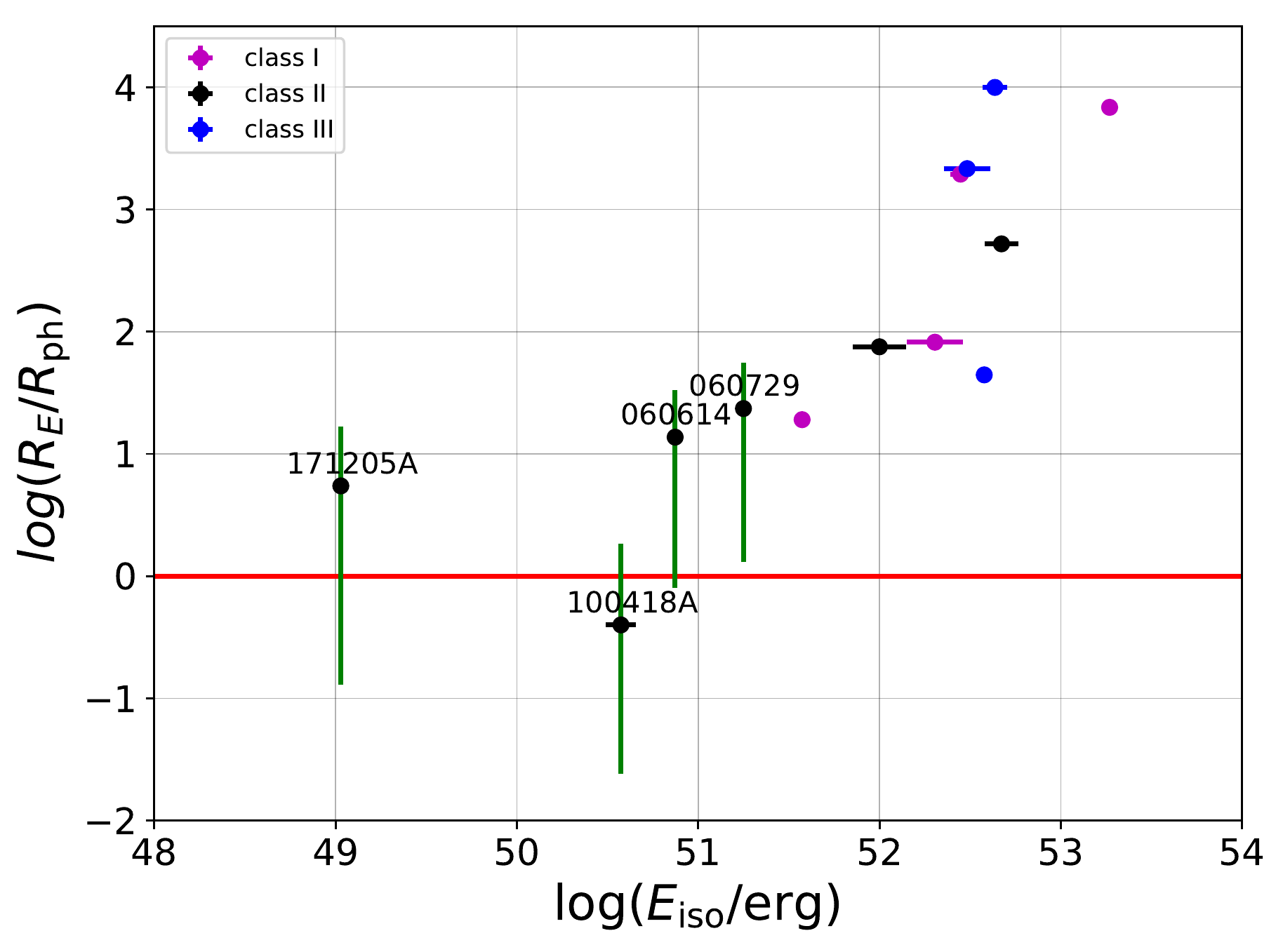}
\caption{\textbf{ Ratio of the prompt emission radius, $R_{\rm E}$ and
the photospheric radius, $R_{\rm ph}$ versus total isotropic energy
$E_{\rm iso}$.} Purple, black and blue points represent GRBs in the
three different classes I, II, III respectively (see Supplementary Tables \ref{tab:ClassI},
\ref{tab:ClassII}, \ref{tab:ClassIII}). The errors correspond to a significance of one sigma.
The low luminous GRBs with lowest Lorentz factor (GRBs 060614, 060729,
100418A and 171205A) are marked by their names. For these GRBs we consider
$\epsilon_B = 0.1$, leading to typical initial jet Lorentz factor $\Gamma_i \approx 10$
and ratio $R_{\rm E}/R_{\rm ph} \sim 5-25$, 
but for GRB 100418A, which is marginally consistent with $R_{\rm E}/ R_{\rm ph} \sim 1$.
For all other GRBs, we assume $\epsilon_B = 10^{-3}$ when making this
plot, and a variability time $\Delta t_{\min}$  taken from
Refs. \protect\cite{Golkhou2014, Sonbas2015}, except for GRB 060714, where
$\Delta t=5$ s is assumed based on data from the {\it Swift}-BAT
light-curve. We point out that a higher value of $\epsilon_B$ increases this ratio.
The vertical green lines associated to each GRB illustrate the possible
ratios of $R_{\rm E}/R_{\rm ph}$ obtained for magnetization in the range
$10^{-3} \leq \epsilon_B \leq 0.7$. The horizontal red line indicates
$R_{\rm E}/R_{\rm ph} = 1$. It shows that in all cases, sufficiently high
value of $\epsilon_B$ within the examined range leads to
$R_{\rm E}>R_{\rm ph}$. The source data to reproduce this figure are
provided as a Source Data file.} 
\label{fig:ratio_of_radii_Eiso}
\end{figure}
~

\begin{figure}[ht!]
\centering
\includegraphics[width=\linewidth]{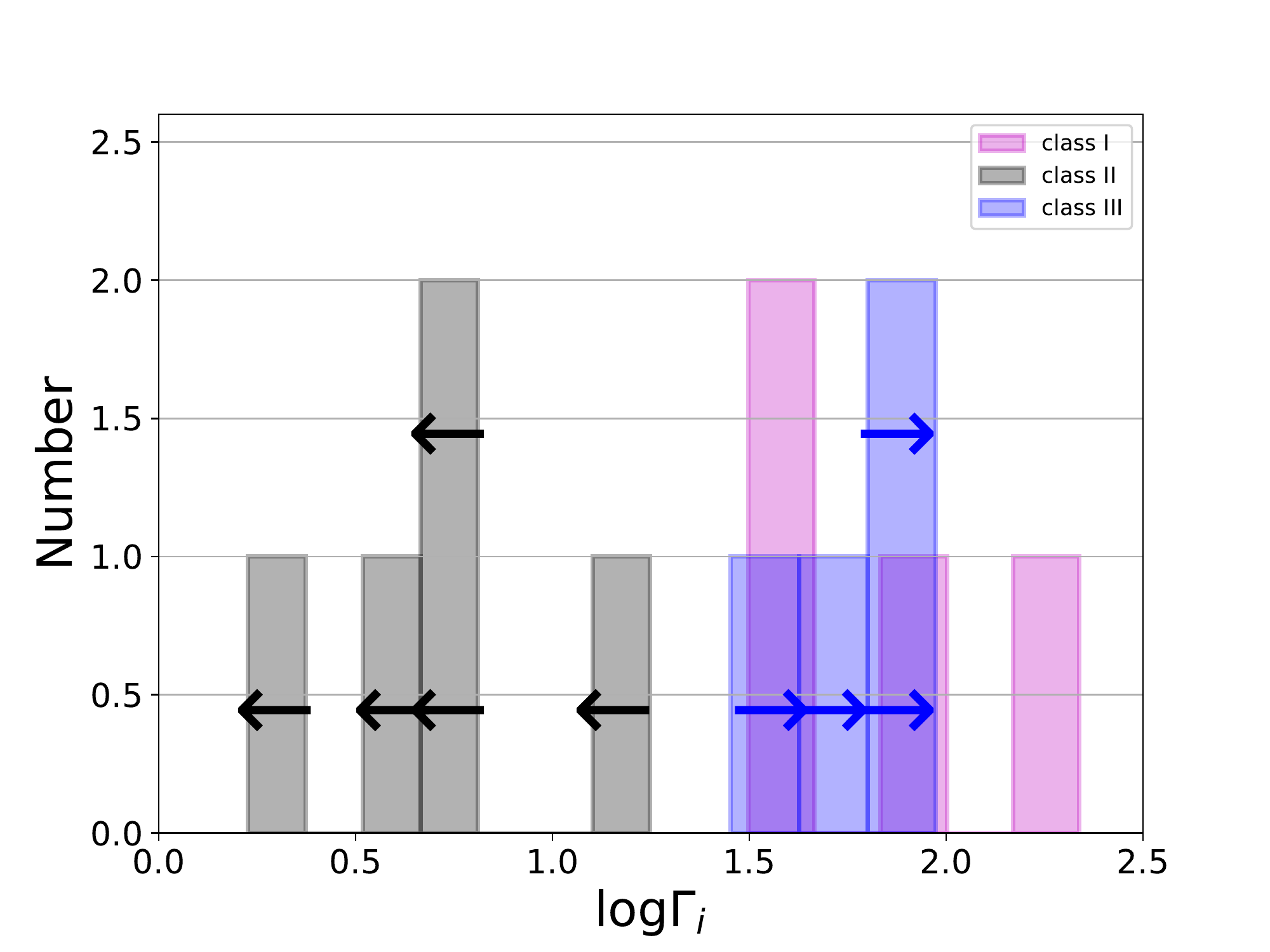}
\caption{\textbf{ Histogram of the initial jet Lorentz factors for
the 13 GRBs in our sample.} These values are obtained by assuming that
the fraction of energy in the magnetic field is 
$\epsilon_{B} = 10^{-3}$. The purple bars represent values deduced directly
from the data (class~I), the black bars are upper limits (class II) and the
blue bars are lower limits (class~III). Upper and lower limits are also marked
by arrows. The average value of the initial GRB jet Lorentz factor is
$\langle \Gamma_i \rangle \approx 51$ (median is 32), although the range span
is between $~ 1.7 \lesssim \Gamma_i \le 218$ (see Table \ref{tab:outflow_parameters}).
We point out that GRB 080607 which has the highest value of $\Gamma_i$ has a
large gap in its X-ray LC between the plateau and self-similar phases.
Furthermore, GRB 171205A which has the lowest value of $\Gamma_i$ is
associated with SN 2017iuk, therefore, both the optical plateau and
self-similar slopes of this burst are effected by the SN bump (see, Supplementary Method \LinkTo{sec:SuppMethod1b} for further discussion). The source data to reproduce this figure are
provided as a Source Data file.}
\label{fig:hist-Gamma}
\end{figure}
~

\begin{figure}[ht!]
\centering
\includegraphics[width=\linewidth]{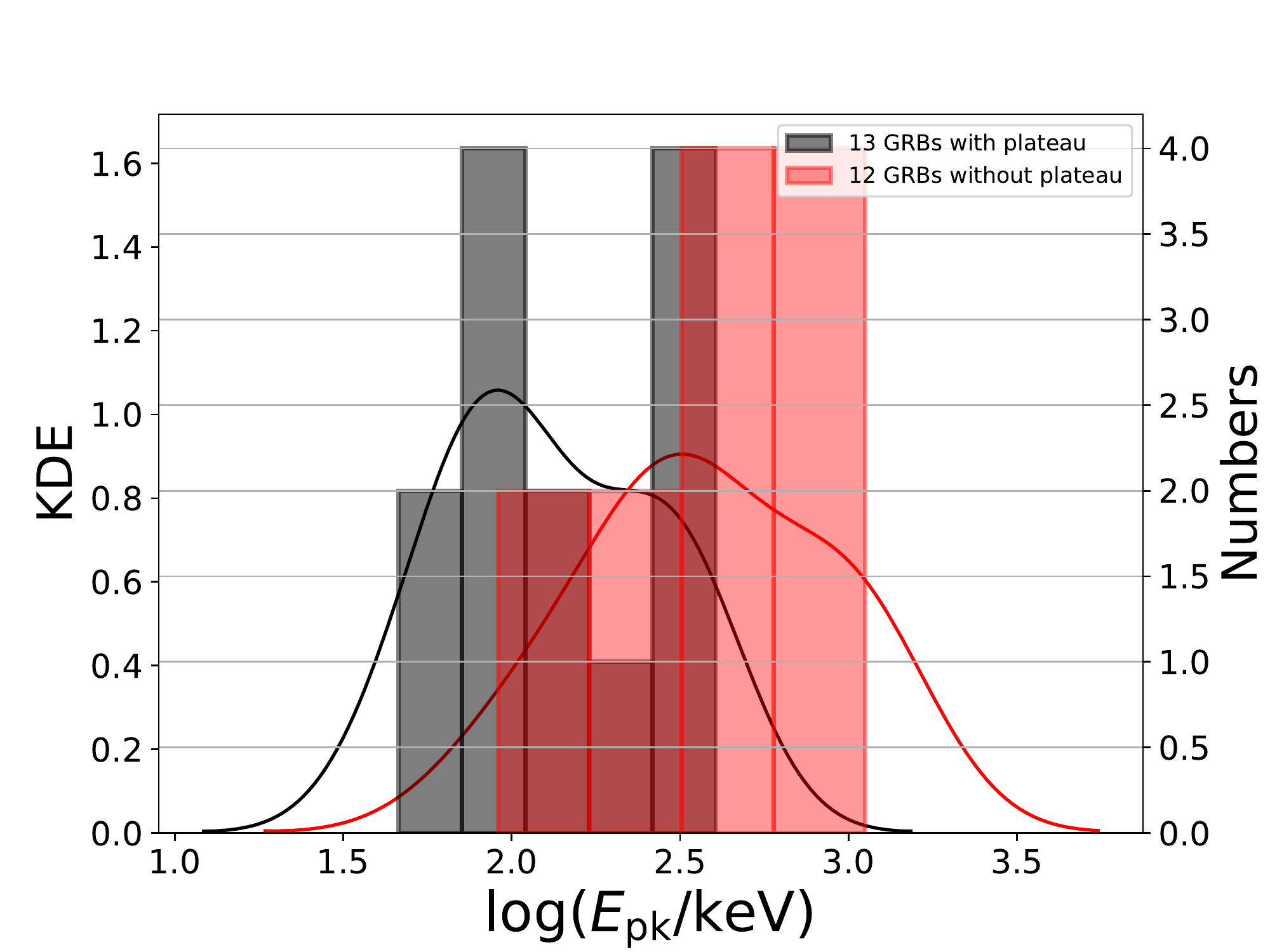}
\caption{\textbf{ Distributions of peak energy, $E_{\rm pk}$.} The black bars
represent the 13 GRBs with plateau phases in our sample, the red bars
represent 12 GRBs without plateau phases presented in Table 6
of Ref. \protect\cite{Liang10}. Note that GRBs 050319, 061021 as well
as early GRBs observed before the launch of \textit{Swift}-BAT in Ref. \protect\cite{Liang10} were discarded.
In the panel the right-hand ordinate is the number of burst in each histogram
bin and the left-hand ordinate is the value of the kernel density estimation
(KDE), which is shown by the black and red curves for each sample respectively.
Clearly, "plateau" bursts have a lower peak energy than
other bursts. This result is consistent with the idea that jets in these bursts
have a lower initial jet Lorentz factor. We also performed a Kolmogorov-Smirnov test
(KS test: D = 0.60 and p = 0.014) which can clearly show if these two samples
originate from the same population. The KS test result shows that there is
only 0.14\% chance for these samples to have originated from the same population
and they differ with 60\%. We view this as another hint towards understanding
the difference between GRBs that do show a plateau and those do not. The source data to reproduce this figure are provided as a Source Data file.}
\label{fig:logEpeak_13GRBs_with_plateau_12GRBs_without_plateau}
\end{figure}
~

\begin{figure}[ht!]
\centering
\includegraphics[width=\linewidth]{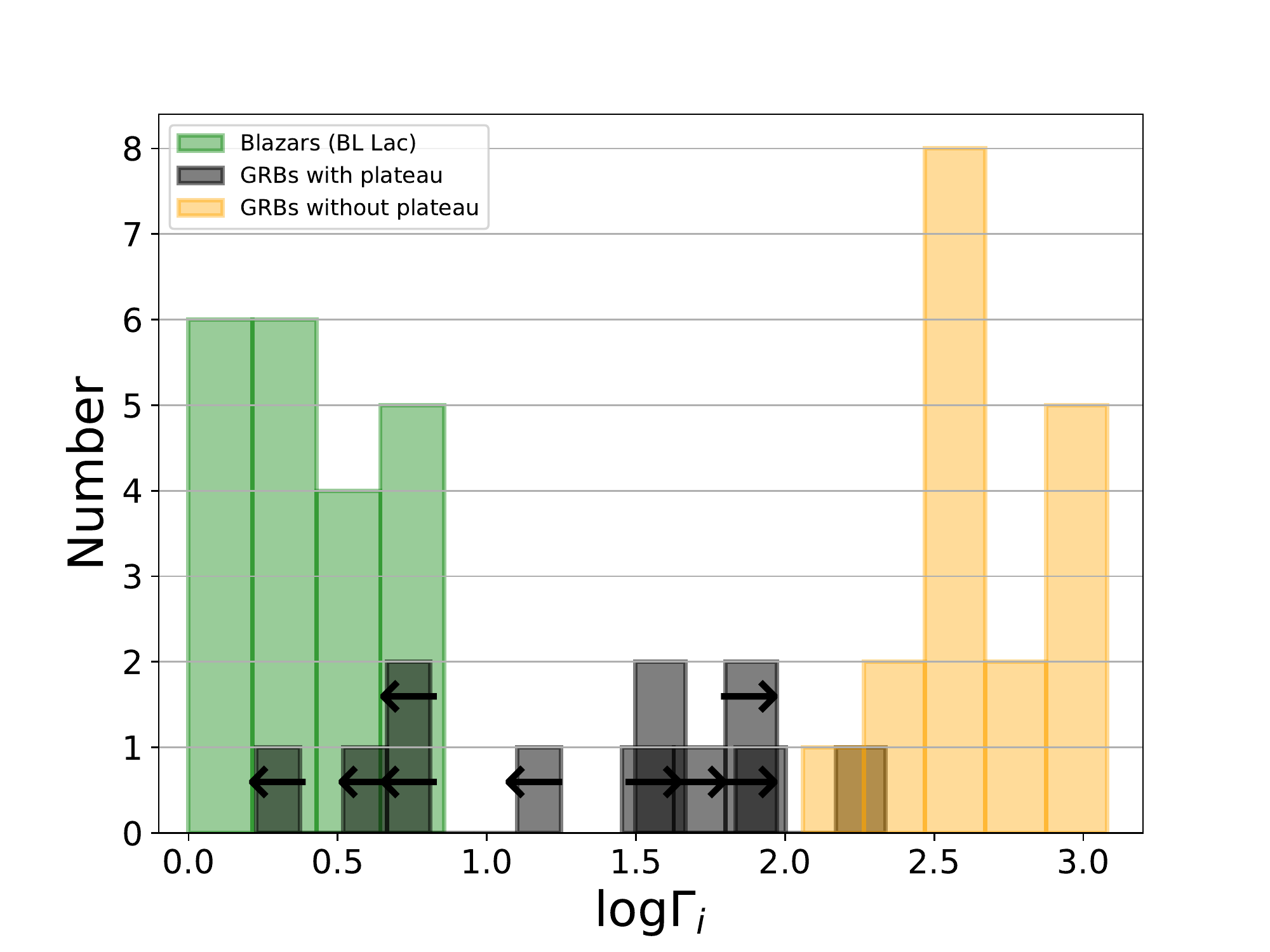}
\caption{\textbf{ Histogram of the initial jet Lorentz factors.} Black
represents the GRBs in our sample, as they
appear in Fig. \ref{fig:hist-Gamma}. Yellow represents GRBs observed
by the Fermi-LAT instrument, taken from Ref.\protect\cite{Racusin11}, whose jet
Lorentz factor is deduced from the opacity argument
(GRBs 090926A, 090902B, 090510, 080916C; see Ref.\protect\cite{Racusin11}, figure 11)
and other GRBs that do not show a plateau phase (except GRBs 050319 and 061021),
taken from Ref.\protect\cite{Liang10} (table 6 therein). Green represents the
inferred lower limit to the Lorentz factor of BL Lac objects (Blazars),
taken from Ref.\protect\cite{Ghisellini1993} (see their table 2A). The values
of the Lorentz factor we found, of few tens, fills the gap that previously
existed in the range of initial jet velocities. The source data to reproduce this figure are
provided as a Source Data file.}
\label{fig:hist-Gamma_BLac}
\end{figure}

~

~

~

~

~

~

~

~

~

~

~

~

~

~

~

~
~

~

~

~

~

~

~

~
~

~

~

~

~

~

~

~
~

~

~

~

~

~

~

~
~

~

~

~

~

~

~

~
~

~

~

~

~

~

~

~
~

~

~

~

~

~

~

~
~

~

~

~

~

~

~

~

~

~

~

~

~

~

~

~
~

~

~

~

~

~

~

~
~

~

~

~

~

~

~

~
~

~

~

~

~

~

~

~
~

~

~

~

~

~

~

~
~

~

~

~

~

~

~

~
~

~

~

~

~

~

~

~

\setcounter{figure}{0}    
\setcounter{table}{0}    

\renewcommand{\figurename}{Supplementary Figure}
\renewcommand{\tablename}{Supplementary Table}
\renewcommand{\refname}{Supplementary References}



\section*{Supplementary Information} 
\CreateLink{sec:SuppInfo}{`Supplementary Information'}


\subsection*{Supplementary Method 1. Sample and data analysis}
\CreateLink{sec:SuppMethod1}{1, `Sample and data analysis'}

We use the sample of 222 GRBs with plateau phase and with known
redshifts\footnote{\url{https://www.mpe.mpg.de/~jcg/grbgen.html}}
defined in Ref.\cite{Srinivasaragavan+20} (see also Refs. \cite{Dainotti+20,
Dainotti2020a, Dainotti2021closure, DainottiOmodei2021}).
These GRBs were detected by the Neil Gehrels \textit{Swift} Observatory \cite{Gehrels+04}
from January 2005 until August 2019. They represent
the 56\% of the GRBs (394) with known redshifts observed
by the \textit{Swift} satellite in this period. To define this sample,
two criteria are used by Ref.\cite{Srinivasaragavan+20}, (1) the
X-ray light curve of GRBs must have at least one data point in the beginning
of the plateau phase and (2) possess a plateau angle smaller than
$ 41^{\circ}$. The choice of the $ 41^{\circ}$ comes from the analysis
performed in Refs. \cite{Dainotti2016, Dainotti2017c} to allow a more
reliable identification of the plateau itself. The plateau phase is identified by
fitting the \textit{Swift}-BAT and XRT data together with the phenomenological
Willingale model \cite{Willingale+07}.  

In order to make our analysis as reliable as possible, we further limit
the sample to burst having the best quality
observations. Therefore, we added three further criteria to the ones used by
Ref.\cite{Srinivasaragavan+20}. We require (i) a long lasting plateau phase spanning several thousands of seconds with
a temporal X-ray slope larger than $-0.7$, followed by a power-law decay
phase at later times (the self-similar phase). (ii) A
sufficiently large number of data points ($\gtrsim$~5 which corresponds to
the number of parameters used to fit the data with one break) during the plateau
and self-similar phases to enable the fits to give well constrained parameter values.
For this to be valid, we excluded all X-ray flares (see figure 1 of
Ref.\cite{Zhang+06})\footnote{Flares share many properties
with prompt emission pulses. They are most likely linked
to the late central engine activities \cite{Ioka+05, Zhang+06}. They 
produce steeper slope during both the plateau and
the self-similar phase \cite{BG00, GGB08, Oates+12, Racusin+16, Dereli+17}.
Moreover, Ref.\cite{Lazzati+07} shows that flares cannot be produced
by the forward shock.} defined in the online \textit{Swift}
repository\footnote{\url{https://www.swift.ac.uk/xrt_live_cat/}} from the analyzed data.
When, after removing the flares we end up having too few data points either during
the plateau or the self-similar phase, we excluded the GRB from the analysis.

After applying those criteria, we ended up with
130 GRBs having a well-defined X-ray plateau emission that is followed
by a power-law decay phase (which can be interpreted as the late
afterglow, or the self-similar phase).
Our last criteria is to require an optical counterpart at
around the same time as X-ray data. We find 24 GRBs in Ref.\cite{Dainotti+20}
consistent with all these criteria. However, we only have
full access to the published optical data. Therefore, we perform the analysis and confront our model only to a representative sample made
of 13 GRBs listed in Supplementary Tables \ref{tab:ClassI}, \ref{tab:ClassII}, \ref{tab:ClassIII}. 


\subsubsection*{Supplementary Method 1a. X-ray data and fitting process}
\CreateLink{sec:SuppMethod1a}{1a, `X-ray data and fitting process'}

The X-ray count rate light curves (hereinafter LCs) \cite{Evans+07} of
each 13 GRBs have been downloaded from the online \textit{Swift} 
repository\footnote{\url{https://www.swift.ac.uk/xrt_curves/}}
, for the full
\textit{Swift}-XRT bandpass ($E_{\rm min}, E_{\rm max}$) = (0.3,10) keV.
We then fit each light-curve with a two or three-segments
broken power-law (BPL) model (i.e. with one or two break times) depending
on the required break during the late afterglow phase (self-similar phase). The latter break generally coincides with a jet break (see the canonical
X-ray afterglow light curve\footnote{The canonical X-ray afterglow light curve shows five distinct components: the steep decay phase, which is the tail of prompt emission; the shallow decay phase (or plateau); the normal decay phase; the late steepening phase; X-ray flares.} in Ref.\cite{Zhang+06}, figure 1 therein). The X-ray light-curve of 13 GRBs together with our fit results are presented in Supplementary Figs. \ref{fig:fit_results_050319_X-ray_opt}a, \ref{fig:fit_results_060605_X-ray_opt}a, \ref{fig:fit_results_060614_X-ray_opt}a, \ref{fig:fit_results_060714_X-ray_opt}a, \ref{fig:fit_results_060729_X-ray_opt}a, \ref{fig:fit_results_061121_X-ray_opt}a, \ref{fig:fit_results_080310_X-ray_opt}a, \ref{fig:fit_results_080607_X-ray_opt}a, \ref{fig:fit_results_091029_X-ray_opt}a, \ref{fig:fit_results_100418A_X-ray_opt}a, \ref{fig:fit_results_110213A_X-ray_opt}a, \ref{fig:fit_results_130831A_X-ray_opt}a, \ref{fig:fit_results_171205A_X-ray_opt}a.

The two-segments BPL model is characterized by the following
simple function
\beq
f(t) = A \times \left\{ \ba{ll} (t/ t_{\rm b})^{-\alpha_1} & {\rm for} ~ t <  t_{\rm b}, \\ 
( t/ t_{\rm b})^{-\alpha_2} & {\rm for} ~ t > t_{\rm b}  \ea.
\right.
\eeq
In this expression t is the time, $t_{\rm b}$ is the break time, A is the
amplitude at the break time, $\alpha_1$ and $\alpha_2$ are the temporal slopes
before and after the break time respectively. The three-segments BPL
model consider of an additional break and segment.

During the fit, we excluded the early steep decay segment and
highly significant emission bumps following the GRB prompt tails (see Ref.\cite{Zhang+06}
figure 1 therein) defined in the online \textit{Swift} repository. In
addition, we removed all flares. Some faint flares are not defined in the online
\textit{Swift} repository, however, they affect the definition of the break time
at the end of the plateau phase ($T_{\rm a}$) in individual GRB light curves. Comparing
the optical and the X-ray break times we found that the X-ray break time
($2.8 \pm 0.6 \times 10^3$ s) in the LC of GRB 091029 is earlier than the optical
one ($1.54 \pm 0.11 \times 10^4$ s). This is in contrast with the theoretical expectation
that, due to the rise in the cooling frequency, the transition time is expected to be
earlier in the optical band than in the X-ray band. Therefore, by checking the
BAT and XRT unabsorbed flux density light curves at 10 keV retrieved from the online \textit{Swift} burst analyser repository, we identified
a flare between $1379 - 2421$s in the X-ray LC
of GRB 091029. Then we objectively removed this flare in the X-ray LC of this burst.
It is also important to note that the flares in the
online \textit{Swift} repository
\footnote{\url{https://www.swift.ac.uk/xrt_live_cat/docs.php}}
using the phenomenological Willingale model \cite{Willingale+07} are indicative
and not precise.

It is also known that the data obtained by window timing (WT) mode can be
effected more 
from the high latitude emission than the data obtained by
photon counting (PC) mode because the WT mode is usually taken at early times
as it measures the flux better when the source is bright (see Ref.\cite{Evans+09}
for more explanation). Light-curves
containing both WT and  PC data require special attention to analyse the effect of different data type on the slopes.
To avoid this issue, we only used data acquired in
PC mode even if WT mode data exist
(namely for GRB 061121 and GRB 060605). Once the data were retrieved and reduced, we perform a fit to determine the X-ray temporal slopes ($\alpha_{\rm p,X}$, $\alpha_{\rm A1,X}$, $\alpha_{\rm A2,X}$) and the time of the transition between the two components, ${ T}_{\rm a, X}$ as well as the time of the jet break ${T}_{\rm b,X}$.

~

\textbf{An example of the fitting process:}
To visualise the fitting process, we illustrate in Supplementary 
Fig. \ref{fig:fit_results_060614_X-ray_opt}a the X-ray count light-curve obtained
from GRB 060614 by the \textit{Swift}-XRT instrument (over the period of PC mode data acquisition). The data are superposed with
the fits results obtained for a three-segments BPL model, i.e.
with two break times. To perform the fit, we used the Bayesian analysis
tool \textit{emcee}\footnote{\textit{emcee}
is an MIT licensed pure-Python implementation of Goodman \& Weare’s Affine
Invariant Markov chain Monte Carlo (MCMC) Ensemble sampler \cite{Goodman2010}}.
We set up the computation with 64 MCMC walkers, a 100
burn-in period and the 5000 MCMC steps. We chose these steps taking into
account the auto-correlation time. The value of the MCMC steps can slightly
change for each GRB depending on the auto-correlation time but the choice of
5000 steps is the one that guarantees that the auto-correlation time is always
taken into account. We employ the following uninformative prior distributions
for the fit parameters
\begin{align}
\left\{
\begin{aligned}
    Pr(N) &= \mathcal{U}(-5,5), \\ 
    Pr(\alpha_{\rm p}) &= \mathcal{U}(-5,5), \\ 
    Pr(\alpha_{\rm A1}) &= \mathcal{U}(-5,5), \\
    Pr(\alpha_{\rm A2}) &= \mathcal{U}(-5,5), \\
    Pr( \log(T_{\rm a}) ) &= \mathcal{U}(2,4.5), \\
    Pr( \log(T_{\rm b}) ) &= \mathcal{U}(4.5,6), \\
    Pr( \log( V ) ) &= \mathcal{U}(-15,12), 
    \end{aligned} \right.
\end{align}
where $N$ is the normalisation, $\alpha_{\rm p}$, $\alpha_{\rm A1}$ and
$\alpha_{\rm A2}$ are the slopes during the plateau, in the self-similar
phase and following the jet break respectively. Here, $\mathcal{U}$ notes the uniform prior. The characteristics times
$T_{\rm a}$ and $T_{\rm b}$ mark the end of the plateau and the jet break
respectively. Finally $V$ is a nuisance parameter measuring the spread of the
data. The parameter range for the normalisation can change depending on the band
(X-ray or optical) in which the fit is performed.



As a Specific example, we show the fit parameters of GRB 060614 in Supplementary Table \ref{tab:fit_paramaters_060614}. The corner plot of the
posterior probability distributions of the fit parameters and the covariances
between the fit parameters is displayed in Supplementary
Fig. \ref{fig:corner_plot_060614}. 

The temporal fit parameters in the X-ray band obtained for all 13 GRBs
in our sample are presented in Supplementary Tables \ref{tab:ClassI},
\ref{tab:ClassII}, \ref{tab:ClassIII}, for different classes I, II, and
III respectively. The definitions of these classes are explained in Results subsection `Sample classification' in the main manuscript.

The X-ray spectral slopes ($\beta_{\rm p,X}$, $\beta_{\rm A1,X}$,
$\beta_{\rm A2, X}$) of twelve GRBs out of thirteen
(all GRBs but GRB 060729) for the time intervals defined by the
characteristic times of the X-ray light-curve are also
obtained from the online \textit{Swift} repository and presented
in Supplementary Tables \ref{tab:ClassI}, \ref{tab:ClassII}, \ref{tab:ClassIII}. 
It is important to note that the photon index (which is equal to
$\beta + 1$ where $\beta$ is the spectral index) is
presented in the online \textit{Swift} repository. For
GRB 060729, we used the result from the spectral
analysis of Ref.\cite{Zaninoni+13} since the value
of the spectral index is not given by the online \textit{Swift} repository.
These X-ray spectral and temporal parameters together with temporal slopes obtained from the temporal fit are combined into the closure relations ($F_{\nu} \propto t^{-\alpha} \nu ^{-\beta}$) as explained in the Result subsection `Closure relations and determination of the electron power-law indices'. This relations are presented in Supplementary Figs. \ref{fig:CR-plateau} and \ref{fig:CR-self-similar}. 

~

\textbf{Caveats for the X-ray analysis for specific GRBs:}
\begin{itemize}
    \item GRB 080607. For this burst, the
    error of the temporal slope during the plateau phase is large. This is
    due to the lack of the data points at the end of
    the X-ray plateau phase. The slope after the plateau phase is
   rather steep. However, the
    later (last) slope is more compatible with the expectations from the self-similar phase, see, Supplementary
    Fig. \ref{fig:fit_results_080607_X-ray_opt}a. Therefore, we used later slope presented in Supplementary Table \ref{tab:ClassI}. The light-curve has two flares
    before the
    onset of the plateau phase. The transition
    between the peak-flux of the second flare and the plateau phase is well
    represented by a BPL, with steep slopes $-2.27^{+0.07}_ {-0.07}$ and 
    $-1.42^{+0.25}_{-0.30}$, separated at a break time of $341^{+31}_{-28}$s. 
     The post-flare slopes are
    similar to the slopes found after the plateau phase.
    Therefore, either another flare might exist during the plateau
    or the plateau slope might be steeper
    with a break later than the
    ${T}_{\rm a,X}$ obtained here.
    \item GRB 130831A: The lack of data in the X-ray LC of this GRB between $10^3$
    and $10^4$s, typical for the duration of the plateau, affects the
    measurement of ${T}_{\rm a,X}$, see Supplementary
    Fig. \ref{fig:fit_results_130831A_X-ray_opt}a. We tried to fit the light
    curve with a model composed of three power-law and two
    breaks. For this model, we found that $\log(T_{\rm a,X}/ \rm s)~=~3.08_{-0.18}^{+0.28}$. This break is
    consistent with being achromatic in both X-ray and optical bands
    (see, Supplementary Method \LinkTo{sec:SuppMethod1b}).
    However, due to the large uncertainty on $T_{\rm a, X}$ we found that fitting the light curve
    with a single break provides a better fit. Therefore, in our analysis we used the parameters obtained for the model with one temporal break. These parameters are given in Supplementary Table \ref{tab:ClassI}. In addition, before the lack of data, we also define a faint flare between $595 - 811$ s in the X-ray LC of this burst. This flare coincides with the beginning of the optical flare, see Supplementary
    Fig. \ref{fig:fit_results_130831A_X-ray_opt}b. Therefore, we also performed a fit without this flare.  We found that the both slopes ($-0.80_{-0.03}^{+0.03}$ and $-0.98_{-0.06}^{+0.06}$ respectively with a break at $1.0_{-0.16}^{+0.25} \times 10^4$~s) are consistent with the expected slope ($\sim - 1.2$) in a self-similar phase. This result is also consistent with a previous study in Ref.\cite{dePasquale+16}. We then conclude that either this GRB is not the best representation of the GRBs that show plateau phase or the lack of the data is misleading the results.
\end{itemize}



~

\subsubsection*{Supplementary Method 1b. Optical data and fitting process}
\CreateLink{sec:SuppMethod1b}{1b, `Optical data and fitting process'}

Below, we describe the details of the fitting procedure in the optical
band for each GRB in our sample. Before making the fits and if needed, the data
were first transformed to the AB magnitude system. The results of the fits (in the temporal regions of interest) are presented in Supplementary Tables \ref{tab:ClassI}, \ref{tab:ClassII} and \ref{tab:ClassIII}. The following detailed explanations for each GRB independently are split depending on the origin of the data.

For GRB 091029, GRB 110213A, GRB 130831A and GRB 171205A, the optical data
were retrieved from all possible published sources listed in Supplementary Table \ref{tab:ClassI}. 
For each of these bursts, we compiled a
large set of data obtained in different bands from different instruments.
The temporal fits are performed simultaneously
in each band, which means that we consider all the bands at the 
same time, only changing the normalisation between the observation bands.
Our aim here is to reduce the dependence on a single instrument and
a single band in the determination of the temporal slopes and the
break times. This method provides tight estimations of the fit parameters.

\begin{itemize}
    \item GRB 091029: The photometric data set was
    retrieved from tables 5, 6, 7, 8, 9 of Ref.\cite{Filgas2012}. The data are displayed alongside our fit results in Supplementary Fig. \ref{fig:fit_results_091029_X-ray_opt}b.
    We fit the data between 427.7s and $2\times10^5$s with a
    three-segment broken power law model (i.e. with two breaks, hereinafter two-break model). Although later data exist, the flattening in the U-band likely indicates an external source of radiation, such as the contribution from the host galaxy. Therefore, data after $2\times10^5$s are excluded from the fit. 
    The first temporal break at $3.7^{+ 0.3}_{-0.2}\times10^3$s is interpreted as 
    the beginning of the plateau, although the earlier (i.e. before the plateau) slope is $-0.60_{-0.01}^{+0.02}$.
    %
    \item GRB 110213A: The photometric observations are obtained from
    table 4 of Ref.\cite{Cucchiara2011}. They are displayed alongside our 
    fit results in Supplementary Fig. \ref{fig:fit_results_110213A_X-ray_opt}b. 
    The optical light-curve of this GRB has two peaks. The brightest peak is at 263s and
    the second peak is at 4827s.
    This flare can be clearly seen in the unabsorbed flux density light curve
    (see \textit{Swift}-BAT-XRT at 10 keV in the online \textit{Swift} burst analyser
    repository). This flare produces a rising slope
    during the X-ray plateau phase, however, it has a negligible effect on the
    value of break time at the end of the plateau phase
    $T_{\rm a,X}$.
    %
    \item GRB 130831A: The photometric data of this burst
    is obtained from the VizieR Online Data Catalog \cite{dePasquale2018}. 
    The data are displayed alongside our 
    fit results in Supplementary Fig. \ref{fig:fit_results_130831A_X-ray_opt}b.
    A flare can clearly be identified between 425s and 3400s, therefore, this time interval is not considered in our fit. The flare interpretation of this re-brightening is further supported by the fact that we found the temporal slopes before 425s and after 3400s to be the same. The light curves in the Rc and Ic bands flatten after $10^5$s, most likely due to an external source. Shortly later, this flattening also appears in all other bands. Therefore, the data after $10^5$s are discarded from our analysis. 
    %
    \item GRB 171205A: The photometric data set is obtained from Ref.\cite{Izzo2019}. 
    The data set is displayed alongside our fit results in Supplementary Fig. \ref{fig:fit_results_171205A_X-ray_opt}b.
    Since this GRB is associated to the supernovae SN 2017iuk, 
    the optical data start to rise at
    ${T_{\rm b,O}} = 1.73\times 10^5$~s. We fit all the data before the
    peak of the supernovae (SN) at $9.51 \times 10^5$ s with a two break model. 
    The first break (${T_{\rm a,O}}$) represents the transition from the plateau to
    the self-similar phase, while the second break,
    at ${T_{\rm b,O}}$ marks the rise of the optical light curve due to the supernovae
    contribution. 
    We find that the temporal slopes are somewhat shallower, and
    even slightly rising than in other bursts. Moreover, the break time ${T_{\rm a,O}}$ occurs earlier in the optical band than the X-ray band. We believe that this may be due to the SN bump after $\approx2$ days \cite{Izzo2019}. 
\end{itemize}

For  GRB 050319, GRB 060605, GRB 080607 and GRB 100418A, we used the optical data from table 8
of Ref.\cite{Zaninoni+13}. In that paper, all available data obtained from different instruments were collected. 
The details of the fitting process is given below for each of
these GRB.

\begin{itemize}
    \item GRB 050319: 
    We use a two-break model to fit the optical light-curve until the observer time $ t = 5.34\times 10^5$~s, after which the observations show a steep flux decrease with an exponent steeper than $-1.2$ (which is theoretically expected from the self-similar phase.)  
    The first break is observed at time $371 ^{+31.6}_{-28.6}$~s, which is too small to be the end of the plateau in the framework of our model. We therefore consider the break at $3.58^{+ 0.30}_{- 0.32}\times 10^4$ s to be the end of the plateau. In any case, the optical slopes of each phase are not very different: 
    the early (before the plateau) slope is $-0.41_{-0.01}^{+0.02}$, the slope of the plateau is  $-0.59_{-0.01}^{+0.01}$ and the slope during the self-similar expansion is $-0.64_{-0.01}^{+0.02}$.
    \item GRB 060605: We fit the data between 185~s and $2\times 10^4$ s with
    with a two-break model. 
    After the time $t = 2\times 10^4$ s, the optical data flattens in
    all bands, see, Supplementary Fig. \ref{fig:fit_results_060605_X-ray_opt}b. Therefore,
    this data is not included in the fit.
    %
    \item GRB 080607: The LC of this has three breaks at around 89~s, at $793^{+77}_{-64}$~s and at
    $T_{\rm a,O} = 2.55^{+ 0.26}_{- 0.12} \times 10^3$~s,
    see Supplementary Fig. \ref{fig:fit_results_080607_X-ray_opt}b. Since we are interested in the late time evolution, we fit the data starting at 89~s.
    The temporal slope before the plateau phase is $-1.22^{+0.01}_{-0.02}$, which
    is similar to the slope of the self-similar phase, see Supplementary Table \ref{tab:ClassIII}.
    Contrary to the X-ray band (see above, Supplementary Method \LinkTo{sec:SuppMethod1a}), the plateau phase in the optical band is
    clearly identified and characterized.
    \item GRB 100418A: In our analysis of the optical data, we excluded the
    data only reported in the GCN (Gamma-ray Coordination Network). For this burst
    specifically, we fit a one-break model to the data.
\end{itemize}

The optical LC of the rest of the GRBs in our sample (namely GRB 060614,
GRB 060714, GRB 060729, GRB 061121, and GRB 080310) are
taken from Ref.\cite{Racusin+11} (see also, Ref.\cite{Oates+09} for the details).
These data were obtained by the \textit{Swift} Ultraviolet/Optical (UVOT) instrument
in bands V, B, U, W1, M2 and W2. In Ref.\cite{Racusin+11}, all the LCs were normalized
to the U band, and we use these normalised LCs in our analysis.
The details of the fitting process are given below for each GRB.

\begin{itemize}
    \item GRB 060614: We fit the data between
    $4.84\times 10^3$ and $8.39 \times 10^5$ s. Indeed, at
    early times between 113.16s and 379.87s, the LC has a steep decay which coincides with
    the steep decay observed in X-ray data,
    see Supplementary Fig. \ref{fig:fit_results_060614_X-ray_opt}a. This decay is
    therefore interpreted as the end of the prompt phase. Moreover,
    no observations were performed between 380s and $4.8\times 10^3$s. At
    times larger than $8.39 \times 10^5$s
    the LC flattens which is interpreted as host galaxy contribution, see Supplementary Fig.
    \ref{fig:fit_results_060614_X-ray_opt}b. 
    %
    \item GRB 060714: The observations of this burst are scarce with large
    errors. We excluded the last data point at $1.49
    \times 10^7$s from the fit since it is a clear outlier. We also did
    not included the non-detection limits in our analysis. With 8 data points
    left, a break can clearly be identified and the parameter of a one-break model be
    constrained, even though the observations after $10^5$s have large errors, see Supplementary Fig. \ref{fig:fit_results_060714_X-ray_opt}b. 
    
    \item GRB 060729: We fit the data between
    $266$s and $1.12 \times 10^6$s.
    At earlier times, the optical flux rises, and at later times it flattens with a
    break at $\log(t) =
    {6.16}_{-0.03}^{+0.02}$, while the slope after the break is
    $-0.16_{-0.07}^{+0.07}$, see Supplementary
    Fig. \ref{fig:fit_results_060729_X-ray_opt}b. 
    %
    \item GRB 061121: We fit the data between
    $ t = 104.97$ and $1.66 \times 10^5$ s. At early
    time between 69.19s and 89.97s, the LC has an optical flare which coincides
    with the X-ray flare defined in the online \textit{Swift} repository. 
    At later times, the data flattens. We further exclude the
    early optical bump seen between 143.45
    and 568.48 s to have a better
    identification of the plateau slope and the break time ${T_{\rm a,O}}$, see
    Supplementary Fig. \ref{fig:fit_results_061121_X-ray_opt}b. If our fit would
    include the decay of this bump (namely, starting at 230~s), its fitted slope and the break time would be
    $-1.01^{+0.05}_{-0.06}$ and $\log(T/\rm s)
    = 3.12 ^{+0.06}_{-0.06}$ respectively. However, we would find that the later slopes ($-0.45^{+0.06}_{-0.05}$, $-0.93^{+0.07}_{-0.08}$ respectively) and the later break time ($\log(T_a/\rm s) =4.08^{+0.10}_{-0.10}$) would be compatible with the fit parameters presented in Supplementary Table \ref{tab:ClassIII}. Therefore, the effect of removing the second optical bump from the analysis is negligible.
    \item GRB 080310: The first data point at 132.39s has a much lower flux than during the plateau. This data point can be part of the rise at early time. Such a rise is clearly seen in some of the optical LC in our sample (e.g., GRB 060605). Therefore, it is discarded from our analysis. Moreover, we also did not include
    in our analysis the observation with no detection, see, Supplementary Fig. \ref{fig:fit_results_080310_X-ray_opt}b. 
\end{itemize}

The optical spectral slopes ($\beta_{\rm p,O}$, $\beta_{\rm A1,O}$,
$\beta_{\rm A2,O}$) of twelve GRBs out of thirteen in our sample (all except GRB 130831A
which does not have a published one) are obtained from the literature:
data of GRBs 050319, 060605, 060614, 060729, 061121, 080310, 080607 and  100418A
are presented in Ref.\cite{Zaninoni+13}; GRB060714 in Ref.\cite{Zafar2011};
GRB091029 in Ref.\cite{Filgas2012}; GRB110213A in Ref.\cite{Cucchiara2011};
and GRB171205A in Ref.\cite{Izzo2019}. All these slopes
are given in Supplementary Tables \ref{tab:ClassI}, \ref{tab:ClassII}, \ref{tab:ClassIII}
for the different classes I, II, III respectively. These spectral slopes are also
combined into the closure relation ($F_{\nu} \propto t^{-\alpha} \nu ^{-\beta}$)
in Supplementary Figs. \ref{fig:CR-plateau} and \ref{fig:CR-self-similar} together with
temporal slopes obtained from the optical temporal fit. This relation is as explained in the Result subsection `Closure relations and determination of the electron power-law indices'.


\subsubsection*{Supplementary Method 1c. Flux Ratio}
\CreateLink{sec:SuppMethod1c}{1c, `Flux Ratio'}

\textbf{X-ray energy flux ($\nu F_{\nu}$):} The unabsorbed
flux density ($F_\nu$) light curves of 
XRT at 10 keV \cite{Evans+09} are obtained from the online
\textit{Swift} burst analyser repository\footnote{\url{https://www.swift.ac.uk/burst_analyser/}}.
These flux density LCs are already corrected for
absorption from both the galactic and host galaxy. In our analysis, we convert
these light curves from flux density (Jy) to energy flux (${\rm erg~cm^{-2}~s^{-1}}$)
at 10 keV ($\nu$ = $2.42 \times 10^{18}$ ~Hz). Then, we used these energy flux
LCs to compute the $\nu F_{\nu}$ X-ray flux at the end of the plateau phase at
${T_{a,\rm X}}$ and at another specific time, typically, $1000$s.

\textbf{Optical energy flux ($\nu F_{\nu}$):} For computing the
energy flux in the optical band, all the LCs are corrected for both the
galactic and the host galaxy extinctions. i) The galactic extinction along
the line of sight is obtained from the IRSA
webpage\footnote{\url{https://irsa.ipac.caltech.edu/applications/DUST/}} for
the U band filter. It is computed with the method introduced in Ref.\cite{Schlegel+98}.
ii) The extragalactic reddening E(B - V) of all GRBs (except GRBs 060714 and 100418A)
are obtained during the plateau phase and the references are listed in Supplementary 
Table \ref{tab:Extinctions}. We first used these values to compute the host
galaxy extinctions in the V-band (except for 060714
for which we used the value provided in Ref.\cite{Schady+10}). The host
galaxy extinction $A_{v,\rm host} = R_v \times E(B - V)$ is computed by using
one of the three visual extinction to reddening ratio, $R_v$, of 3.08,
3.16, 2.93 for the Milky Way, Large and Small Magellanic Clouds (MW, LMC and SMC)
extinction laws\cite{Pei+92} respectively. The type of extinction law
is defined in the reference papers and
they are also given in Supplementary Table \ref{tab:Extinctions}. Second,
we use these extinctions together
with $A_{\rm u,host} = 1.666 \times A_{\rm v,host}$ see, details in Ref.\cite{Pei+92},
to compute the host galaxy extinction in the U band. Both extinctions
(galactic and host galaxy) for each GRBs are given in Supplementary 
Table \ref{tab:Extinctions}. Note that the flux density of four GRBs
(namely GRB 050319, GRB 060605, GRB 080607, GRB 100418A) taken from Ref.\cite{Zaninoni+13}
are already corrected for both galactic and host extinction.

In details, we proceed as follows:
\begin{itemize}
    \item For six GRBs, namely GRB 060614, GRB 060714, GRB 060729, GRB 061121,
    GRB 080310 and GRB 091029, the optical count rate LCs from Ref.\cite{Racusin+11}
    are used to compute the U-band energy flux LCs. For the conversion, we used the
    U-band count rate to flux conversion factor ($1.628\times10^{-16} ~{\rm erg~
    cm^{-2}~Angstrom^{-1}~counts^{-1}}$), as provided in the UVOT calibration
    website\footnote{\url{https://heasarc.gsfc.nasa.gov/docs/heasarc/caldb/swift/docs/uvot/index.html}}.
    Then, we computed the energy flux ($\nu F_{\nu}$) LCs, where $\nu$ is taken at the
    U-band central frequency equal to $8.65 \times 10^{14}$~Hz. 
    These U-band energy flux LCs are used to compute the
    $\nu F_{\nu}({\rm U})$ optical flux at a specific time namely $\sim 1000$s hereinafter dubbed reference time in the U-band $T_{\rm ref.,\rm U}$ and at the end of the plateau phase at ${T_{\rm a, X}}$. 
\end{itemize}

For all other GRBs, we followed a different procedure presented below: 
\begin{enumerate}
    \item[i)] For the GRBs in Ref.\cite{Zaninoni+13}, namely
    GRB 050319, GRB 060605, GRB 080707, and GRB
    100418), the $\nu F_{\nu}$ optical flux at the end of the X-ray plateau
    phase ($T_{\rm a,X}$) are retrieved from the flux
    density LC in table 7 of the online material in Ref.\cite{Zaninoni+13}.
    In these cases, the frequency $\nu$ is different for each GRBs (V band
    $\nu = 5.48\times 10^{14}$ Hz, CR and RC bands $\nu = 4.55\times 10^{14}$ Hz,
    white band $\nu = 7.79\times 10^{14}$ Hz, see Supplementary Table \ref{tab:Extinctions}).
    For consistency, whenever possible we used the U-band in calculating the LC.
    When the U-band data was not available, we used other bands.
    In such case, we note that, the differences between different optical bands ($\sim 10^{15}$ Hz) are small.
    
    \item[ii)] For all other, namely GRB 110213A, GRB
    130831A, and GRB 171205A, we followed the same
    procedure as in appendix A of Ref.\cite{Zaninoni+13} to convert the
    optical data from the AB magnitude system
    to the optical energy flux.
    Therefore, the final conversion formula is
    $f_{\nu} = 10^{-0.4(m_{AB}+48.585)}$ where $f_{\nu}$ is the flux density in frequency, $m_{AB}$ is magnitude in AB system.
\end{enumerate}

\textbf{Energy flux ratio:}
The specific energy fluxes $\nu F_{\nu}({\rm X})$ at ${T_{\rm a,X}}$
in the X-ray band and $\nu F_{\nu}({\rm U})$ at $T_{\rm ref.,U}$ in
the optical band are used to compute the energy flux ratio between
the X-ray and the optical bands. The ratio is computed by
[$1/(\nu F_{\nu} (\rm U)/\nu F_{\nu} (\rm X))$] and is given in the last column of
Table \ref{tab:outflow_parameters} in the main text.
~


\textbf{Isotropic energy:} The \textit{Swift}-BAT observations are carried in a
relatively narrow energy range (15–150 keV) which prevents from measuring
the peak energy. To obtain a lower limit on the
isotropic energy ${E_{\rm iso}}$, we followed the same method presented in
Ref.\cite{Tang+19}. In this method, a simple power-law
spectral model is adopted, i.e. $\phi (E) \propto E^{\alpha_{\gamma}}$,
where $\phi(E)$ is the source photon spectrum and $\alpha_{\gamma}$ is the
photon spectral index
The estimate on the
isotropic energy in the BAT band $E_{\rm iso}$ is then computed as
\begin{equation}
\centering
    E_{\rm iso} = \frac{4 \pi {d_{\rm L}^2 S}}{(1+{z})^{3-{\alpha_{\gamma}}}}.
\end{equation}
where $S$ is the BAT fluence between 15 and 150 keV. Both $S$ and $\alpha_{\gamma}$ are taken from the
\textit{Swift} GRB table\footnote{\url{https://swift.gsfc.nasa.gov/archive/grb_table/}},
${d_{\rm L}(z)}$ is the luminosity distance calculated assuming a flat $\Lambda$CDM
cosmological model with cosmological parameters $\Omega_m = 0.286$ and ${H_{\rm 0}} =
70 ~{\rm km~s^{-1}~Mpc^{-1}}$, 
and $z$ is the redshift.  


The various parameters relevant to our study are listed in Table \ref{tab:key_parameters} (in main text):
(i) the GRB name, (ii) its redshift ($z$), (iii) the luminosity distance (${d_L}$), (iv)
fluence ($S$), (v) photon spectral index ($\alpha_{\gamma}$), (vi) isotropic energy
(${E_{\rm iso}}$) in the \textit{Swift}-BAT band, (vii) burst duration $T_{90}$
which is obtained from the \textit{Swift} GRB table, (viii) the time at the end of the X-ray
plateau phase ($T_{\rm a,X}$), (ix) $\nu F_{\nu}$ X-ray flux and (xi) $\nu F_{\nu}$
optical (U band) flux at $T_{\rm a,X}$, (x) a reference time in the optical band
called $T_{\rm ref., U}$ (typically at around 1000 s), (xii) $\nu F_{\nu}$ X-ray
flux and (xiii) $\nu F_{\nu}$ optical (U band) flux at the reference time
$T_{\rm ref., U}$.
The isotropic energy ($E_{\rm iso}$) and a time at the
end of the X-ray plateau phase (${T_{\rm a,X}}$), firstly discovered by Ref.
\cite{Dainotti2011b, Dainotti2015b, Dainotti2019book}, are presented in Fig.
\ref{fig:EisoandTa} (main text).


\subsection*{Supplementary Method 2. Theoretical model}
\CreateLink{sec:SuppMethod2}{2, `Theoretical model'}

The key to understanding the observed signal within the framework of our model
is the realization that the end of the plateau corresponds to the transition from a
coasting phase to a self-similar (decaying) expansion phase of the expanding plasma.
The observed signal originates entirely from ambient electrons collected and heated by
the forward shock wave, propagating at relativistic speeds inside a "wind" (decaying
density) ambient medium. During the transition from the coasting to the
self-similar phases a reverse shock crosses the expanding
plasma. However, the contribution from electrons heated by the reverse
shock is suppressed due to (i) the declining ambient density which implies that
the ratio of plasma density to ambient density remains constant (in a conical
expansion), and (ii) its slower speed, which translates into less energetic electrons
that emit at much lower frequencies than forward shock heated electrons, implying
that the contribution to the optical and X-ray bands is negligible. 

To understand how this transition affects the observed spectra, we first describe
the radiative mechanism adopted, which is the classical synchrotron emission from
a power-law distribution of electrons, injected into the radiation zone with a power-law index $p$, namely $N_{el}(\gamma) d \gamma \propto \gamma^{-p}$ above a minimum
value $\gamma_m$. Both theory \cite{Spit08a} and observations \cite{WG99} suggest a typical value of $p$ such that $2.0 \leq p \lesssim 2.4$. Below the injection Lorentz factor $\gamma_m$,  it is safe to assume that the electrons have
a Maxwellian (or quasi-Maxwellian) energy distribution \cite{MR93, SPN98}. This
particle distribution leads to a broken power-law spectrum, whose shape, in the relevant observed
bands (frequency $\nu$), depends on whether the peak frequency, $\nu_m$ (defined
in Supplementary Equation (\ref{eq:nu_m})) is above or below the cooling frequency, $\nu_c$
(defined in Supplementary Equation (\ref{eq:nu_c})). At low frequencies, one needs to consider
the self-absorption process, whose contribution
is neglected here since it is observed at frequencies lower than
the optical frequency.

For $\nu_m > \nu_c$ (the so-called "fast cooling" regime, expected at early times),
the possibilities (marked A--C) are 
\beq
F_\nu = F_{\nu_{\max}} \times \left\{ \ba{ll} \nu_c^{-1/3} \nu^{1/3} &
\nu < \nu_c < \nu_m~~~A \\ \nu_c^{1/2} \nu^{-1/2} & \nu_c < \nu < \nu_m~~~B
\\ \nu_m^{(p-1)/2} \nu_c^{1/2} \nu^{-p/2} & \nu_c < \nu_m < \nu~~~C \ea
\right.
\label{eq:fast}
\eeq
At later times, $\nu_m < \nu_c$  and the plasma enters the "slow cooling" regime, in which (regions D--F)
\beq
F_\nu = F_{\nu_{\max}} \times \left\{ \ba{ll} \nu_m^{-1/3} \nu^{1/3} &
\nu < \nu_m < \nu_c~~~D \\ \nu_m^{(p-1)/2} \nu^{-(p-1)/2} & \nu_m < \nu <
\nu_c~~~E \\ \nu_m^{(p-1)/2} \nu_c^{1/2} \nu^{-p/2} & \nu_m < \nu_c < \nu~~~F
\ea \right.
\label{eq:slow}
\eeq
Here, $\nu_m$ is the typical emission frequency from electrons at the peak of the distribution, namely having Lorentz factor $\gamma_m$, and is given by (for an on-axis observer)
\beq
\nu_m^{ob} = \frac{3}{4 \pi} \frac{q B} { m_e c} \gamma_{m}^2
{\Gamma \over (1 +z)} = 4.196 \times 10^{6} B \gamma_{m}^2 {\Gamma
  \over (1+ z)} \Hz,
\label{eq:nu_m}
\eeq
where $q$ is the electron's charge, $m_e$ is the electron's mass, $c$ is the speed of light, $B$ is the magnetic field, $\Gamma$ is the bulk motion Lorentz factor and $z$ is the redshift.
The cooling frequency $\nu_c$ is the frequency of emission from electrons whose radiative cooling time is equal to the dynamical time, 
\beq
\nu_c^{ob} = {3 \over 4 \pi}  {q \over m_e c} \left({6
    \pi m_e c^2 \over \sigma_T}\right)^2 {\Gamma^3
  \over B^3 r^2 (1 +z)} 
= 2.26\times 10^{45}  {\Gamma^3
  \over B^3 r^2 (1 +z)} \Hz.
\label{eq:nu_c}
\eeq
Here, $\sigma_T$ is Thomson's cross section, and $r$ is the plasma radius, which is
related to the dynamical time (in the comoving frame) by $r \sim \Gamma c t_{\rm
dyn}$\footnote{For simplicity, we neglect a possible factor of the order unity, as
well as contribution form inverse-Compton (IC) cooling, which is negligible for the
parameters used in this work.}. 
The peak flux is estimated by $F_{\nu, peak}^{ob} ={1 \over 4 \pi d_L^2} N_e
P_{\nu,\max}^{ob}$, where $d_L$ is the luminosity distance, $N_e$ is the number
of radiating particles and $P_{\nu, \max}^{ob} = P_{tot}^{ob} /
\nu_{peak}^{ob}$, where $P_{tot}^{ob}$ is the total power radiated by synchrotron
emission by a single electron at $\gamma_m$, and $\nu_{peak}^{ob} = \nu_m^{ob}$.
This gives
\beq
P_{\nu, max}^{ob} = {P_{tot}^{ob}\over \nu_{m}^{ob}} = { {4 \over 3} c \sigma_T
  \gamma_{m}^2 {B^2 \over 8 \pi} \Gamma^2 \over { 3 \over 4 \pi} {q B
    \over m_e c} \gamma_{m}^2 {\Gamma \over (1+z)}} = { 2 \over 9} {m_e c^2 \sigma_T
  \over q} B \Gamma (1+z) = 2.53 \times 10^{-22} B \Gamma (1+z)~~{\rm erg~ s^{-1}~Hz^{-1}}
\label{eq:P_nu}
\eeq
We assume that all the ambient particles collected by the forward shock wave radiate. By assumption, the mass density is given by $\rho(r) = A/r^2$, implying that 
\beq
N_e(r) = {4 \pi \over m_p} \int_0^r \rho(r') r'^2 dr' = {4 \pi A r \over m_p}
\label{eq:2.9}
\eeq
The proportionality constant $A$ is calculated assuming that prior to its final explosion, the progenitor star ejects mass at a constant rate and at a constant velocity, resulting in $\rho(r) = n(r) m_p = {\dot{M} \over 4 \pi v_w r^2} \equiv A r^{-2}.$
For a Wolf-Rayet progenitor, the typical values are \cite{CL00}, $\dot{M} = 10^{-5} \, M_\odot \, {\rm
  yr}^{-1}$, and wind velocity $v_w = 10^8\,{\rm cm\, s^{-1}}$. These
values lead to $A = 5 \times 10^{11} A_\star \,{\rm gr~cm^{-1}}$.

\subsubsection*{Supplementary Method 2a. Dynamics, magnetic field and electron's energy}
\CreateLink{sec:SuppMethod2a}{2a, `Dynamics, magnetic field and electron's energy'}

Following an initial acceleration phase, the plasma coasts at a
nearly steady Lorentz factor $\Gamma_i$.  Once it collects sufficient material from
the ambient medium, $m_{ISM} \gtrsim m_i/ \Gamma_i$ where $m_i$ is the initial ejected
mass, the flow becomes similar, and its evolution is described by the well-known
self-similar solution \cite{BM76}. For an instantaneous explosion releasing energy
$E$ that occurs into a density gradient this solution reads 
\beq
\Gamma(E; r) = \left({9 E \over 16 \pi \rho(r) c^2 r^3} \right)^{1/2} = \left({9 E \over 16 \pi A c^2 r} \right)^{1/2}.
\label{eq:3.1}
\eeq
The relation between the Lorentz factor, radius and observed time in this case was
calculated by Ref.\cite{PW06}, $t^{\rm ob} \simeq (1+z) r / ( 2 \Gamma^2(r) c)$,
enabling to express the Lorentz factor as a function of the observed time during
this phase,
\beq
\Gamma(E, A; t^{\rm obs.}) = \left( {9 E (1+z) \over 32 \pi A c^3 t^{\rm
    ob}} \right)^{1/4}. 
\label{eq:3.5}
\eeq
Transition from the initial (coasting) to the later (self-similar expansion) phases occurs once $\Gamma(E, A; t^{\rm ob}) < \Gamma_i$\footnote{A second condition is that $t > t_{GRB}$, which is always met.}. Thus,
\beq
T_{a}= (1+z){ 9 E \over 32 \pi A c^3 \Gamma_i^4}. 
\label{eq:t_trans}
\eeq

It is explicitly assumed that both the generation of magnetic field and acceleration
of particles to high energies occur at the shock wave. The energy density behind the
shock is given by the shock jump conditions, $u \simeq 4 \Gamma^2 n m_p c^2$
where $m_p$ is the proton mass and $n$ is the particle number density in the
surrounding medium. Therefore, during the initial (coasting) phase where
$\Gamma = \Gamma_i$, this energy density is $u(t^{\rm obs.})_{\rm coasting} =  (1+z)^2 {A / (\Gamma_i^2 t_{\rm obs.}^2) }$. We adopt the standard assumption that a fraction $\epsilon_B$ of this energy is used in generating a magnetic field, and a fraction $\epsilon_e$ is used in heating (accelerating) the electrons. During the initial (coasting) phase, this gives a magnetic field strength $B_{\rm initial} = (8 \pi \epsilon_B u)^{1/2} = 12 ~ (1+z)~A_\star^{1/2}
\Gamma_{i,1.5}^{-1} {t^{\rm obs.}_{3}}^{-1}~\eB^{1/2} ~$G and typical electron Lorentz factor, $\gamma_{m, \rm initial} = \epsilon_e \Gamma_i \left({ m_p / m_e} \right) = 5.5
\times 10^3~\Gamma_{i,1.5} \ee$. Here and below $Q_{,x} = Q/10^x$.

During the deceleration phase, a similar calculation with the use of Supplementary Equation (\ref{eq:3.1}) gives 
\beq
B_{late} = \left({2048 \pi^3 A^3 c^3 (1+z)^3 \epsilon_B^2
  \over 9 E {t^{\rm ob}}^3} \right)^{1/4} = 0.74~\left({1+z \over
  2}\right)^{3/4} \E^{-1/4} A_\star^{3/4} {t^{\rm ob}_{\rm
    day}}^{-3/4} \epsilon_{B,-2}^{1/2}~{\rm G},
\eeq 
and 
\beq
\gamma_{m, late} = \epsilon_e \Gamma {m_p \over m_e} =
\epsilon_e \left( {m_p \over m_e} \right) \left( {9 E (1+z) \over 32
  \pi A c^3 t^{\rm ob}} \right)^{1/4} = 2040 \, \left({1+z \over
  2}\right)^{1/4} \E^{1/4} A_\star^{-1/4} {t^{\rm ob}_{\rm
    day}}^{-1/4}~\epsilon_{e,-1}.
\eeq

\subsubsection*{Supplementary Method 2b. Temporal and spectral signal}
\CreateLink{sec:SuppMethod2b}{2b, `Temporal and spectral signal'}

Using these results in Supplementary Equations (\ref{eq:nu_m}), (\ref{eq:nu_c}) and (\ref{eq:P_nu}) gives the parametric dependence of the key observed frequencies and flux:
\beq
\nu_m^{\rm obs.} = \left\{ \ba{ll}
4.6 \times
10^{16} ~A_\star^{1/2} \Gamma_{i,1.5}^2 {t^{\rm
    obs.}_{3}}^{-1}~\ee^2~\eB^{1/2} ~{\rm Hz} & {\rm (coasting)}, \\
    7.2 \times 10^{13} \,\left({1+z \over
  2}\right)^{1/2} \E^{1/2} {t^{\rm ob}_{\rm day}}^{-3/2}
\ee^2 \eB^{1/2} \Hz & {\rm (decay)},
\ea
\right.
\label{eq:nu_m_final}
\eeq
\beq
\nu_c^{\rm obs.} = \left\{ \ba{ll}
 3 \times 10^{12}
   \left({1+z \over 2}\right)^{-2}~ A_\star^{-3/2} \Gamma_{i,1.5}^2
        {t^{\rm obs.}_{3}}~\eB^{-3/2} ~{\rm Hz}& {\rm (coasting)}, \\
 3.7 \times 10^{13} \, \left({1+z \over
  2}\right)^{-3/2} \E^{1/2} {t^{\rm ob}_{\rm day}}^{1/2}
\eB^{-3/2} A_\star^{-2} \Hz  & {\rm (decay)},
\ea
\right.
\label{eq:nu_c_final}
\eeq
and 
\beq
F_{\nu,peak}^{\rm obs.} = \left\{ \ba{ll}
6.1 \times 10^{-24}
\left({1 + z \over 2} \right) d_{L,28.3}^{-2}~A_\star^{3/2}
\Gamma_{i,1.5}^2 \eB^{1/2} ~{\rm erg~cm^{-2}~s^{-1}~Hz^{-1}} & {\rm (coasting)}, \\
 9.9 \times 10^{-25} \, \left({1+z \over 2}\right)^{3/2}
d_{L,28.3}^{-2} \E^{1/2} A_\star \eB^{1/2} {t^{\rm ob}_{\rm
    day}}^{-1/2} \, {\rm erg~cm^{-2}~s^{-1}~Hz^{-1}} & {\rm (decay)}.
\ea
\right.
\label{eq:nu_F_peak_final}
\eeq
The results of Supplementary Equations (\ref{eq:nu_m_final}) and (\ref{eq:nu_c_final}) imply that both at early (coasting) and later (self-similar decay) phases, the peak frequency decreases with observed time, while the cooling frequency increases with time. This have two important consequences on the spectral and temporal evolution of the observed signal: (i) At high enough frequencies, the transition always occurs from region F ($\nu_c < \nu$) to region E ($\nu < \nu_c$);\footnote{At lower frequencies, it occurs from region D to region E} and (ii) the transition always occurs at lower frequencies first- it will occur in the optical band before the X-ray band, regardless of whether the flow is in the coasting or in the decaying phase. 

Using these results in Supplementary Equations (\ref{eq:fast}) and (\ref{eq:slow}) give the temporal and spectral dependence in each of the 6 possible regimes considered, both during the coasting and during the self-similar decay phases (see for consistency with Ref. \cite{Gao+13}). These are summarized in Supplementary Table \ref{tab:summary} and Supplementary Fig. \ref{fig:summary_regions_plateau}.

These results can now be used to explain the observed signal, which is in an excellent agreement with the theoretical model. Provided that the transition to region E occurs after the end of the coasting phase, the light curve during the coasting phase is $\approx$ flat, $F_\nu \propto t_{\rm obs.}^{(2-p)/2}\sim
t_{\rm obs.}^{0.0 .. -0.2}  $, which is similar to the observed signal in the X-ray band
in classes I and II. Since the transition to region E occurs earlier at longer wavelength,
in some of the bursts in class I the optical band is already in region E, in
which case a decay in the optical light curve,  $F_\nu \propto t_{\rm obs.}^{(1-p)/2} \sim t_{\rm obs.}^{-0.5 .. -0.7}$ is expected. The only other option is that the transition to region E occurs at earlier times, in which case both the X-ray and the optical light curves show a decay, which is the relevant scenario for class III here.
   
Once the flow shifts to the self-similar decay phase, both the X-ray and the optical light curves decay as $F_\nu \propto
t_{\rm obs.}^{(2-3p)/4}\sim t_{\rm obs.}^{-1.0 .. -1.3}$ if in region F, or alternatively $F_\nu
\propto t_{\rm obs.}^{(1-3p)/4} \sim t_{\rm obs.}^{-1.25 .. -1.55}$ if in region E. However, the difference in the temporal index in between these two cases is only 1/4, and may not be easily identified due to the noise. 

\subsubsection*{Supplementary Method 2c. Determining the physical properties of the outflow}
\CreateLink{sec:SuppMethod2c}{2c, `Determining the physical properties of the outflow'}

A flat X-ray plateau indicates that the flux in the X-ray band is in region F. Assuming,
for simplicity a power-law index $p=2$, Supplementary Equation (\ref{eq:slow}) gives 
\beq
\nu F_\nu ({\rm X})^{\rm obs.} = F_{\nu, \rm peak}^{\rm obs.} \nu_c^{1/2} \nu_m^{1/2} = {1 \over d_L^2} c^3 A \Gamma_i^4 \epsilon_e
\label{eq:6.2}
\eeq
where we made use of Supplementary Equations (\ref{eq:nu_m_final}), (\ref{eq:nu_c_final}) and (\ref{eq:nu_F_peak_final}). Combining this result with Supplementary Equation (\ref{eq:t_trans}), one finds that for bursts with known redshift, the transition time and X-ray flux provide a direct measurement of $\epsilon_e$: 
\beq
\epsilon_e = {d_L^2 \over (1+z)} {32 \pi T_{a} \over 9 E} \nu F_\nu ({\rm X})^{\rm obs.}
\label{eq:ee_final}
\eeq

Furthermore, the transition time provides a strong constraint on the ambient medium and the initial Lorentz factor. This is done by writing Supplementary Equation (\ref{eq:t_trans})
\beq
A \Gamma_i^4 = (1+z) {9 E \over  32 \pi c^3 T_{a}},
\label{eq:6.4}
\eeq
which is a very robust result.

A further constrain can be put by using the temporal behaviour of the optical data. In class I, the optical light curve decays, namely the optical band is in region E, and $\nu F_\nu(\rm U)^{\rm obs.} = F_{\nu, \rm peak}^{\rm obs.} \nu_m^{1/2} \nu_{\rm U}^{1/2}$ (for $p=2$).
Using Supplementary Equations (\ref{eq:nu_m_final}), (\ref{eq:nu_F_peak_final}), (\ref{eq:ee_final}) and using an observed optical band $\nu_{\rm U} = 8.65 \times 10^{14}$~Hz as well as fiducial values for  $z = 1$, $ E = 3.16\times10^{52} \rm erg$, $\nu F_\nu ({\rm X})^{\rm obs.} = \rm 3.16\times10^{-12}~erg~cm^{-2}~s^{-1}$ at $T_{a}= 3000 ~\rm s$ and $\nu F_\nu(U)^{\rm obs.} = \rm 10^{-12}~erg~cm^{-2}~s^{-1}$ at 1000~s, one finds 
\beq
A_\star^{7/4} \Gamma_{i, 1.5}^3 \eB^{3/4} = 4 \times 10^{-3}.
\label{eq:eB_classI}
\eeq
Thus, for a given density parameter ($A_\star$), there is a corresponding magnetic parameter ($\epsilon_B$). One may therefore use an external knowledge of $\epsilon_B$ (e.g., an absolute upper limit of $\epsilon_B = 1$) to completely determine the values of $A_\star$ and therefore of $\Gamma_i$.
We further note that in this case, the ratio of fluxes,
\beq
R_I \equiv {\nu F_\nu ({\rm X}) \over \nu F_\nu ({\rm U})} = \left({\nu_c \over \nu_{\rm U}} \right)^{1/2} \left( {\nu_{\rm X} \over \nu_{\rm U}} \right)^{(2-p)/2} 
\eeq
is expected in the range $10 \leq R \leq 300$ for power law indices in the range $1.8 \leq p \leq 2.4$.

In class II, the optical light curve is flat, namely it is in region F. Supplementary Equations (\ref{eq:ee_final}) and (\ref{eq:6.4}) are valid, and thus a direct measurement of $\epsilon_e$ and of $A \Gamma_i^4$ exist. However, as $\nu_c < \nu_{\rm U}$, the ratio of the optical and X-ray fluxes is constant,  
\beq
R_{II} \equiv {\nu F_\nu ({\rm X}) \over \nu F_\nu ({\rm U})} = \left( {\nu_{\rm X} \over \nu_{\rm U}} \right)^{(2-p)/2} \approx 1
\eeq
Instead, we use the fact that $\nu_c$ increases with time (Supplementary Equation (\ref{eq:nu_c_final})) to argue that in this case at the transition time (Supplementary Equation (\ref{eq:t_trans})), $\nu > \nu_c$. This gives a lower limit on the magnetization,
\beq
\eB^{3/2} > 3.1 \times 10^{-3} \left({1+z \over 2} \right)^{-1}   E_{52.5} A_\star^{-5/2} \Gamma_{i,1.5}^{-2}.
\label{eq:6.10}
\eeq
Thus, in this case, for a given $A_\star$ there is a minimum value of the magnetic field, which is inversely proportional to $A_\star$. Thus, again, one may use an external constraint (such as $\epsilon_B< 1$) to constrain a minimum value of $A_\star$ which translates to a maximum value of $\Gamma_i$. 

Finally, in class III both the X-ray and the optical light curves decay, namely both are in region E. In this case, Supplementary Equation (\ref{eq:6.4}) is valid, but not Supplementary Equation (\ref{eq:ee_final}). It is therefore not possible to obtain a direct measure of $\epsilon_e$ in this case. The best constraint can be put by using the requirement that at the beginning of the plateau, at $\sim 100$~s, the X-ray frequency is already below the cooling frequency, $\nu_{\rm X} < \nu_c$. This translates into an upper limit, 
\beq
\left({1+z \over 2} \right) A_\star^{3/2} \Gamma_{i,1.5}^{-2} \eB^{3/2} < 2.5 \times 10^{-6}
\label{eq:6.13}
\eeq
This provides an upper limit on the value of $\epsilon_B$ for a given $A_\star$, which increases as $A_\star$ decrease. Using external constraint, e.g., $\epsilon_B = 0.01$ is therefore useful in providing an upper limit on $A_\star$, and a lower limit on $\Gamma_i$. Combined with a measurement of the optical flux, this also gives a lower limit on the value of $\epsilon_e$.

The ratio of the X-ray to optical fluxes in this case is intermediate, 
\beq
R_{III} \equiv {\nu F_\nu ({\rm X}) \over \nu F_\nu ({\rm U})} = \left( {\nu_{\rm X} \over \nu_{\rm U}} \right)^{(3-p)/2} \approx 10 ... 150
\eeq
depending on the value of $p$.


\subsection*{Supplementary Discussion: Comparison with other models aimed at explaining the X-ray plateau}
\CreateLink{sec:SuppDisc}{`Comparison with other models aimed at explaining the X-ray plateau'}

As discussed in the main text, a plethora of models were proposed in the
literature in attempts to explain the X-ray plateau.  Here we discuss some
of the most recent works, focusing on three topics: energy injection, reverse
shock, and viewing angle. 

Metzger et al. (2011) propose a millisecond protomagnetar model for GRBs
\cite{Metzger+11}. In this magnetar model, the energy released by dipole radiation
is predicted to be transferred to the surroundings via a magnetar wind. This wind
is heated by neutrinos just after the launch of the supernova shock from a core
collapse of a massive star. The outflow is collimated into a bipolar jet by its
interaction with the progenitor star. As the magnetar cools, the wind becomes
ultrarelativistic and Poynting flux dominated on a time-scale comparable to that
required for the jet to clear a cavity through the star. Therefore, in this model
the magnetic dissipation and shocks explain the prompt emission and the
steep decay phase that follows. The late time flares and the X-ray plateau are thought to be powered by continuous dissipation of magnetic energy or 
by extraction of the residual rotational energy.
Indeed, assuming a magnetic field and a spin period typical of a fast
rotating neutron star, Rowlinson et al. (2014) \cite{Rowlinson2014} explain
 the correlation between luminosity and duration of the
plateau phase within the $1\sigma$ uncertainties \cite{Dainotti2017c}
in the framework of this model.
A later paper by Rea et al. (2015) \cite{Rea2015}
shows that the magnetar model can be reconciled within the GRB
emission of the plateau only if supermagnetars
with high magnetic field strength are allowed.

In a successive paper of Stratta et al. (2018) \cite{Stratta2018} for the first
time a non-ideal modelling of spindown magnetar is fitted to the afterglow data
for samples of 40 long and 13 short GRBs with a well-defined plateau.
The conclusion reached in that paper is that the data of both short and long GRBs can be
explained within the magnetar model. However, the difference between the
short and long bursts is that the long GRBs are characterized by a
lower magnetic field and longer spin period compared to the short GRBs.
The correlation between magnetic field and
spin period follows the established physics of the spin-up
line for accreting neutron star (NS) in galactic binary systems. The $B-P$ relation
obtained with this sample of 53 GRBs matches spin-up line predictions
for the magnetar model with mass accretion rates expected in the GRB prompt phase.
The latter are $\sim$ $11-14$ orders of magnitude higher than those inferred for
the galactic accreting NSs.

Matsumoto et al. 2020 \cite{Matsumoto+20} show a link between extended emission
seen in some short GRBs and the plateau phase. They assume a continuous energy injection from a central compact object, such as a magnetar,
for the origin of both emission phases. Such an
extended emission lasting up to $10^2-10^3$~s is also reported in several long GRBs
\cite{BKG13, Kaneko+15}. In our study, when analyzing the full sample of 222 GRBs
with known redshift and a plateau phase, we found that when such an extended emission
exists, it affects the slopes of both the plateau and the following self-similar phases.
The slopes for those bursts that show extended emission are steeper in both the
plateau and self similar phases ($<-0.7$ and $<-1.2$ respectively) than the limit
we take as an indication for a plateau (temporal slope of $>-0.7$). These are
therefore excluded from our final sample. 

Zhao et al. 2020 \cite{Zhao+20} argued that in a few cases, there is a steep
decay following the plateau phase, after which a second plateau may exist. They
interpreted this as an outcome of a central engine consisting
of a rapidly spinning magnetar that collapses to a newborn black hole. When
analyzing the 3 GRBs in their sample, we found that the evidence for a steep decay
is not strong: the X-ray LC of these 3 bursts either contain flares, or
very few data points exist. 

In contrast to previous works, {\c{C}}{\i}k{\i}nto{\u{g}}lu et al. (2020)
\cite{Cikintoglu+20} argue that a clear plateau phase may not be
realized if the magnetic field of the nascent magnetar is in a transient
rapid decay stage. Due to this stage, the spin-down power may decline too
fast and cause a lack of the plateau phase. With this idea, they analyze the X-ray
light curve of 6 GRBs without plateau and conclude that these GRBs might be hosting
millisecond magnetars.

A different type of model was proposed by Lyutikov et al. (2017)
\cite{Lyutikov+17} and Barkov et al. (2021) \cite{Barkov+21}, in which
the plateau phase, flares and possible steep slopes immediately after the
plateau phase, are all explained by a dominant reverse shock. This reverse
shock is different than the "classical" reverse shock \footnote{Emission from such a reverse shock can be easily dominated
by that of the external forward shock in the X-ray band. In fact, the difference
in magnitude between the two components is expected to be at least one order
of magnitude \cite{KZ15}.} predicted as part of the "fireball" model
\cite{UB07, Genet+07} because it is assumed to propagate through ultrarelativistic,
highly magnetized pulsar-like winds produced by long-lasting central engines.
While this is an interesting idea, similar to the models discussed above, only
very few GRBs observed until now seem to fit the predictions of this model.
Moreover, those GRBs have several flares during their X-ray data which makes
it difficult to draw a firm conclusion. Within the realm of
the classical "fireball" model, contribution from the reverse shock is expected
during the transition from the coasting to the similar phase. Indeed when
analyzing the light curves of the GRBs in our sample we did see some evidence
for a possible contribution from a reverse shock in the optical light curve
(especially in GRB 110213A). However, a full analysis of this effect
is beyond the scope of this manuscript, and will be presented elsewhere.

Steep slopes ($<-3$) immediately after the plateau phase
(in this case called internal plateau) were first identified by Troja et
al. 2007 \cite{Troja+2007} in the X-ray light curve of GRB 070110.
The analysis was done by fitting a flat X-ray light-curve up
to ~16000 s, followed by a sharp decline. However, when looking at the data, one
can find that the flux between 8000s and 16000s may be due
to a flare, implying that the data is consistent with a single slope (no plateau
phase) and a large late time flare. If this interpretation is correct, there is
a single decay slope, which is naturally much shallower than the one found by
Troja et al. (2007) \cite{Troja+2007}. Recently, Tang et al. (2019) \cite{Tang+19}
increased the sample to 10 GRBs showing an internal plateau. A
close look into this sample reveals that one can interpret the X-ray
afterglow light-curve of those GRBs in more than one way. The
first point is the effect of flares on the identification of the plateau, as
well as on the light curve decline. In the analysis carried by Tang et
al. (2019) \cite{Tang+19}, no flares were assumed to exist in the X-ray light
curve of those GRBs. However, when including the effects of flares, the temporal
decline index changes. This is further enhanced by the fact that in some bursts
the data is very sparse – in a few cases, in fact no data exists during the
plateau at all, and the existence of a plateau is inferred only by the brightness of the flare (e.g., see the flare identification in
the X-ray light curve of GRB 050730 by Chincarini at al. (2007) \cite{Chincarini+07}). 

Alternatively, internal plateaus might have a different origin
- they may be powered by a long-lasting central engine, in a similar way to
the X-ray flares, that might be powered by an extended
central engine activity \cite{Ghisellini+2007}. Such a model
is commonly used to explain the "internal plateau" seen in the \textit{Swift}-BAT or
\textit{Swift}-XRT light curve of short GRBs around a
few hundred seconds ($\sim 180$ s) \cite{Lu2015}. 

Clearly, once the effects of flares are isolated, the conclusions of the two analysis are different. Thus, overall, we conclude that the “internal plateaus” found in the X-ray light curve of some long GRBs are a matter of interpretation, and more data / analysis is needed. We thus suggest here a different interpretation, which is not inconsistent with the data.

Another suggestion, recently promoted by Oganesyan et al. (2020) \cite{Oganesyan+20}
and Beniamini et al. (2020) \cite{Beniamini+20} is that of a structured jet viewed
 off-axis. These works were motivated by the observation of an
off-axis jet detected from the very low luminosity GRB 170817A, 
associated to GW 170817, the first gravitational waves associated to the merger of
two neutron stars. While Oganesyan et al. (2020)
\cite{Oganesyan+20} explained the plateau phase by high-latitude emission,
Beniamini et al. (2020) \cite{Beniamini+20} argued in favour of a structured jet on
the near-core lines of sight. The addition of degrees of
freedom in these models, might enable good explanation to
bursts which show a plateau.

In this work, we adopted and developed a much simpler model
than all of the above. This model does not require any
addition or modification to the classical GRB "fireball" model. Instead, we
simply consider a different region of parameter space: a flow having an initial
Lorentz factor of the order of few tens, propagating into a "wind" environment,
as proposed earlier by Shen \& Matzner (2012)\cite{ShM12}. We find that this model can naturally
explain the observed plateau, both in the optical and X-rays, within the
framework of synchrotron emission from particles accelerated to a power-law
distribution by the forward shock wave. We further find that these assumptions
lead to a typical ambient density of up to 2 orders of magnitude below the
expectation from a wind produced by a Wolf-Rayet star. In our work, we show
that this simple theoretical idea matches
many observations obtained in both X-ray and optical bands. We further extended
the theory to show how it can be used to extract meaningful information on
the outflow properties, and showed that there is no contradiction between
the parameter values required by our model
and those known from GRBs which do not show a plateau.







~

\begin{table}[ht!]
\centering
\begin{tabular}{cccccccccccc}
\hline
& GRB names & 080607 & 091029 & 110213A & 130831A \\
\hline
\multirow{8}{*}{\rotatebox[origin=c]{90}{X-rays}} &  $\log(T_{\rm a,X}/\rm s)$ & $3.35_{-0.05}^{+0.06}$  &$4.15_{-0.10}^{+0.05}$ &$3.13_{-0.06}^{+0.05}$ &$2.87_{-0.04}^{+0.05}$  \\
& $log(T_{\rm b,X}/\rm s)$ & $4.07_{-0.04}^{+0.05}$  & ... &$3.97_{-0.03}^{+0.03}$ & ...  \\
&$\alpha_{\rm p,X}$ & $0.08_{-0.23}^{+0.23}$  &$-0.22_{-0.06}^{+0.09}$ &$0.30_{-0.11}^{+0.14}$ &$0.04_{-0.18}^{+0.19}$  \\
&$\alpha_{\rm A1,X}$&  $-2.24_{-0.12}^{+0.11}$ &$-1.14_{-0.03}^{+0.03}$ &$-1.06_{-0.05}^{+0.05}$ &$-1.07_{-0.04}^{+0.03}$  \\
&$\alpha_{\rm A2,X}$ &  $-1.44_{-0.06}^{+0.06}$ & ... &$-1.94_{-0.03}^{+0.04}$ & ...  \\
&$\beta_{\rm p,X}$ & $0.82_{-0.15}^{+0.15}$  & $1.06_{-0.12}^{+0.12}$ &$0.98_{-0.10}^{+0.11}$ &$0.83_{-0.11}^{+0.14}$ \\
&$\beta_{\rm A1,X}$ & $1.20_{-0.15}^{+0.16}$  &$1.05_{-0.12}^{+0.12}$ &$1.02_{-0.05}^{+0.06}$ &$0.72_{-0.11}^{+0.12}$  \\
&$\beta_{\rm A2,X}$& $1.13_{-0.19}^{+0.20}$ & ... &$1.14_{-0.19}^{+0.21}$ & ... & \\ \hline
\multirow{8}{*}{\rotatebox[origin=c]{90}{Optical}} & $\log(T_{\rm a,O}/\rm s)$ &  $3.40_{-0.02}^{+0.05}$  &$4.19_{-0.03}^{+0.03}$ &$3.33_{-0.01}^{+0.003}$ & $3.69_{-0.02}^{+0.01}$  \\
& $\log(T_{\rm b,O}/\rm s)$ & ...  & ... & $3.71_{-0.01}^{+0.01}$ & 931.0   \\ 
& $\alpha_{\rm p,O}$ &  $-0.49_{-0.02}^{+0.01}$ &$-0.36_{-0.02}^{+0.01}$ & $-0.61_{-0.02}^{+0.02}$ &$-0.4_{-0.01}^{+0.01}$  \\
& $\alpha_{\rm A1,O}$ & $-1.22_{-0.01}^{+0.02}$ & $-1.06_{-0.01}^{+0.01}$ & $0.68_{-0.07}^{+0.07}$ &$-1.49_{-0.03}^{+0.04}$  \\
& $\alpha_{\rm A2,O}$& ...  & ... & $-1.56_{-0.02}^{+0.02}$ & ...  \\ 
& $\beta_{\rm p,O}$ & $-0.84_{-0.01}^{+0.01}$  & $-0.46_{-0.06}^{+0.06}$ & ...  & ...  \\
& $\beta_{\rm A1,O}$& $-0.80_{0.0}^{0.0}$ & $-0.32_{-0.06}^{+0.05}$ & $-1.12_{-0.24}^{+0.24}$ & ...  \\
& $\beta_{\rm A2,O}$ & ... & $-0.34_{-0.06}^{+0.06}$ &$-1.22_{-0.18}^{+0.18}$ & ... \\ \hline
\multicolumn{2}{c}{\begin{tabular}{c}References \\ for optical \\ spectral indices: \end{tabular}} & Ref.\cite{Zaninoni+13}&Ref.\cite{Filgas2012} &  Ref.\cite{Cucchiara2011} &
\end{tabular}
\caption{\label{tab:ClassI} \textbf{Class I (flat X-ray plateau and decaying optical
plateau).} Row 1: GRB names in this class. Row 2 and 3:
temporal breaks in the X-ray light curves at the end of the plateau phase
and later time respectively. Row 4 to 9: temporal and spectral slopes
in the X-ray band. Row 10 and 11: temporal breaks
in the optical light curves at the end of the plateau phase and at
later time respectively.  Row 12 to 17: temporal and spectral
slopes in the optical band. The errors correspond to a significance of one sigma. 
See additional details in Supplementary Method \LinkTo{sec:SuppMethod1a} as well as in Supplementary Method \LinkTo{sec:SuppMethod1b}.} 
\end{table}

~

\begin{table}[ht!]
\centering
\begin{tabular}{cccccccccccc}
\hline
& GRB name & 060605 & 060614 & 060729 & 080310 & 100418A & 171205A \\
\hline 
\multirow{8}{*}{\rotatebox[origin=c]{90}{X-rays}}&$\log(T_{\rm a,X}/\rm s)$ &$3.69_{-0.16}^{+0.11}$ &$4.53_{-0.03}^{+0.03}$  & $4.58_{-0.02}^{+0.04}$ & $4.04_{-0.04}^{+0.04}$  & $4.90_{-0.07}^{+0.11}$ & $4.96_{-0.04}^{+0.04}$  \\
&$\log(T_{\rm b,X}/\rm s)$ & $4.18_{-0.06}^{+0.09}$ &$5.13_{-0.09}^{+0.09}$  & $5.15_{-0.04}^{+0.04}$ & ... & ... & ... & \\
&$\alpha_{\rm p,X}$ &$-0.41_{-0.04}^{+0.08}$ & $0.02_{-0.05}^{+0.05}$  & $-0.14_{-0.02}^{+0.02}$ & $-0.21_{-0.10}^{+0.10}$ & $-0.09_{-0.06}^{+0.06}$ & $0.21_{-0.09}^{+0.09}$ \\
 &$\alpha_{\rm A1,X}$ &$-1.24_{-0.25}^{+0.15}$ &$-1.41_{-0.11}^{+0.13}$  & $-0.87_{-0.06}^{+0.05}$ & $-1.55_{-0.06}^{+0.06}$  & $-1.39_{-0.13}^{+0.09}$ & $-1.06_{-0.05}^{+0.05}$ \\
&$\alpha_{\rm A2,X}$ & $-2.25_{-0.17}^{+0.14}$ &$-2.23_{-0.15}^{+0.12}$  & $-1.41_{-0.02}^{+0.02}$ & ... & ... & ... \\
&$\beta_{\rm p,X}$ &$0.90_{-0.07}^{+0.14}$ &$0.74_{-0.05}^{+0.11}$   & $0.97_{-0.04}^{+0.04}$ & $1.14_{-0.07}^{+0.12}$ & $0.9_{-0.3}^{+0.4}$ & $0.55_{-0.10}^{+0.21}$\\
&$\beta_{\rm A1,X}$ &$1.16_{-0.17}^{+0.17}$ &$0.92_{-0.15}^{+0.16}$  & $0.98_{-0.04}^{+0.04}$ & $0.89_{-0.15}^{+0.16}$  & $0.84_{-0.29}^{+0..32}$ & $0.94_{-0.22}^{+0.23}$ \\
&$\beta_{\rm A2,X}$ &$1.08_{-0.15}^{+0.20}$ &  ... &  ... &  ...  & ... & ... \\
\hline
\multirow{9}{*}{\rotatebox[origin=c]{90}{Optical}}&$\log(T_{\rm a,O}/\rm s)$ & $2.77_{-0.03}^{+0.02}$ &$4.47_{-0.04}^{+0.02}$  & $4.78_{-0.05}^{+0.05}$ & $3.54_{-0.12}^{+0.10}$ & $4.67_{-0.14}^{+0.10}$  &$4.55_{-0.04}^{+0.02}$   \\
&$\log(T_{\rm b,O}/\rm s)$ & $3.47_{-0.02}^{+0.05}$ & $4.84_{-0.03}^{+0.05}$   & ... & $4.60_{-0.62}^{+0.27}$  & ... &$5.25_{-0.04}^{+0.04}$  \\ 
&$\alpha_{\rm p,O}$ & $0.18_{-0.02}^{+0.01}$ &$0.10_{-0.03}^{+0.04}$  & $-0.07_{-0.03}^{+0.03}$ & $-0.05_{-0.08}^{+0.08}$ & $0.22_{-0.08}^{+0.07}$ &$0.15_{-0.01}^{+0.01}$  \\
&$\alpha_{\rm A1,O}$ &$-1.02_{-0.02}^{+0.01}$ &$-0.96_{-0.15}^{+0.16}$  & $-1.49_{-0.08}^{+0.07}$ & $-1.02_{-0.12}^{+0.16}$  & $-0.70_{-0.05}^{+0.05}$ &$-0.65_{-0.01}^{+0.01}$  \\
&$\alpha_{\rm A2,O}$ &$-1.62_{-0.01}^{+0.02}$ &$-2.02_{-0.08}^{+0.08}$ & ... & $-1.26_{-0.33}^{+0.19}$  & ... &$0.44_{-0.01}^{+0.01}$  \\
&$\beta_{\rm p1,O}$ &$-1.32_{-0.03}^{+0.03}$ &$-0.30_{0.0}^{0.0}$  & $-0.59_{0.0}^{0.0}$ & $-0.97_{-0.04}^{+0.04}$   & ... &$-0.65_{-0.39}^{+0.39}$ \\
&$\beta_{\rm p2,O}$ & ... & ...  & $-0.88_{0.01}^{0.0}$ & ...  & ... & $-0.83_{-0.19}^{+0.19}$  \\
&$\beta_{\rm A1,O}$ &$-1.14_{-0.02}^{+0.02}$ &$-0.35_{0.0}^{0.0}$  & $-0.54_{0.0}^{0.0}$ & $-0.88_{-0.04}^{+0.04}$  & $-1.19_{-0.02}^{+0.02}$ &$-0.77_{-0.15}^{+0.15}$  \\
&$\beta_{\rm A2,O}$& ... & ...   & ... & ... &  ... &  $-0.88_{-0.60}^{+0.60}$ \\
\hline
\multicolumn{2}{c}{\begin{tabular}{c}References \\ for optical \\ spectral indices: \end{tabular}} & Ref.\cite{Zaninoni+13}&Ref.\cite{Zaninoni+13} &  Ref.\cite{Zaninoni+13} & Ref.\cite{Zaninoni+13}& Ref.\cite{Zaninoni+13}&
\end{tabular}
\caption{\label{tab:ClassII} \textbf{Class II (both X-ray and optical plateaus are flat).} Columns and Rows are as defined in Table \ref{tab:ClassI}.
}
\end{table}
~

\begin{table}[ht!]
\centering
\begin{tabular}{ccccccccccc}
\hline
&GRB name & 050319 & 060714 & 061121  \\
\hline 
\multirow{6}{*}{\rotatebox[origin=c]{90}{X-rays}}&$\log(T_{\rm a,X}/\rm s)$ &$4.50_{-0.06}^{+0.06}$ & $3.66_{-0.10}^{+0.13}$ & $3.96_{-0.03}^{+0.03}$ \\
&$\alpha_{\rm p,X}$ &$-0.55_{-0.02}^{+0.02}$  & $-0.42_{-0.11}^{+0.10}$  & $-0.59_{-0.01}^{+0.02}$  \\
&$\alpha_{\rm A1, X}$ &$-1.44_{-0.12}^{+0.11}$  & $-1.24_{-0.04}^{+0.03}$ &  $-1.42_{-0.02}^{+0.02}$  \\
&$\beta_{\rm p,X}$ &$1.00_{-0.06}^{+0.09}$  &$0.88_{-0.14}^{+0.15}$ &  $0.84_{-0.28}^{+0.30}$  \\
&$\beta_{\rm A1,X}$ &$1.18_{-0.20}^{+0.21}$ &$1.08_{-0.19}^{+0.20}$ & $0.85_{-0.07}^{+0.07}$  \\
&$\beta_{\rm A2,X}$ & ...  & ...  & $0.77_{-0.10}^{+0.10}$  \\
\hline
\multirow{5}{*}{\rotatebox[origin=c]{90}{Optical}}&$\log(T_{\rm a,O}/\rm s)$ &$4.55_{-0.04}^{+0.04}$ &$3.77_{-0.18}^{+0.16}$ & $4.14_{-0.32}^{+0.35}$   \\
&$\alpha_{\rm p,O}$ &$-0.59_{-0.01}^{+0.01}$ &$-0.27_{-0.09}^{+0.09}$ & $-0.48_{-0.05}^{+0.05}$   \\
&$\alpha_{\rm A1,O}$ &$-0.64_{-0.011}^{+0.012}$  &  $-0.61_{-0.28}^{+0.27}$ & $-1.08_{-0.64}^{+0.27}$  \\
&$\beta_{\rm p,O}$ &$-0.36_{0.0}^{0.0}$  &$-0.44_{-0.04}^{+0.04}$  & $-0.68_{-0.06}^{+0.06}$  \\
&$\beta_{\rm A1,O}$ &$-0.61_{0.0}^{0.0}$ & $-0.98_{-0.11}^{+0.11}$ & $-0.68_{-0.02}^{+0.02}$  \\
\hline
\multicolumn{2}{c}{\begin{tabular}{c}References \\ for optical \\ spectral indices: \end{tabular}} & Ref.\cite{Zaninoni+13}& Ref.\cite{Zafar2011} &  Ref.\cite{Zaninoni+13}
\end{tabular}
\caption{\label{tab:ClassIII} \textbf{Class III (both X-ray and optical plateaus are decaying).} Columns and Rows are as defined in Table \ref{tab:ClassI}.
} 
\end{table}
~

\begin{table}[ht!]
\centering
\begin{tabular}{ccccccccccc}
\hline
$\log(T_a/s)$ &  $\log(T_b/s)$ & N & $\alpha_{\rm p}$ & $\alpha_{\rm A1}$ &  $\alpha_{\rm A2}$ & $\log(V)$\\
\hline
4.53 $_{-0.03}^{+0.03}$ & 5.13 $_{-0.09}^{+0.09}$  & -0.68 $_{-0.02}^{+0.02}$  & 0.02 $_{-0.05}^{+0.05}$  & -1.41 $_{-0.11}^{+0.13}$  & -2.23 $_{-0.15}^{+0.12}$  &-11.54$_{-2.55}^{+2.34}$ \\
\hline
\end{tabular}
\caption{\label{tab:fit_paramaters_060614} \textbf{X-ray temporal fit parameters of GRB 060614 with the Bayesian analysis tool \textit{emcee}.} ${T_{\rm a}}$ is the break time at the end of the plateau phase presented in a log scale and ${T_{\rm b}}$ is the jet break time presented in a log scale. N is the normalization of the fit. The slopes $\alpha_{\rm p}$, $\alpha_{\rm A1}$, $\alpha_{\rm A2}$ are obtained during the plateau phase, self-similar phase and after the jet break time respectively. V is the intrinsic scatter of the data presented in a log scale. The errors correspond to a significance of one sigma.}
\end{table}
~

\begin{table}[ht]
\centering
\begin{tabular}{|l|l|l|}
\hline
Scenario & Coasting & Self-Similar Decay \\
\hline
A~~~$\nu < \nu_c < \nu_m$ &  $F_\nu \propto t_{\rm obs.}^{-1/3} \nu^{1/3}$  & $F_\nu \propto t_{\rm obs.}^{-2/3}  \nu^{1/3}$ \\

B~~~$\nu_c < \nu < \nu_m$ &  $F_\nu \propto t_{\rm obs.}^{1/2} \nu^{-1/2}$ & $F_\nu \propto t_{\rm obs.}^{-1/4} \nu^{-1/2}$ \\

C~~~$\nu_c < \nu_m < \nu$ & $F_\nu \propto t_{\rm obs.}^{(2-p)/2} \nu^{-p/2} \sim
   t_{\rm obs.}^{0 .. -0.2}$ & $F_\nu \propto
t_{\rm obs.}^{(2-3p)/4}  \nu^{-p/2}  \sim t_{\rm obs.}^{-1 .. -1.3}$ \\ 

D~~~$\nu < \nu_m < \nu_c$ & $F_\nu \propto t_{\rm obs.}^{1/3} \nu^{1/3}$ & $F_\nu \propto t_{\rm obs.}^{0} \nu^{1/3}$ \\

E~~~$\nu_m < \nu < \nu_c$ &  $F_\nu \propto t_{\rm obs.}^{(1-p)/2}
\nu^{-(p-1)/2} \sim t_{\rm obs.}^{-0.5 .. -0.7}$ & $F_\nu
\propto t_{\rm obs.}^{(1-3p)/4}  \nu^{-(p-1)/2} \sim t_{\rm obs.}^{-1.25 .. -1.55}$ \\

F~~~$\nu_m < \nu_c < \nu$ &  $F_\nu \propto t_{\rm obs.}^{(2-p)/2} \nu^{-p/2} \sim
   t_{\rm obs.}^{0 .. -0.2}$ & $F_\nu \propto
t_{\rm obs.}^{(2-3p)/4}  \nu^{-p/2} \sim t_{\rm obs.}^{-1 .. -1.3}$ \\
\hline
\end{tabular}
\caption{\label{tab:summary} \textbf{Temporal and spectral evolution in each of the
possible six spectral regimes with the corresponding regions indicated
in the first column}. Here, $p$ is the power-law index of the accelerated electrons.
Expected values based on both theory \protect\cite{Spit08a} and observations \protect\cite{WG99}
suggest $2.0 \leq p \lesssim 2.4$. We indicate the expected temporal evolution
in regions C, E and F for power-law index in this range. 
}
\end{table}
~

\begin{table}[ht!]
\centering
\begin{tabular}{ccccccccccc}
\hline
GRB name & $A_{\rm U,Gal.}$  & $A_{\rm U,host}$ & Extinction law & note\\
\hline
050319 & 0.052 & 0.53$\pm$0.012 & MW & (V-band) \cite{Zaninoni+13} \\
060605 & 0.255 &  1.33$\pm$0.10 & MW & (CR-band) \cite{Zaninoni+13}\\
060614 & 0.107 & 0.23$\pm$0.18 & MW & (U-band) \cite{Zaninoni+13}\\
060714 & 0.378 & $1.53^{+0.65}_{-0.58}$ & ... & $Av_{host}$ = $0.79^{+0.39}_{-0.35}$ from Ref.\cite{Schady+10}\\
060729 & 0.272 & 0.433$\pm$0.18 & SMC & (U-band) \cite{Zaninoni+13}\\
061121 & 0.226 & 0.92$\pm$0.30 & MW & (U-band)  \cite{Zaninoni+13}\\
080310 & 0.196 &  0.433$\pm$0.18 & SMC & (U-band)  \cite{Zaninoni+13}\\
080607 & 0.111 & 2.98$\pm$0.37 & MW & (Rc-band)  \cite{Zaninoni+13}\\
091029 & 0.08 & ... & ... & (U-band) no host galaxy extinction \cite{Filgas2012} \\
100418A & 0.362 & 1.2$\pm$0.43 & SMC & (white-band)  \cite{Zaninoni+13}\\
110213A & 1.612 & ... & ... & (U-band) no host galaxy extinction \cite{Cucchiara2011} \\
130831A & 0.223 & 0.103$\pm$0.0513 & MW & (U-band) low host galaxy extinction \cite{dePasquale2018}\\
171205A & 0.251 &  0.103 & MW & (U-band) \cite{Izzo2019} \\
\hline
\end{tabular}
\caption{\label{tab:Extinctions} \textbf{Galactic and host galaxy extinctions of
the 13 GRBs in our sample.} Column 1: GRB name, Columns 2 and 3:
Galactic and host galaxy extinctions in the U band. The errors correspond to a significance of one sigma. Column 4: The
 extinction law used to compute the extragalactic
reddening. Column 5: band passes used in our analyses and corresponding
references.}
\end{table}
~

\begin{figure}[ht!]
\centering
\begin{tabular}{cc}
\Huge{a} &  \raisebox{-.5\height}{\includegraphics[width=0.8\linewidth]{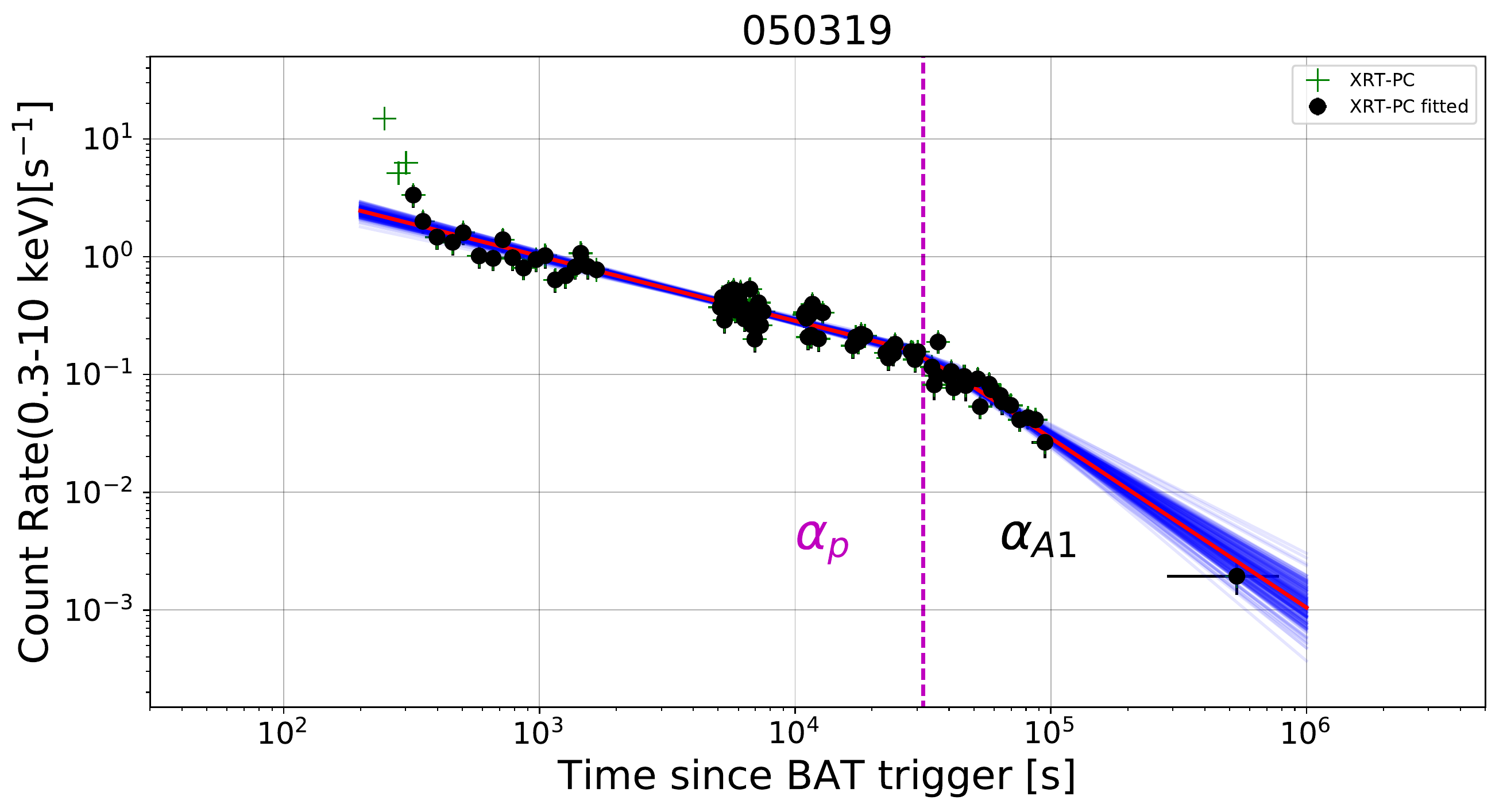}} \\
\Huge{b} & \raisebox{-.5\height}{\includegraphics[width=0.8\linewidth]{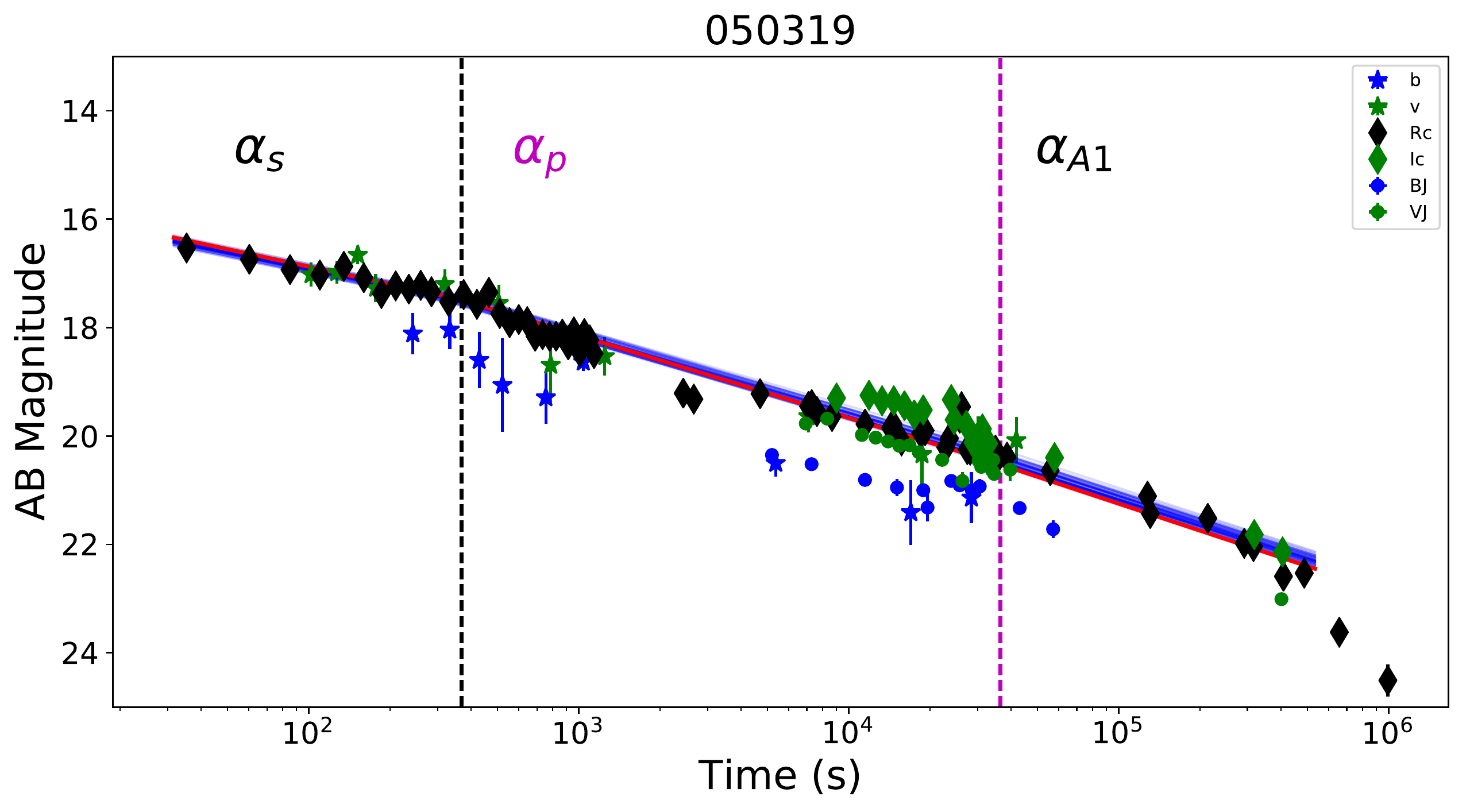}}
\end{tabular}
\caption{\textbf{\large (a)} \textbf{The X-ray LC of GRB 050319.} The green
crosses represent the XRT-PC mode data which are excluded from the fit. The black dots represent XRT-PC mode fitted data. The errors correspond to a significance of one sigma. The dashed vertical (purple) line represents the break times $T_{\rm a}$ (at the end of the plateau phase). The $\alpha_{\rm p}$ and $\alpha_{\rm A1}$ are the slopes during the plateau phase and the self-similar phase respectively.
\textbf{\large (b)} \textbf{The optical LC of GRB 050319.} The blue and green colors together with stars, diamonds and dots represent the data in b, v, Rc, Ic, BJ, VJ bands respectively. The errors correspond to a significance of one sigma. For the details of the excluded data, see, Supplementary Method \LinkTo{sec:SuppMethod1b}.
Although all the bands are jointly fitted, for the demonstration purpose, only the Rc band fit result is displayed. The dashed vertical lines (black and
purple) represent the break times $T_{\rm s}$ and $T_{\rm a}$ (at the end of the plateau
phase) respectively. The $\alpha_{\rm As}$, $\alpha_{\rm p}$, $\alpha_{\rm A1}$, are the slopes before the plateau phase, during the plateau phase and self-similar
phase respectively. In both panels \textbf{\large (a)} and \textbf{\large (b)}, the red line shows the mean of the posterior distribution and blue lines are 200 randomly selected samples from the MCMC sampling.}
\label{fig:fit_results_050319_X-ray_opt}
\end{figure}
~

\begin{figure}[ht!]
\centering
\begin{tabular}{cc}
\Huge{a} &  \raisebox{-.5\height}{\includegraphics[width=0.8\linewidth]{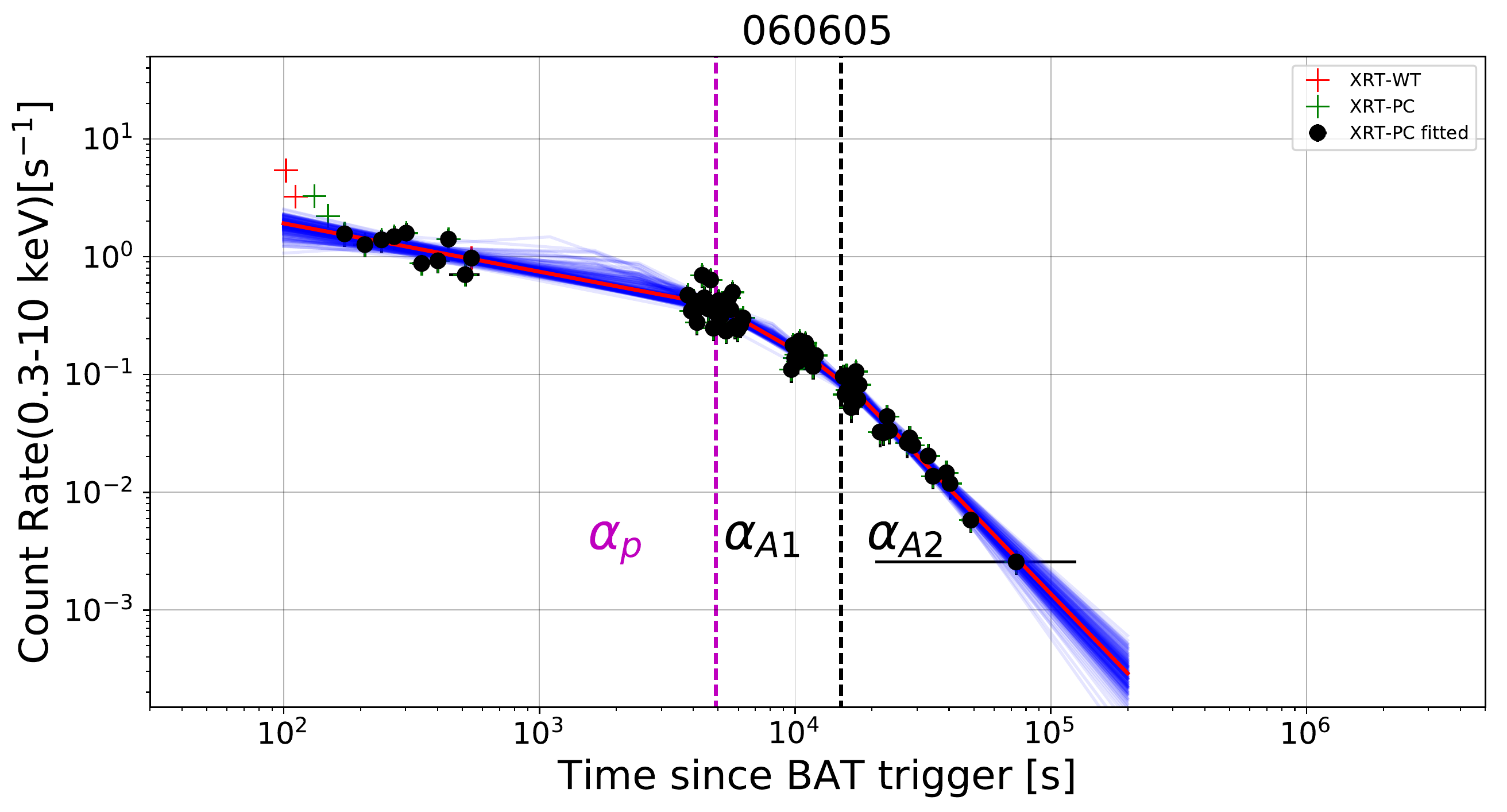}} \\
\Huge{b} & \raisebox{-.5\height}{\includegraphics[width=0.8\linewidth]{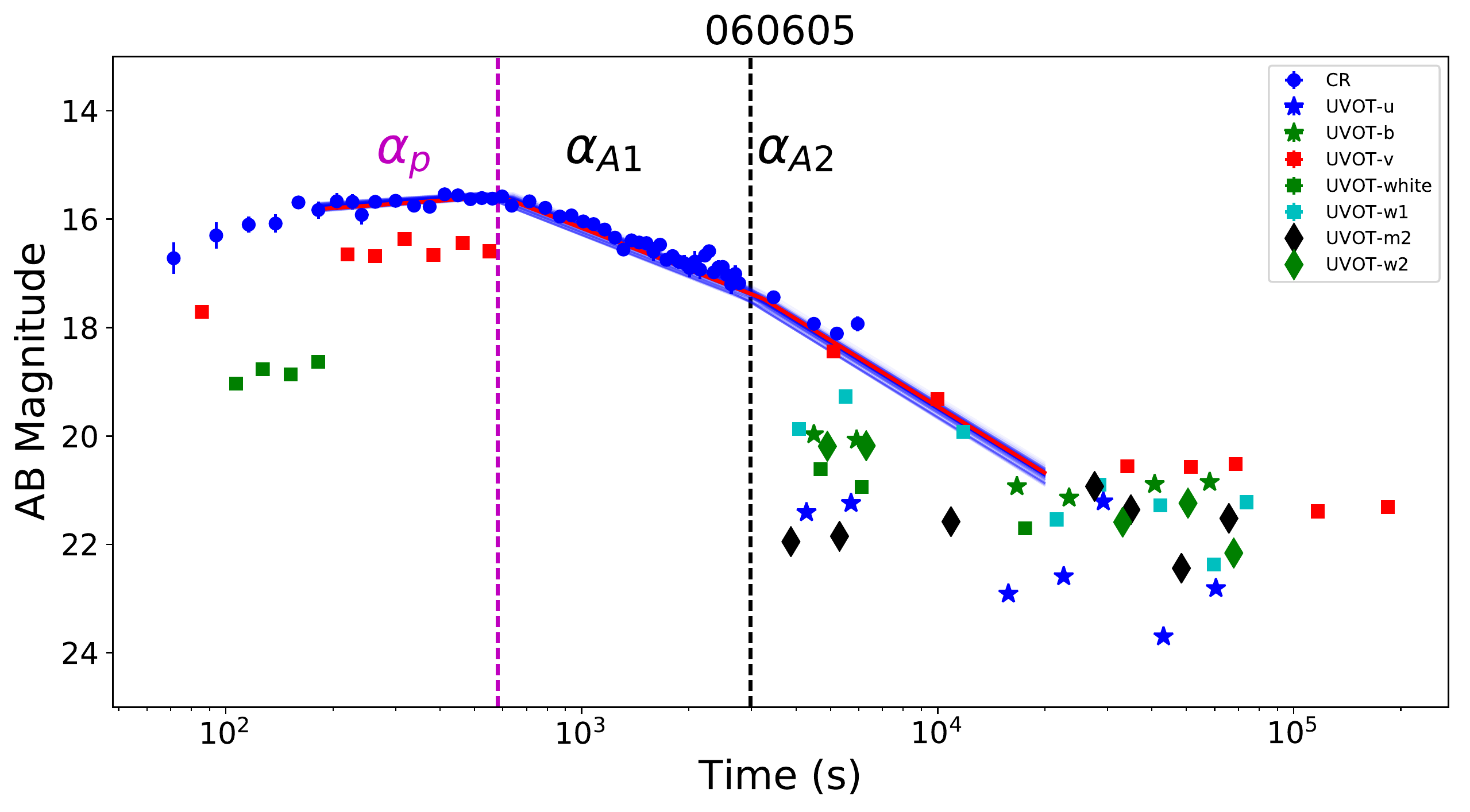}}
\end{tabular}
\caption{\textbf{\large (a)} \textbf{The X-ray LC of GRB 060605.} The red and green crosses represent the XRT-WT and the XRT-PC mode data respectively which are excluded from the fit. The black dots represent the XRT-PC mode fitted data. The errors correspond to a significance of one sigma.
\textbf{\large (b)} \textbf{The optical LC of GRB 060605.} The blue, green, red, cyan, black colors together with points, stars, squares diamonds represent the data in CR, UVOT-u, UVOT-b, UVOT-v, UVOT-white, UVOT-w1, UVOT-m2, UVOT-w2 bands respectively. The errors correspond to a significance of one sigma. For the details of the excluded data, see, Supplementary Method \LinkTo{sec:SuppMethod1b}. Although all the bands jointly fitted, for the demonstration purpose only CR band fit result displayed. In both panels \textbf{\large (a)} and \textbf{\large (b)}, the red line shows the mean of the posterior distribution and blue lines are 200 randomly selected samples from the MCMC sampling. The dashed vertical lines (purple and black) represent the break times $T_{\rm a}$ (at the end of the plateau phase) and $T_{\rm b}$ respectively. The $\alpha_{\rm p}$, $\alpha_{\rm A1}$, $\alpha_{\rm A2}$ are slopes during the plateau phase, self-similar phase and after the jet break time respectively.}
\label{fig:fit_results_060605_X-ray_opt}
\end{figure}
~

\begin{figure}[ht!]
\centering
\begin{tabular}{cc}
\Huge{a} &  \raisebox{-.5\height}{\includegraphics[width=0.8\linewidth]{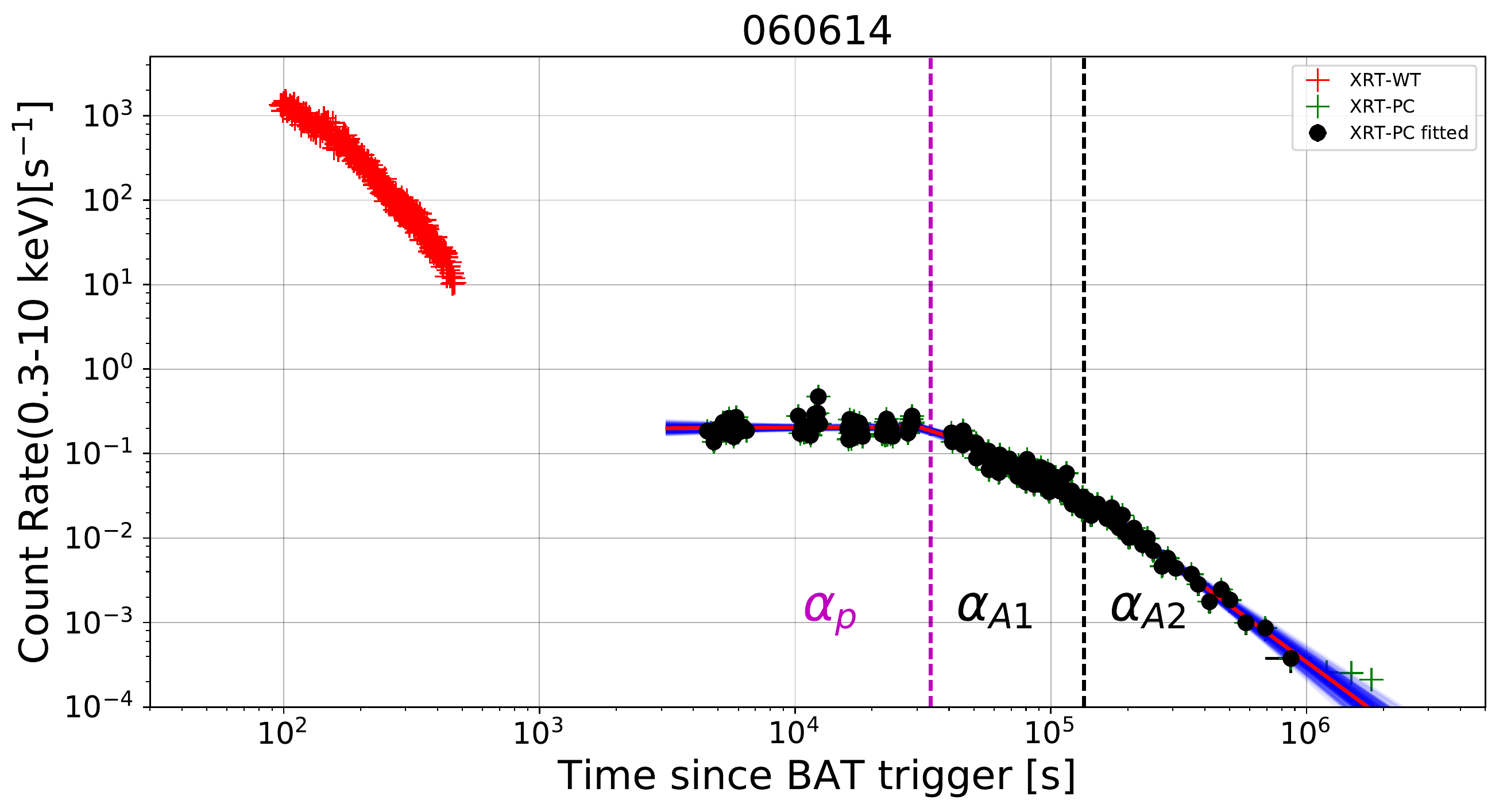}} \\
\Huge{b} & \raisebox{-.5\height}{\includegraphics[width=0.8\linewidth]{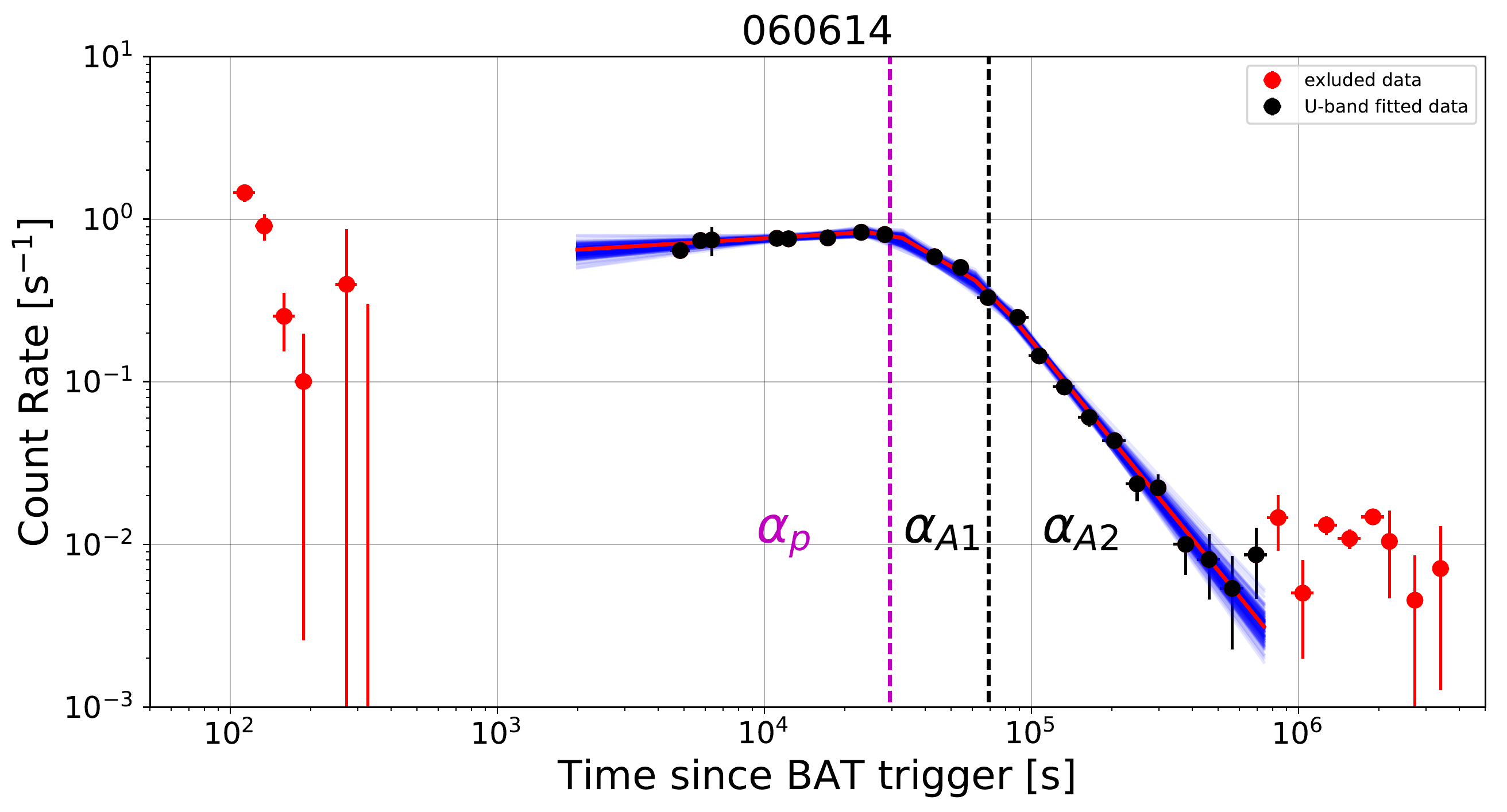}}
\end{tabular}
\caption{\textbf{\large (a)} \textbf{The X-ray LC of GRB 060614.} The red and green crosses represent the XRT-WT and the XRT-PC mode data respectively which are excluded from the fit. The black dots represent the XRT-PC mode fitted data. The errors correspond to a significance of one sigma.
\textbf{\large (b)} \textbf{The optical LC of GRB 060614.} The black and red dots represent the data in UVOT-u band. The errors correspond to a significance of one sigma. While the black dots are considered in the fitting process, the red dots are excluded from the fit (see, Supplementary Method \LinkTo{sec:SuppMethod1b} for explanations). In both panels \textbf{\large (a)} and \textbf{\large (b)}, the red line shows the mean of the posterior distribution and blue lines are 200 randomly selected samples from the MCMC sampling. The dashed vertical lines (purple and black) represent the break times $T_{\rm a}$ (at the end of the plateau phase) and $T_{\rm b}$ respectively. The $\alpha_{\rm p}$, $\alpha_{\rm A1}$, $\alpha_{\rm A2}$ are slopes during the plateau phase, self-similar phase and after the jet break time respectively.}
\label{fig:fit_results_060614_X-ray_opt}
\end{figure}
~

~

\begin{figure}[ht!]
\centering
\begin{tabular}{cc}
\Huge{a} &  \raisebox{-.5\height}{\includegraphics[width=0.8\linewidth]{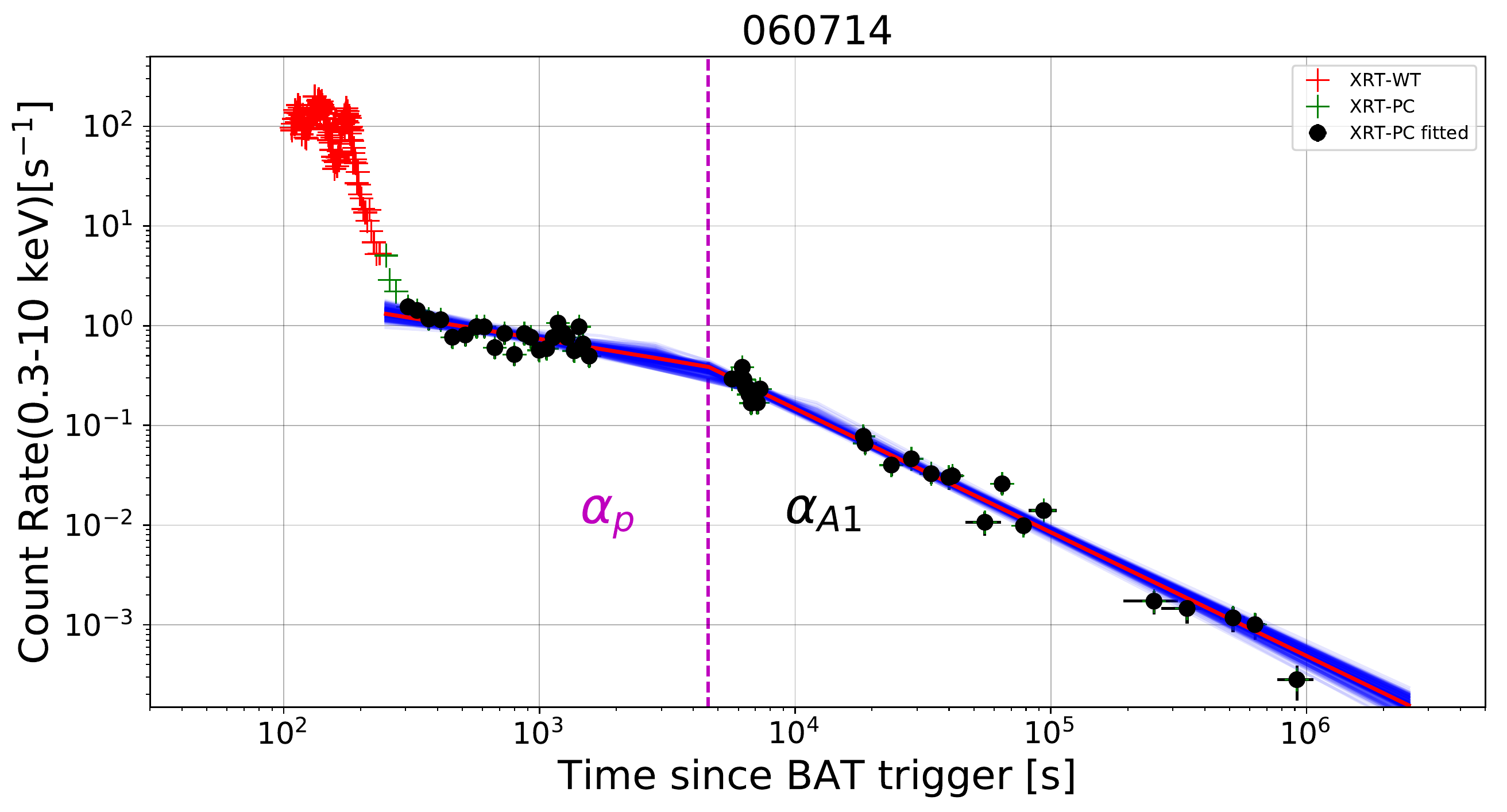}} \\
\Huge{b} & \raisebox{-.5\height}{\includegraphics[width=0.8\linewidth]{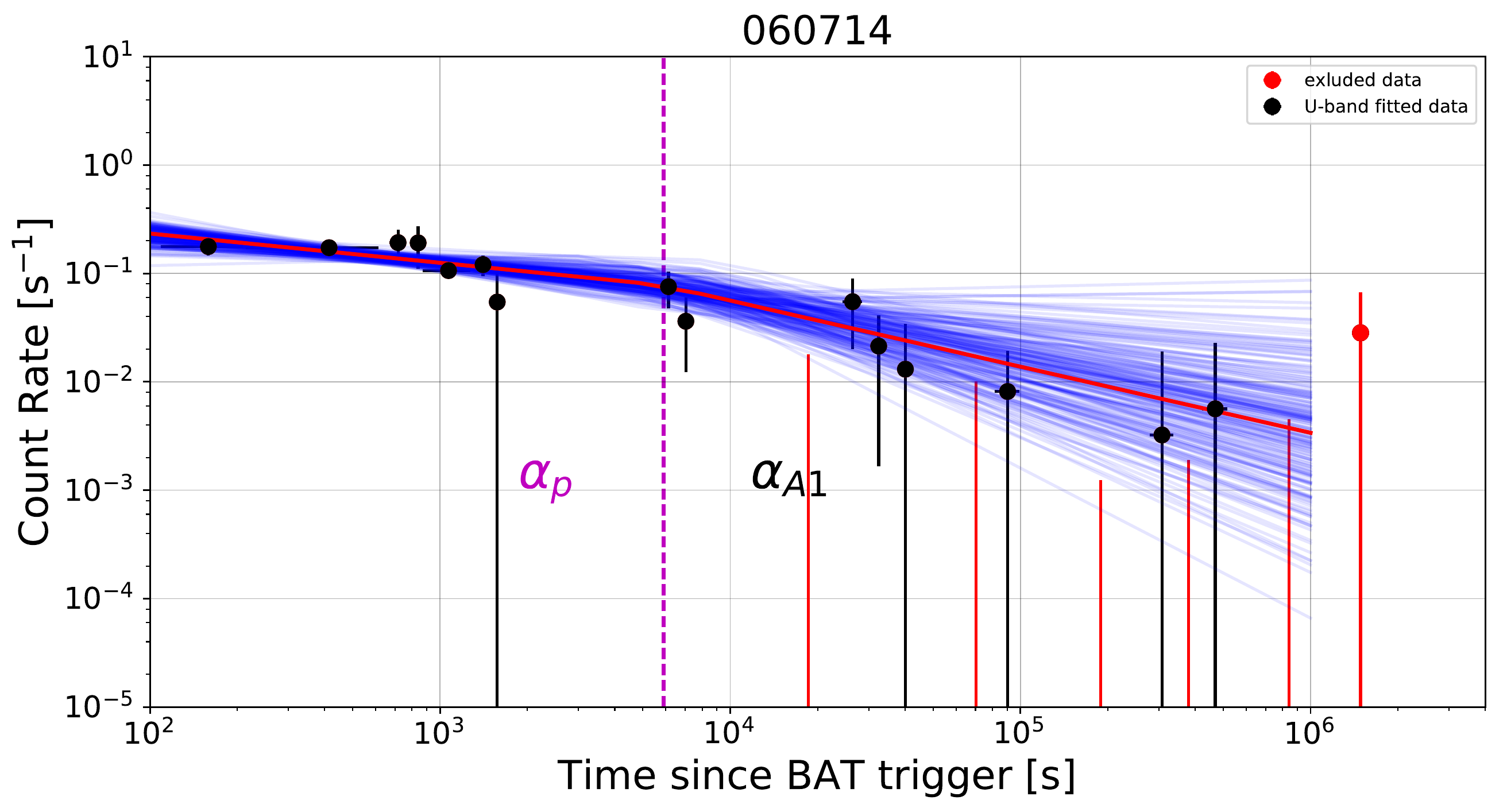}}
\end{tabular}
\caption{\textbf{\large (a)} \textbf{The X-ray LC of GRB 060714.} The red and green crosses represent the XRT-WT and the XRT-PC mode data respectively which are excluded from the fit. The black dots represent the XRT-PC mode fitted data. The errors correspond to a significance of one sigma. 
\textbf{\large (b)} \textbf{The optical LC of GRB 060714.} The black and red dots represent the data in UVOT-u band. The errors correspond to a significance of one sigma. While the black dots are considered in the fitting process, the red dots are excluded from the fit (see, Supplementary Method \LinkTo{sec:SuppMethod1b} for explanations). In both panels \textbf{\large (a)} and \textbf{\large (b)}, the red line shows the mean of the posterior distribution and blue lines are 200 randomly selected samples from the MCMC sampling. The dashed vertical line (purple) represents the break times $T_{\rm a}$ (at the end of the plateau phase). The $\alpha_{\rm p}$, $\alpha_{\rm A1}$ are slopes during the plateau phase and self-similar phase respectively.}
\label{fig:fit_results_060714_X-ray_opt}
\end{figure}
~

\begin{figure}[ht!]
\centering
\begin{tabular}{cc}
\Huge{a} &  \raisebox{-.5\height}{\includegraphics[width=0.8\linewidth]{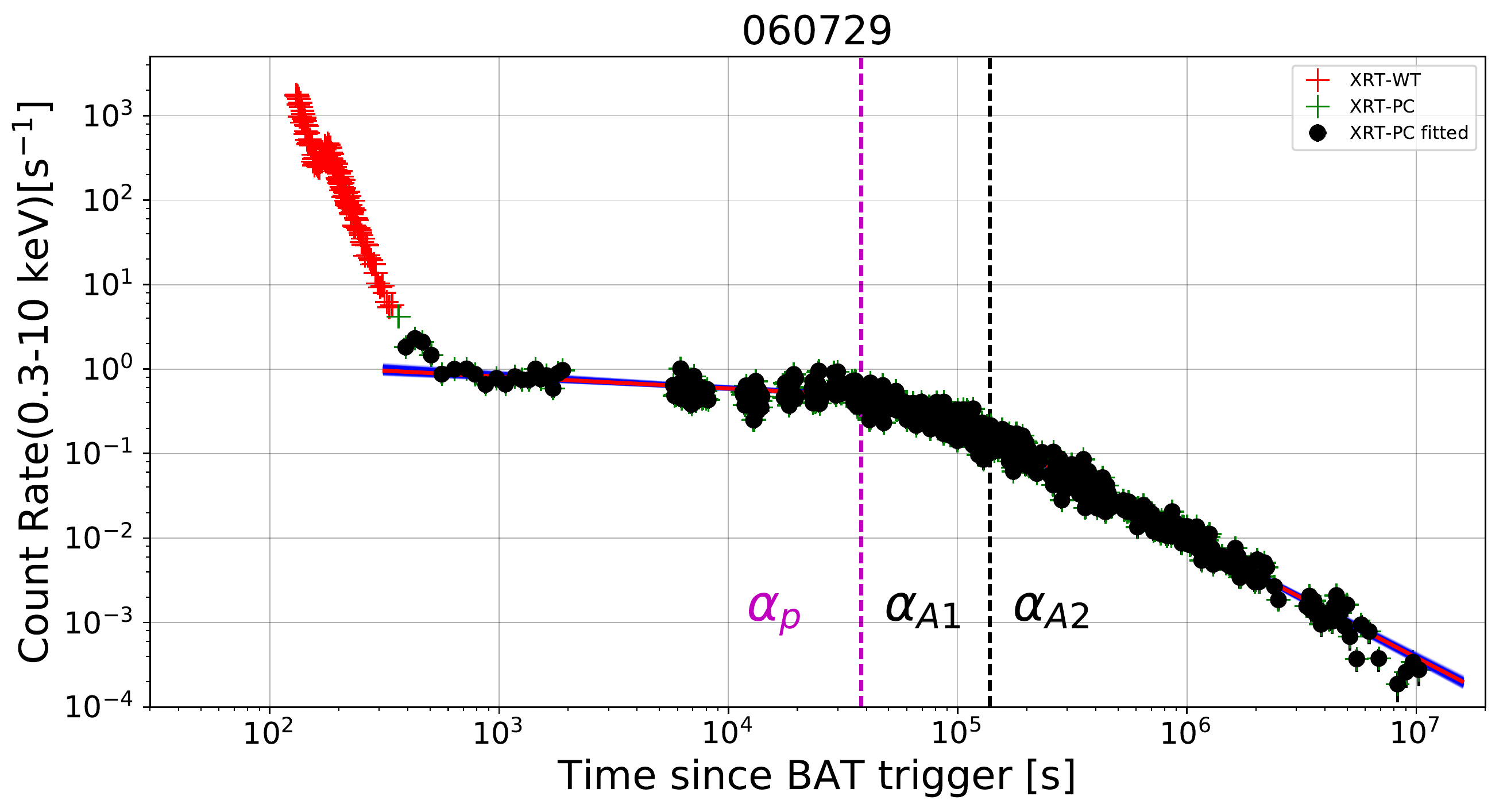}} \\
\Huge{b} & \raisebox{-.5\height}{\includegraphics[width=0.8\linewidth]{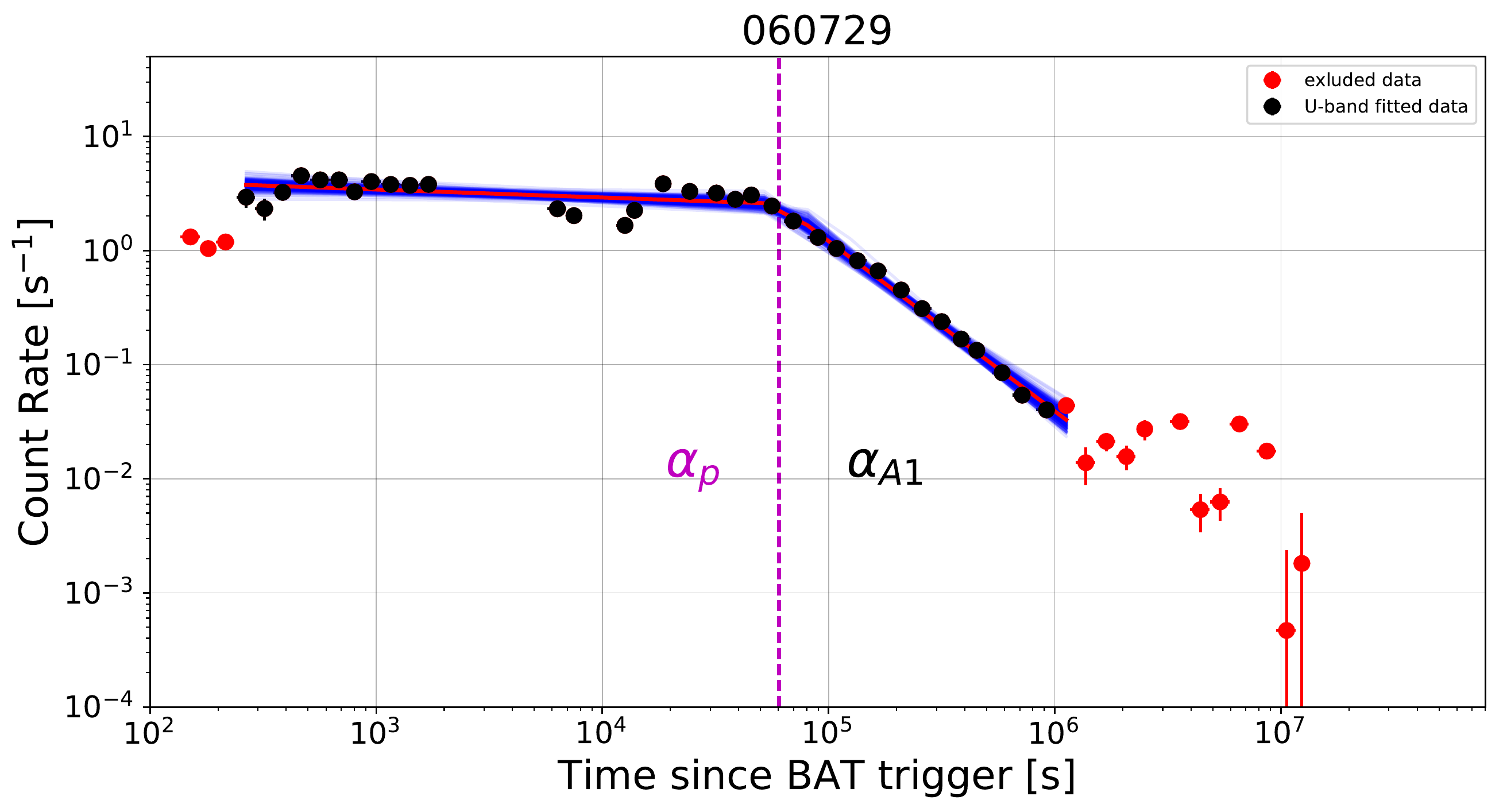}}
\end{tabular}
\caption{\textbf{\large (a)} \textbf{The X-ray LC of GRB 060614.} The red and green crosses represent the XRT-WT and the XRT-PC mode data respectively which are excluded from the fit. The black dots represent the XRT-PC mode fitted data. The errors correspond to a significance of one sigma. The dashed vertical lines (purple and black) represent the break times $T_{\rm a}$ (at the end of the plateau phase) and $T_{\rm b}$ respectively. The $\alpha_{\rm p}$, $\alpha_{\rm A1}$, $\alpha_{\rm A2}$ are slopes during the plateau phase, self-similar phase and after the jet break time respectively. 
\textbf{\large (b)} \textbf{The optical LC of GRB 060614.} The black and red dots represent the data in UVOT-u band. The errors correspond to a significance of one sigma. While the black dots are considered in the fitting process, the red dots are excluded from the fit (see, Supplementary Method \LinkTo{sec:SuppMethod1b} for explanations). The dashed vertical line (purple) represents the break times $T_{\rm a}$ (at the end of the plateau phase). The $\alpha_{\rm p}$, $\alpha_{\rm A1}$ are slopes during the plateau phase and self-similar phase respectively. In both panels \textbf{\large (a)} and \textbf{\large (b)}, the red line shows the mean of the posterior distribution and blue lines are 200 randomly selected samples from the MCMC sampling.}
\label{fig:fit_results_060729_X-ray_opt}
\end{figure}
~

\begin{figure}[ht!]
\centering
\begin{tabular}{cc}
\Huge{a} &  \raisebox{-.5\height}{\includegraphics[width=0.8\linewidth]{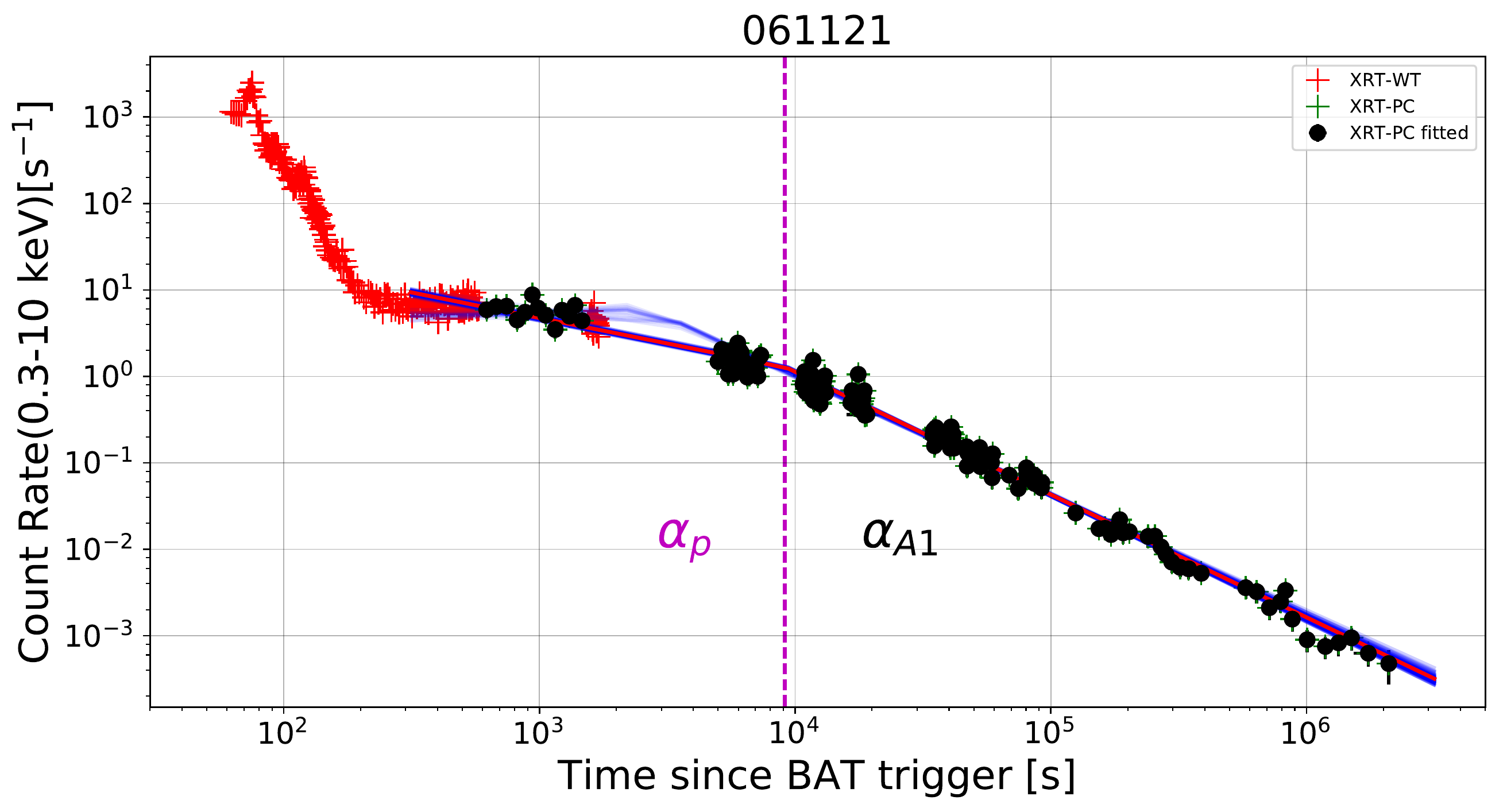}} \\
\Huge{b} & \raisebox{-.5\height}{\includegraphics[width=0.8\linewidth]{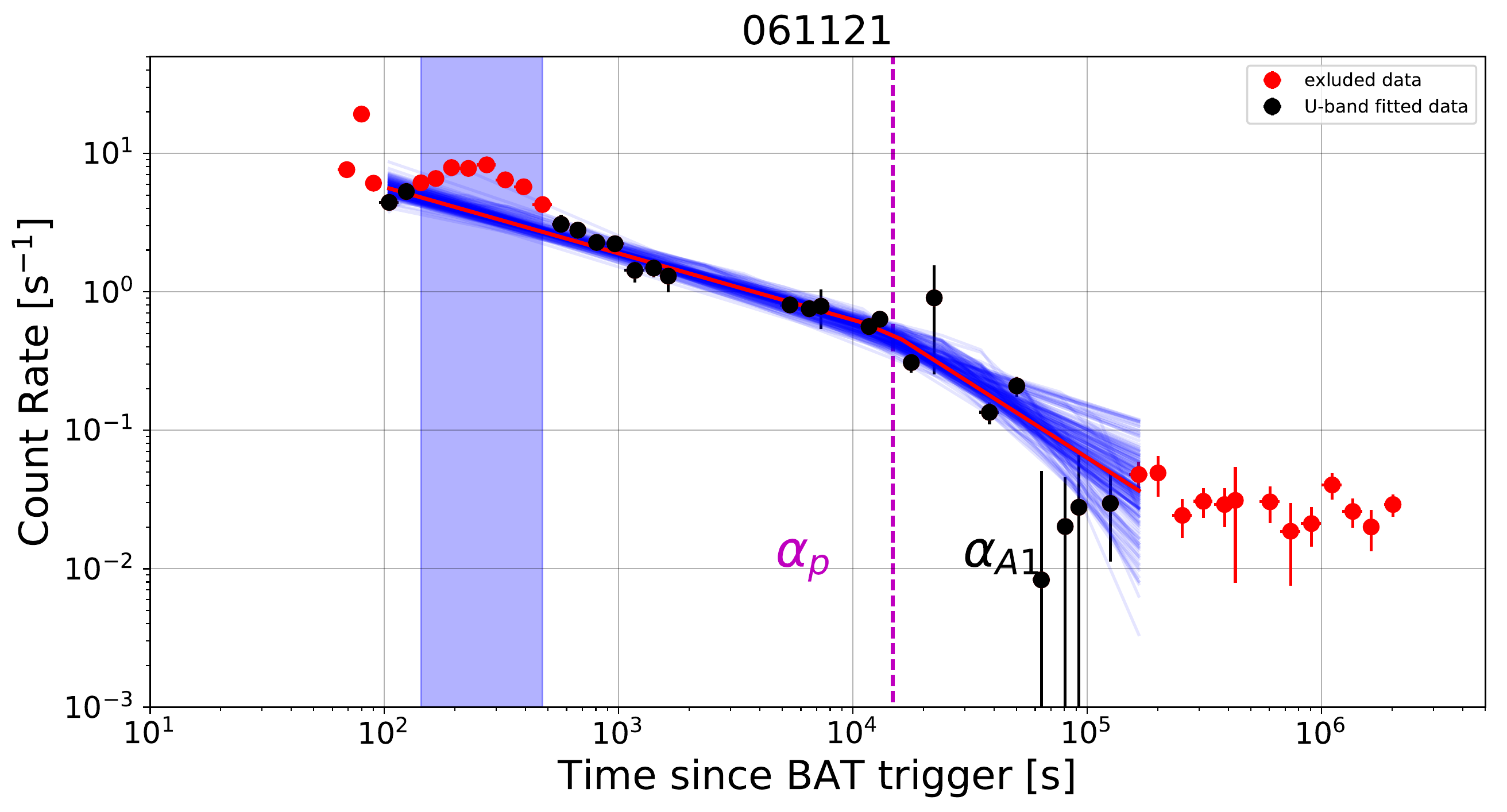}}
\end{tabular}
\caption{\textbf{\large (a)} \textbf{The X-ray LC of GRB 061121.} The red crosses represent the XRT-WT mode data which are excluded from the fit. The black dots represent the XRT-PC mode fitted data. The errors correspond to a significance of one sigma.
\textbf{\large (b)} \textbf{The optical LC of GRB 061121.} The black and red dots represent the data in UVOT-u band.The errors correspond to a significance of one sigma. While the black dots are considered in the fitting process, the red dots (including blue shaded region) are excluded from the fit (see, Supplementary Method \LinkTo{sec:SuppMethod1b} for explanations). In both panels \textbf{\large (a)} and \textbf{\large (b)}, the red line shows the mean of the posterior distribution and blue lines are 200 randomly selected samples from the MCMC sampling. The dashed vertical line (purple) represents the break times $T_{\rm a}$ (at the end of the plateau phase). The $\alpha_{\rm p}$, $\alpha_{\rm A1}$ are slopes during the plateau phase and self-similar phase respectively.}
\label{fig:fit_results_061121_X-ray_opt} 
\end{figure}
~

\begin{figure}[ht!]
\centering
\begin{tabular}{cc}
\Huge{a} &  \raisebox{-.5\height}{\includegraphics[width=0.8\linewidth]{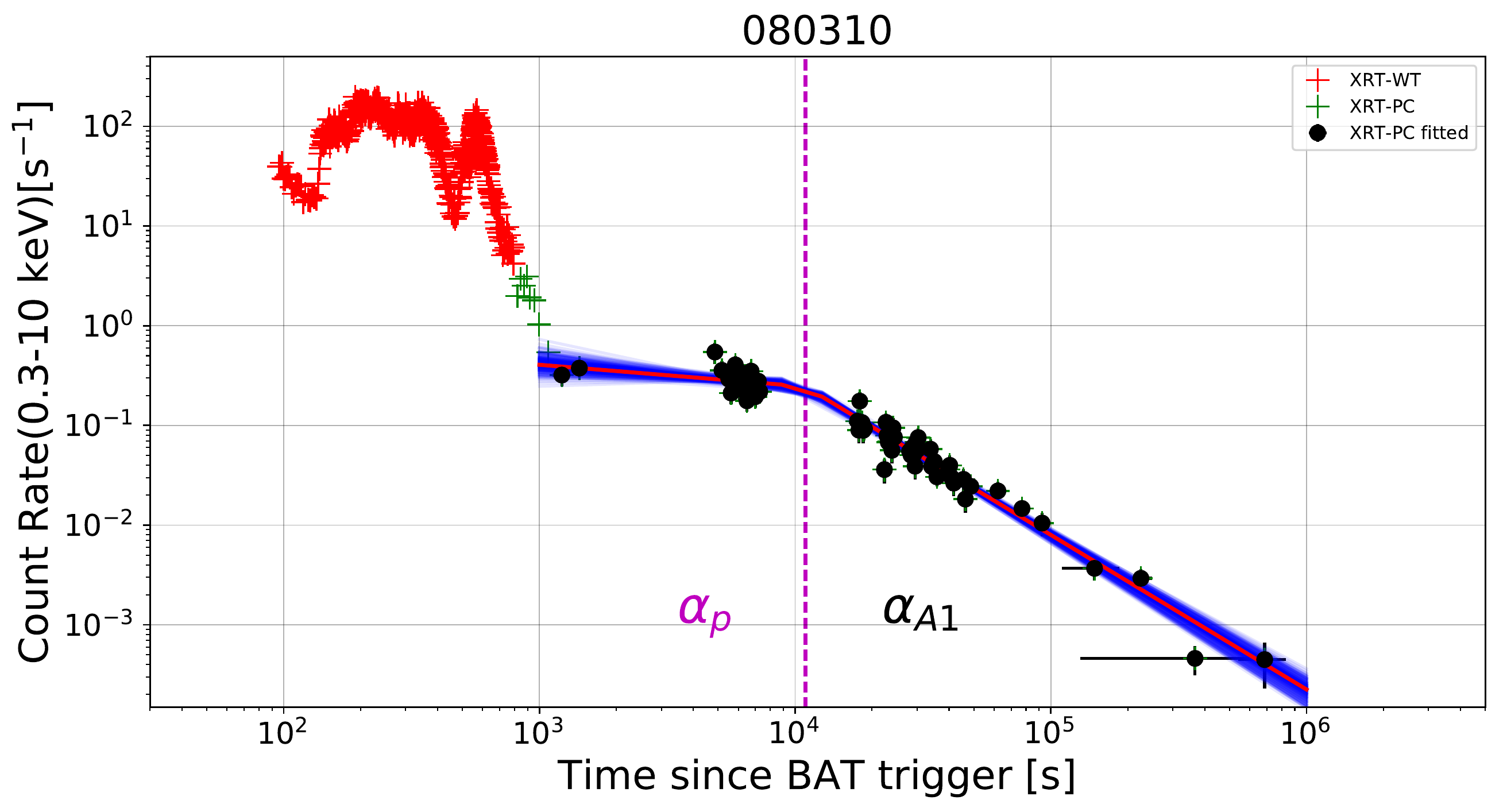}} \\
\Huge{b} & \raisebox{-.5\height}{\includegraphics[width=0.8\linewidth]{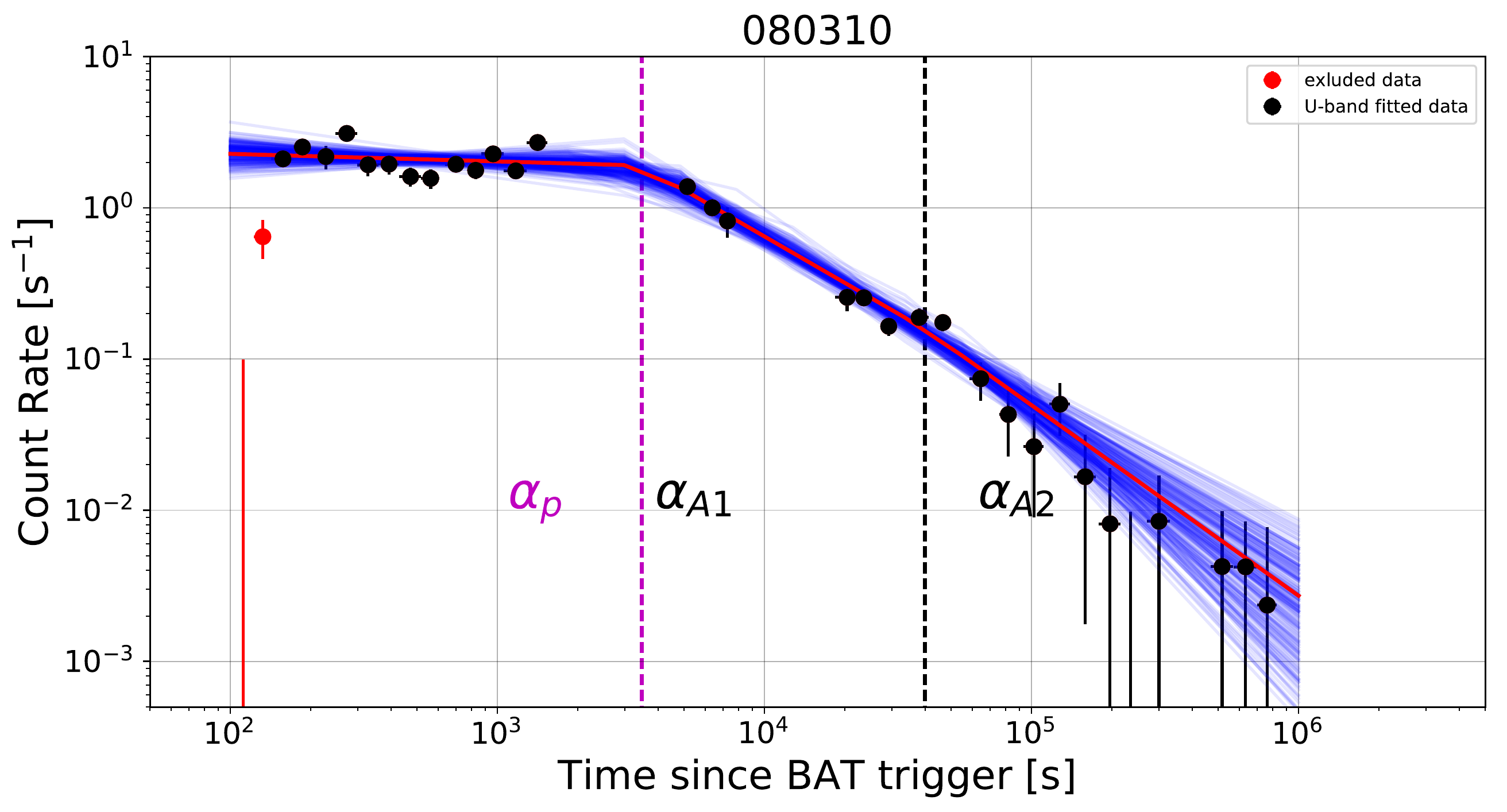}}
\end{tabular}
\caption{\textbf{\large (a)} \textbf{The X-ray LC of GRB 061121.} The red crosses represent the XRT-WT and XRT-PC mode data respectively which are excluded from the fit. The black dots represent the XRT-PC mode fitted data. The errors correspond to a significance of one sigma. The dashed vertical line (purple) represents the break times $T_{\rm a}$ (at the end of the plateau phase). The $\alpha_{\rm p}$, $\alpha_{\rm A1}$ are slopes during the plateau phase and self-similar phase respectively. 
\textbf{\large (b)} \textbf{The optical LC of GRB 061121.} The black and red dots represent the data in UVOT-u band. The errors correspond to a significance of one sigma. While the black dots are considered in the fitting process, the red dots (including blue shaded region) are excluded from the fit (see, Supplementary Method \LinkTo{sec:SuppMethod1b} for explanations). The dashed vertical lines (purple and black) represent the break times $T_{\rm a}$ (at the end of the plateau phase) and $T_{\rm b}$ respectively. The $\alpha_{\rm p}$, $\alpha_{\rm A1}$, $\alpha_{\rm A2}$ are slopes during the plateau phase, self-similar phase and after the jet break time respectively. In both panels \textbf{\large (a)} and \textbf{\large (b)}, the red line shows the mean of the posterior distribution and blue lines are 200 randomly selected samples from the MCMC sampling.}
\label{fig:fit_results_080310_X-ray_opt} 
\end{figure}
~

\begin{figure}[ht!]
\centering
\begin{tabular}{cc}
\Huge{a} &  \raisebox{-.5\height}{\includegraphics[width=0.8\linewidth]{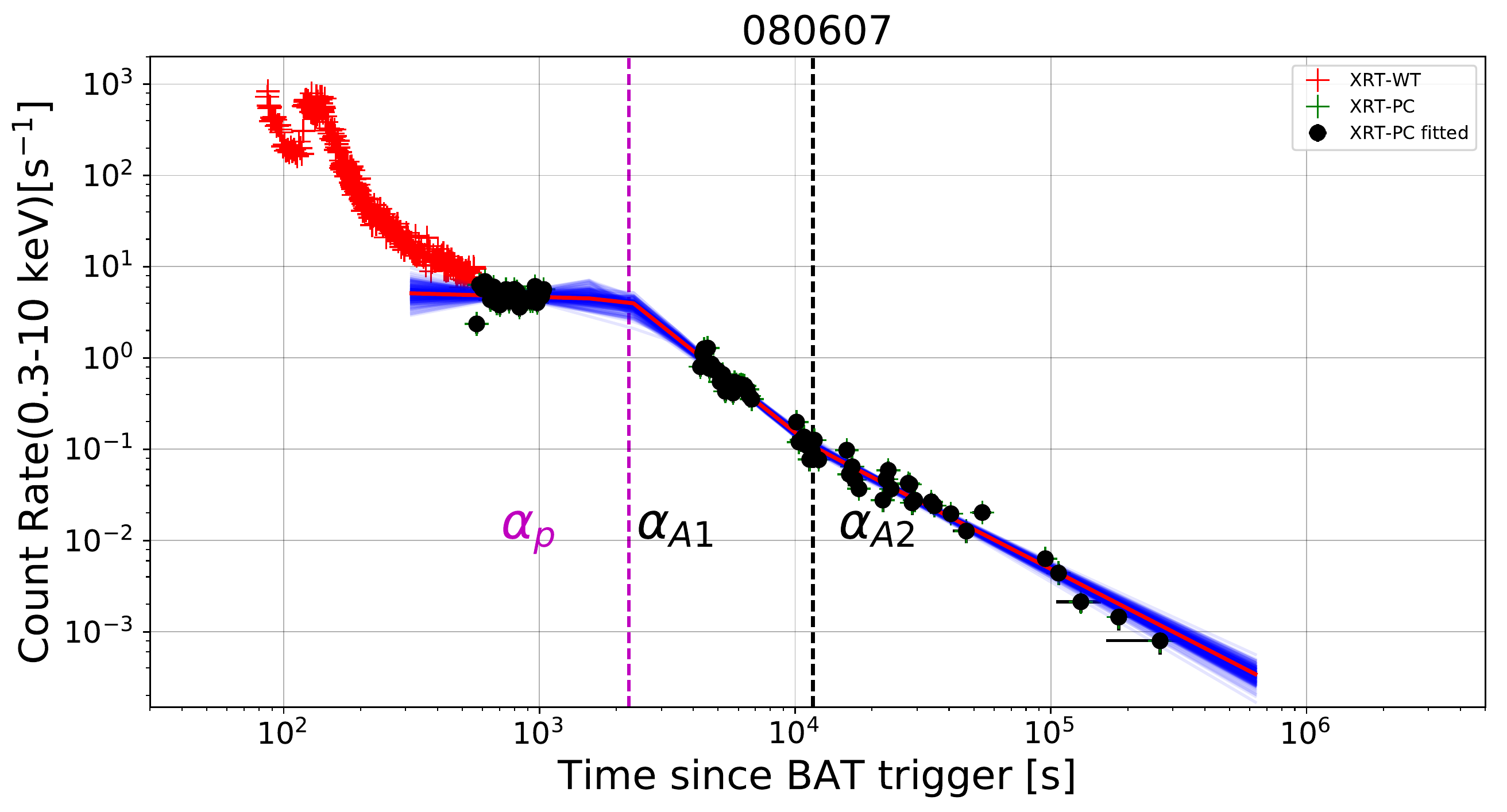}} \\
\Huge{b} & \raisebox{-.5\height}{\includegraphics[width=0.8\linewidth]{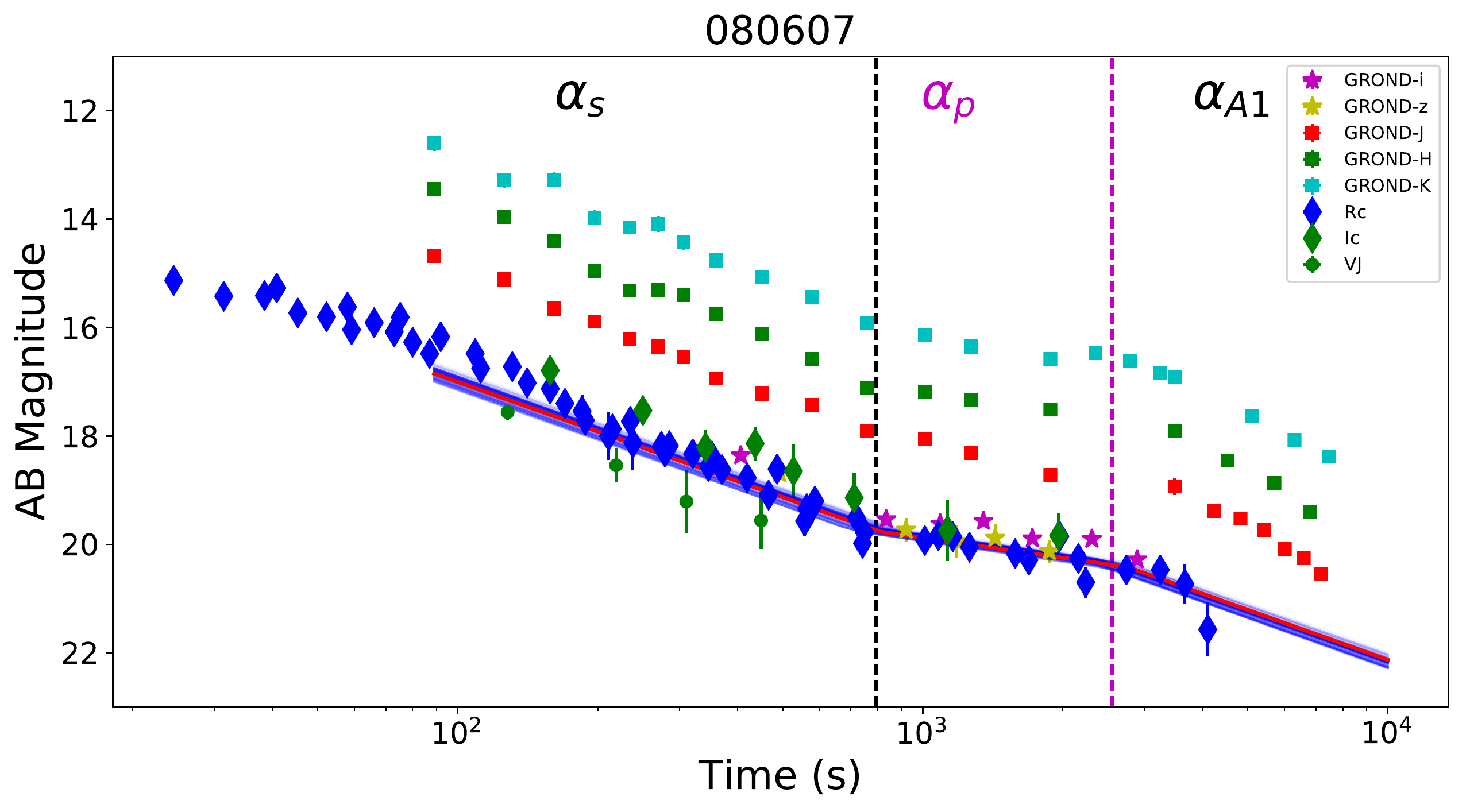}}
\end{tabular}
\caption{\textbf{\large (a)} \textbf{The X-ray LC of GRB 080607.} The red
crosses represent the XRT-WT mode data which are excluded from the fit. The black dots represent XRT-PC mode fitted data. The errors correspond to a significance of one sigma. The dashed vertical lines (purple and black) represent the break times $T_{\rm a}$ (at the end of the plateau phase) and $T_{\rm b}$ respectively. The $\alpha_{\rm p}$, $\alpha_{\rm A1}$, $\alpha_{\rm A2}$ are slopes during the plateau phase, self-similar phase and after the jet break time respectively.
\textbf{\large (b)} \textbf{The optical LC of GRB 080607.} The purple, yellow, red, green, cyan, blue, black colors together with stars, squares, diamonds and point represent the data in GROND-i, GROND-z, GROND-J, GROND-H, GROND-K, Rc, Ic, VJ bands respectively. The errors correspond to a significance of one sigma. For the details of the excluded data, see, Supplementary Method \LinkTo{sec:SuppMethod1b}.
Although all the bands are jointly fitted, for the demonstration purpose, only the Rc band fit result is displayed. The dashed vertical lines (black and
purple) represent the break times $T_{\rm s}$ and $T_{\rm a}$ (at the end of the plateau
phase) respectively. The $\alpha_{\rm As}$, $\alpha_{\rm p}$, $\alpha_{\rm A1}$, are the slopes before the plateau phase, during the plateau phase and self-similar
phase respectively. In both panels \textbf{\large (a)} and \textbf{\large (b)}, the red line shows the mean of the posterior distribution and blue lines are 200 randomly selected samples from the MCMC sampling.}
\label{fig:fit_results_080607_X-ray_opt}
\end{figure}
~

\begin{figure}[ht!]
\centering
\begin{tabular}{cc}
\Huge{a} &  \raisebox{-.5\height}{\includegraphics[width=0.8\linewidth]{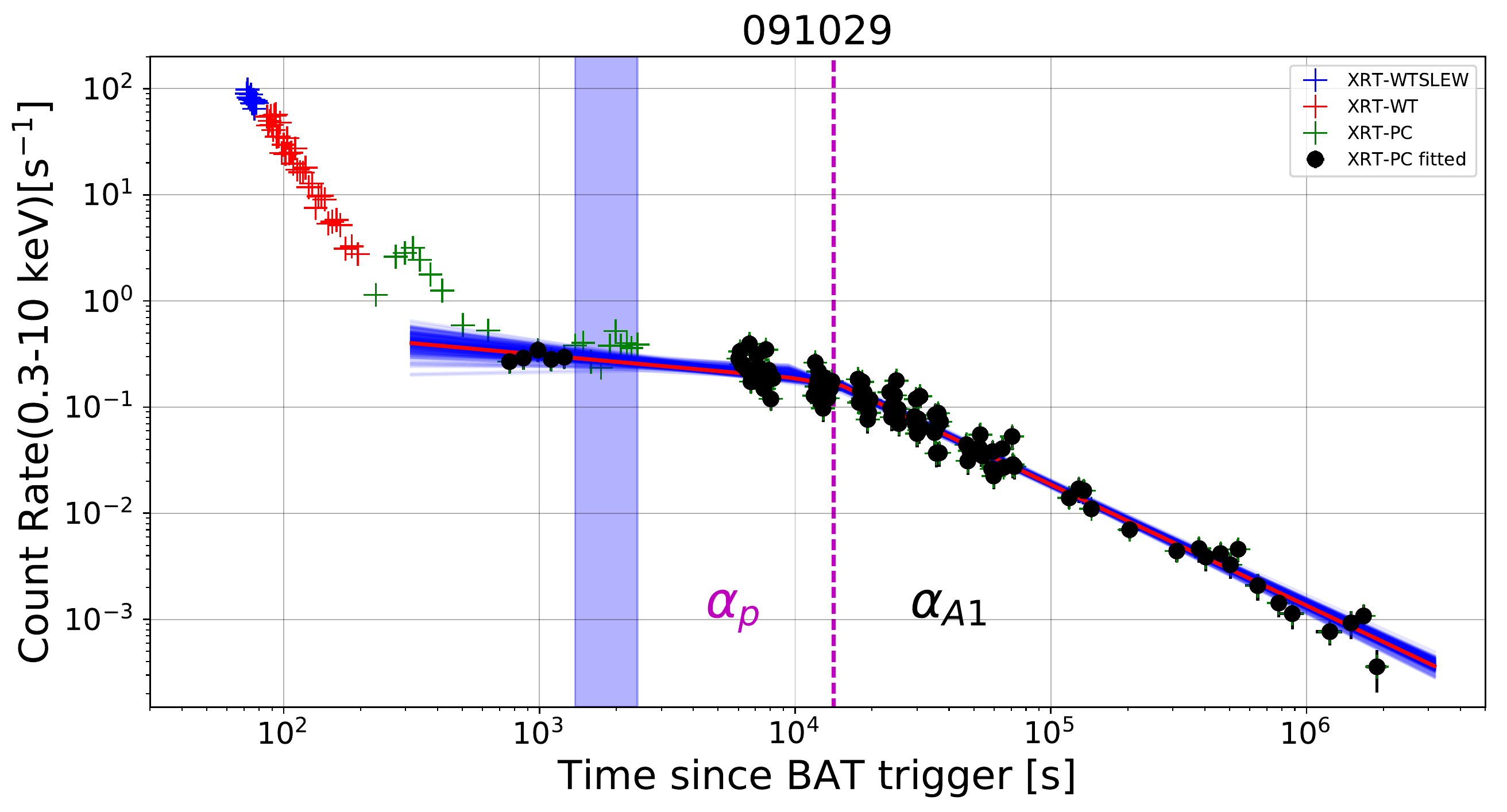}} \\
\Huge{b} & \raisebox{-.5\height}{\includegraphics[width=0.8\linewidth]{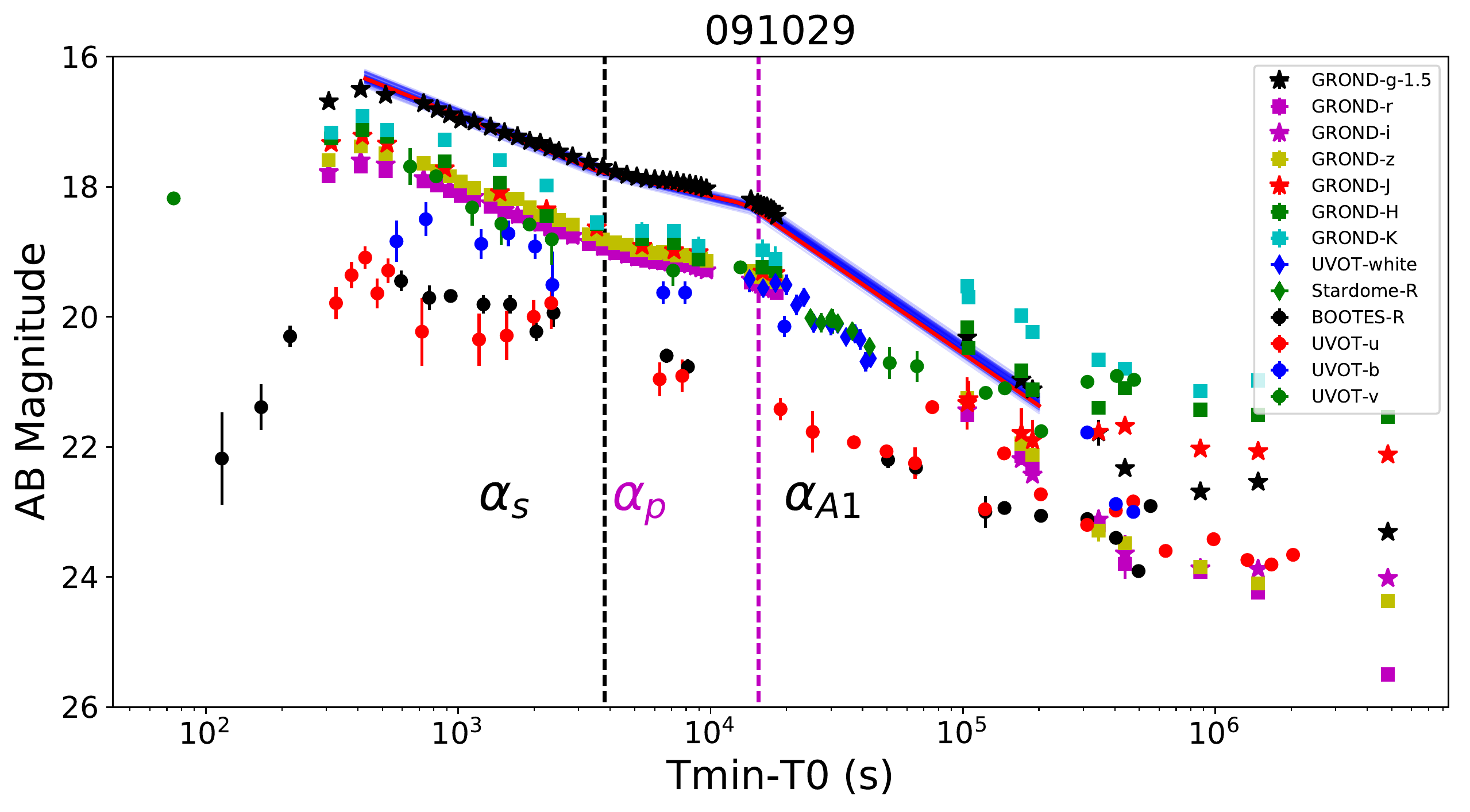}}
\end{tabular}
\caption{\textbf{\large (a)} \textbf{The X-ray LC of GRB 091029.} The blue, red and green
crosses represent the XRT-WTSLEW, XRT-WT, XRT-PC (including blue shaded region, see, Supplementary Method \LinkTo{sec:SuppMethod1a} for explanation) mode data which are excluded from the fit. The black dots represent XRT-PC mode fitted data. The errors correspond to a significance of one sigma. The dashed vertical line (purple) represents the break times $T_{\rm a}$ (at the end of the plateau phase). The $\alpha_{\rm p}$, $\alpha_{\rm A1}$ are slopes during the plateau phase and self-similar phase respectively.
\textbf{\large (b)} \textbf{The optical LC of GRB 091029.} The black, purple, yellow, red, green, cyan, blue, black colors together with stars, squares, diamonds and points represent the data in GROND-g, GROND-r, GROND-i, GROND-z, GROND-J, GROND-H, GROND-K, UVOT-white, Stardome-R, BOOTES-R, UVOT-u, UVOT-b, UVOT-v bands respectively. The errors correspond to a significance of one sigma. For the details of the excluded data, see, Supplementary Method \LinkTo{sec:SuppMethod1b}. 
Although all the bands are jointly fitted, for the demonstration purpose, only the GROND-g band fit result is displayed and GROND-g band light curve distinguished by subtracting 1.5 from the AB magnitude. The dashed vertical lines (black and
purple) represent the break times $T_{\rm s}$ and $T_{\rm a}$ (at the end of the plateau
phase) respectively. The $\alpha_{\rm As}$, $\alpha_{\rm p}$, $\alpha_{\rm A1}$, are the slopes before the plateau phase, during the plateau phase and self-similar
phase respectively. In both panels \textbf{\large (a)} and \textbf{\large (b)}, the red line shows the mean of the posterior distribution and blue lines are 200 randomly selected samples from the MCMC sampling.}
\label{fig:fit_results_091029_X-ray_opt} 
\end{figure}
~

\begin{figure}[ht!]
\centering
\begin{tabular}{cc}
\Huge{a} &  \raisebox{-.5\height}{\includegraphics[width=0.8\linewidth]{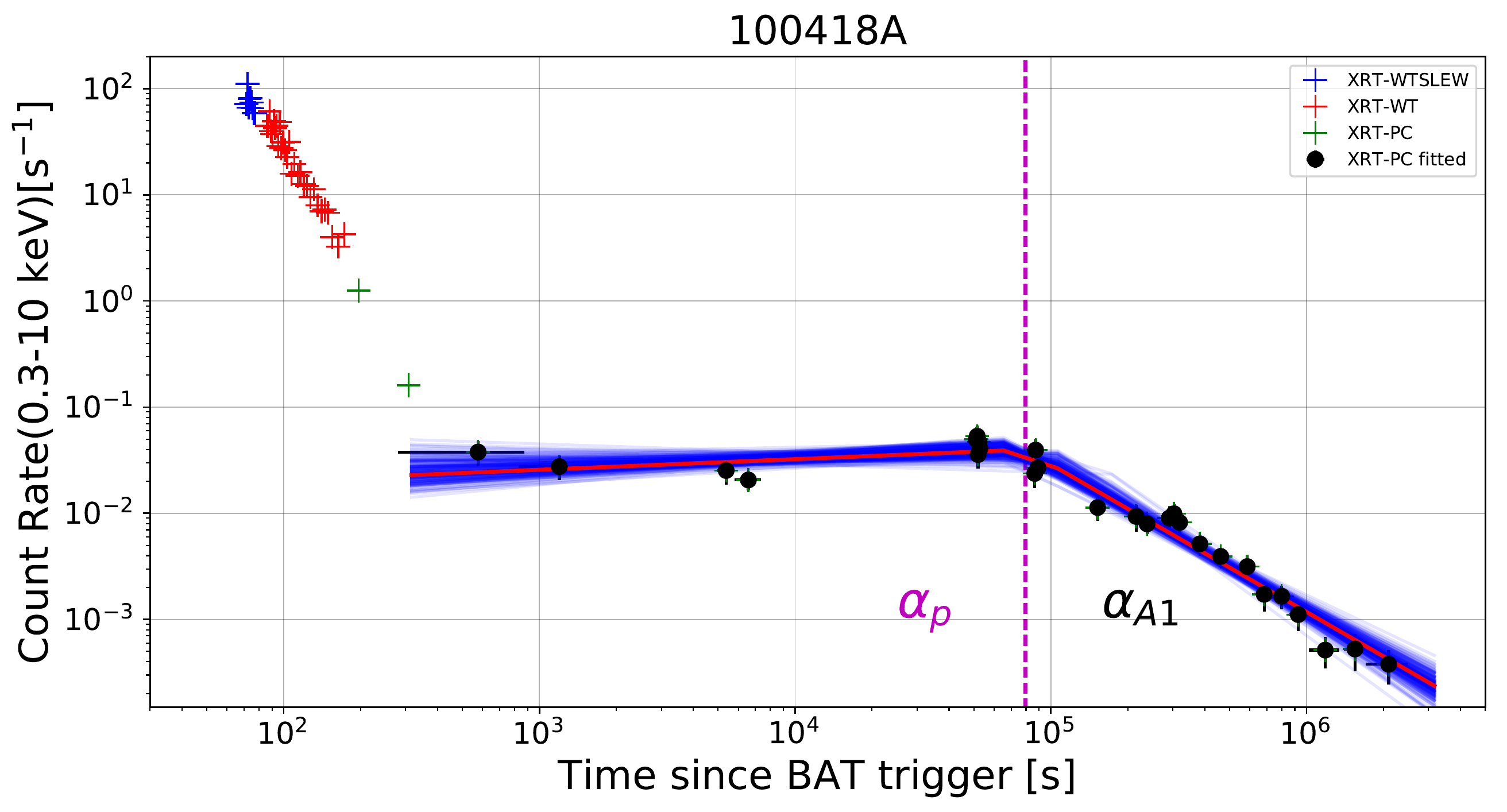}} \\
\Huge{b} & \raisebox{-.5\height}{\includegraphics[width=0.8\linewidth]{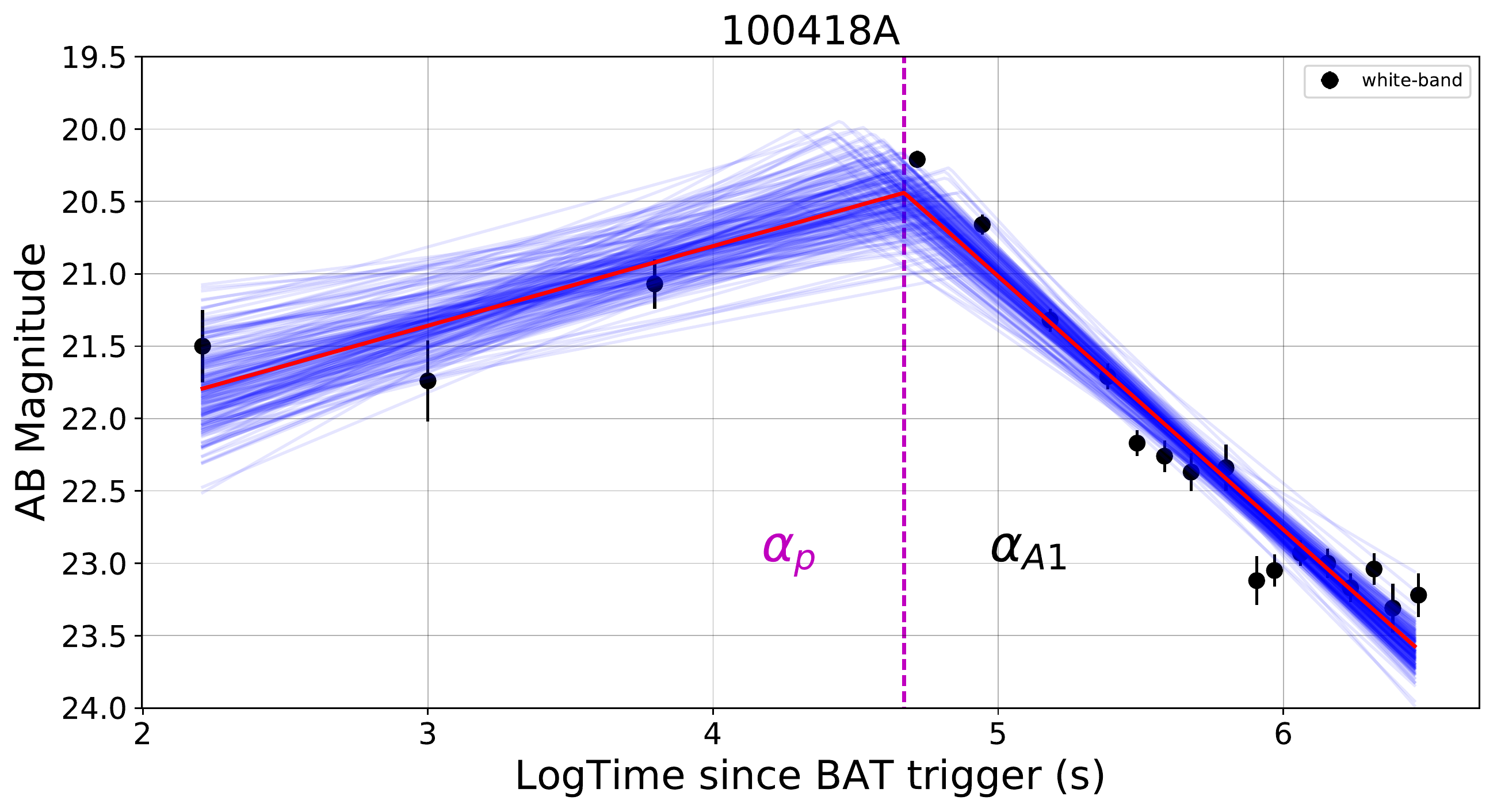}}
\end{tabular}
\caption{\textbf{\large (a)} \textbf{The X-ray LC of GRB 100418A.} The blue, red and green
crosses represent the XRT-WTSLEW, XRT-WT, XRT-PC mode data which are excluded from the fit. The black dots represent XRT-PC mode fitted data. The errors correspond to a significance of one sigma.
\textbf{\large (b)} \textbf{The optical LC of GRB 100418A.} The black points represent the data in UVOT-white band. The errors correspond to a significance of one sigma. In both panels \textbf{\large (a)} and \textbf{\large (b)}, the red line shows the mean of the posterior distribution and blue lines are 200 randomly selected samples from the MCMC sampling. The dashed vertical line (purple) represents the break times $T_{\rm a}$ (at the end of the plateau phase). The $\alpha_{\rm p}$, $\alpha_{\rm A1}$ are slopes during the plateau phase and self-similar phase respectively.}
\label{fig:fit_results_100418A_X-ray_opt} 
\end{figure}
~

\begin{figure}[ht!]
\centering
\begin{tabular}{cc}
\Huge{a} &  \raisebox{-.5\height}{\includegraphics[width=0.8\linewidth]{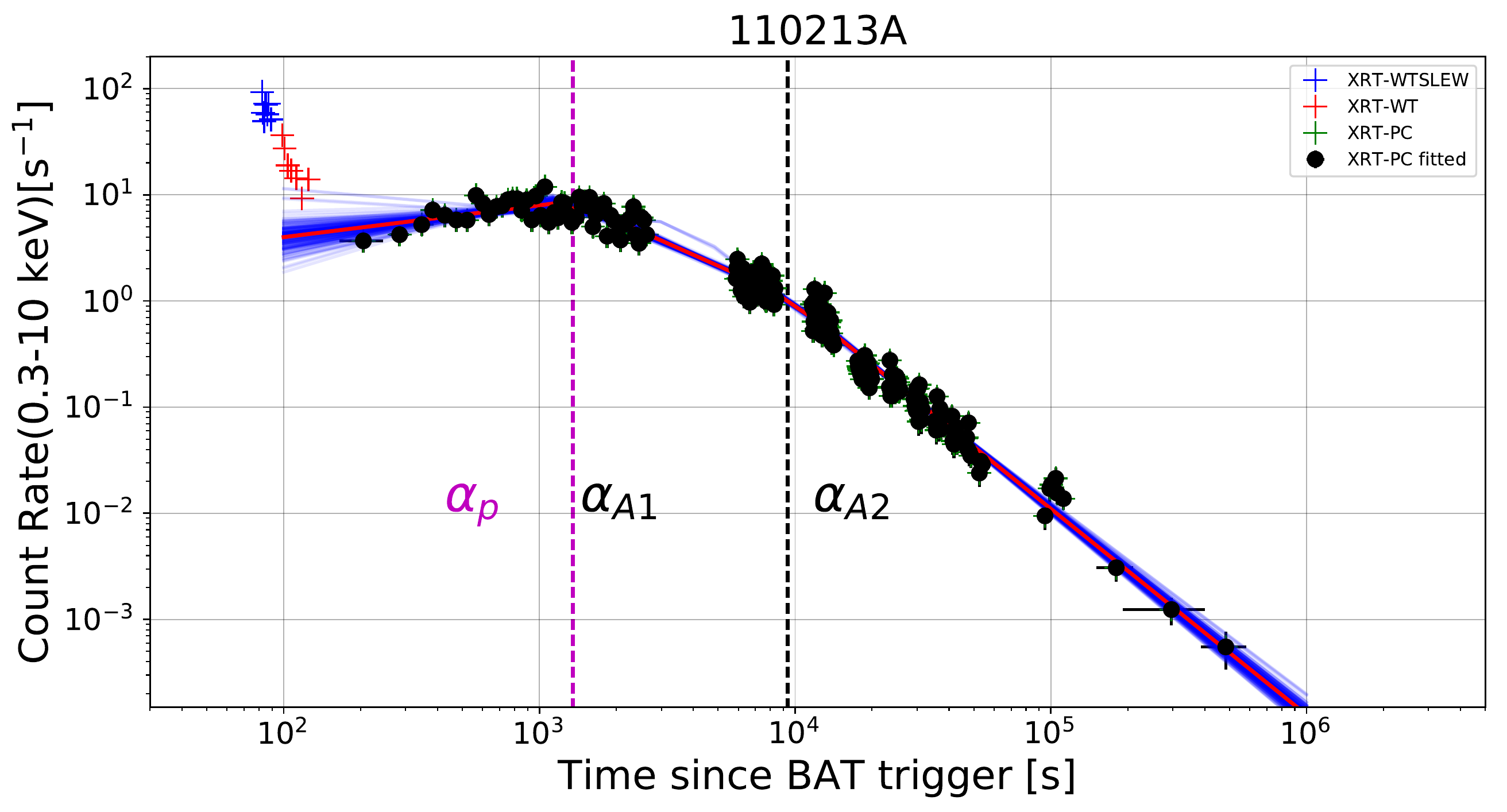}} \\
\Huge{b} & \raisebox{-.5\height}{\includegraphics[width=0.8\linewidth]{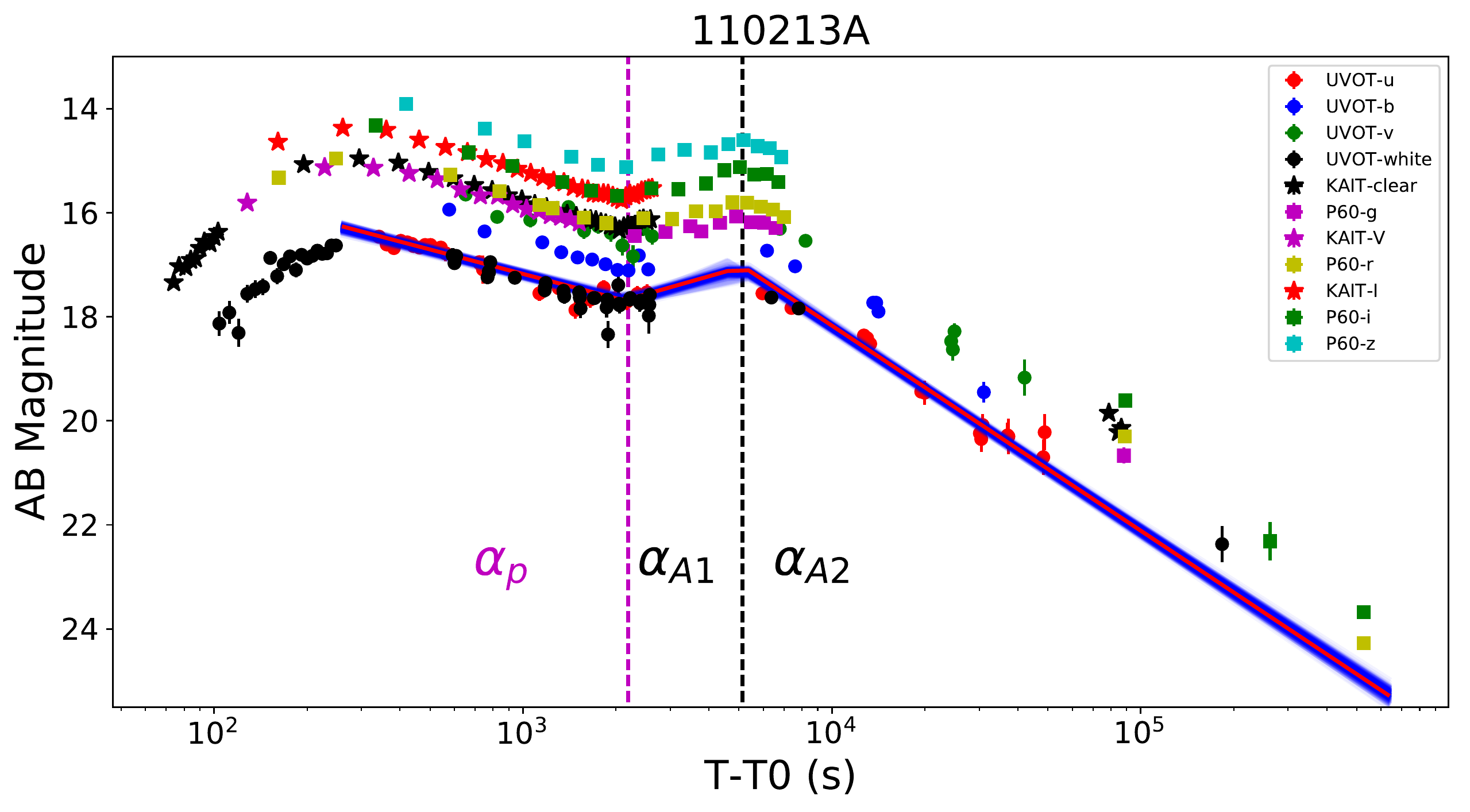}}
\end{tabular}
\caption{\textbf{\large (a)} \textbf{The X-ray LC of GRB 110213A.} The blue, red and green crosses represent the XRT-WTSLEW and XRT-WT mode data respectively which are excluded from the fit. The black dots represent the XRT-PC mode fitted data. The errors correspond to a significance of one sigma.
\textbf{\large (b)} \textbf{The optical LC of GRB 110213A.} The red, blue, green, black, purple, yellow, cyan colors together with points, stars, squares represent the data in UVOT-u, UVOT-b, UVOT-v, UVOT-white, KAIT-clear, P60-g, KAIT-V, P60-r, KAIT-I,  P60-i, P60-z bands respectively. The errors correspond to a significance of one sigma. For the details of the excluded data, see, Supplementary Method \LinkTo{sec:SuppMethod1b}. Although all the bands jointly fitted, for the demonstration purpose only UVOT-u band fit result displayed. In both panels \textbf{\large (a)} and \textbf{\large (b)}, The red line shows the mean of the posterior distribution and blue lines are 200 randomly selected samples from the MCMC sampling. The dashed vertical lines (purple and black) represent the break times $T_{\rm a}$ (at the end of the plateau phase) and $T_{\rm b}$ respectively. The $\alpha_{\rm p}$, $\alpha_{\rm A1}$, $\alpha_{\rm A2}$ are slopes during the plateau phase, self-similar phase and after the jet break time respectively.}
\label{fig:fit_results_110213A_X-ray_opt}
\end{figure}
~

\begin{figure}[ht!]
\centering
\begin{tabular}{cc}
\Huge{a} &  \raisebox{-.5\height}{\includegraphics[width=0.8\linewidth]{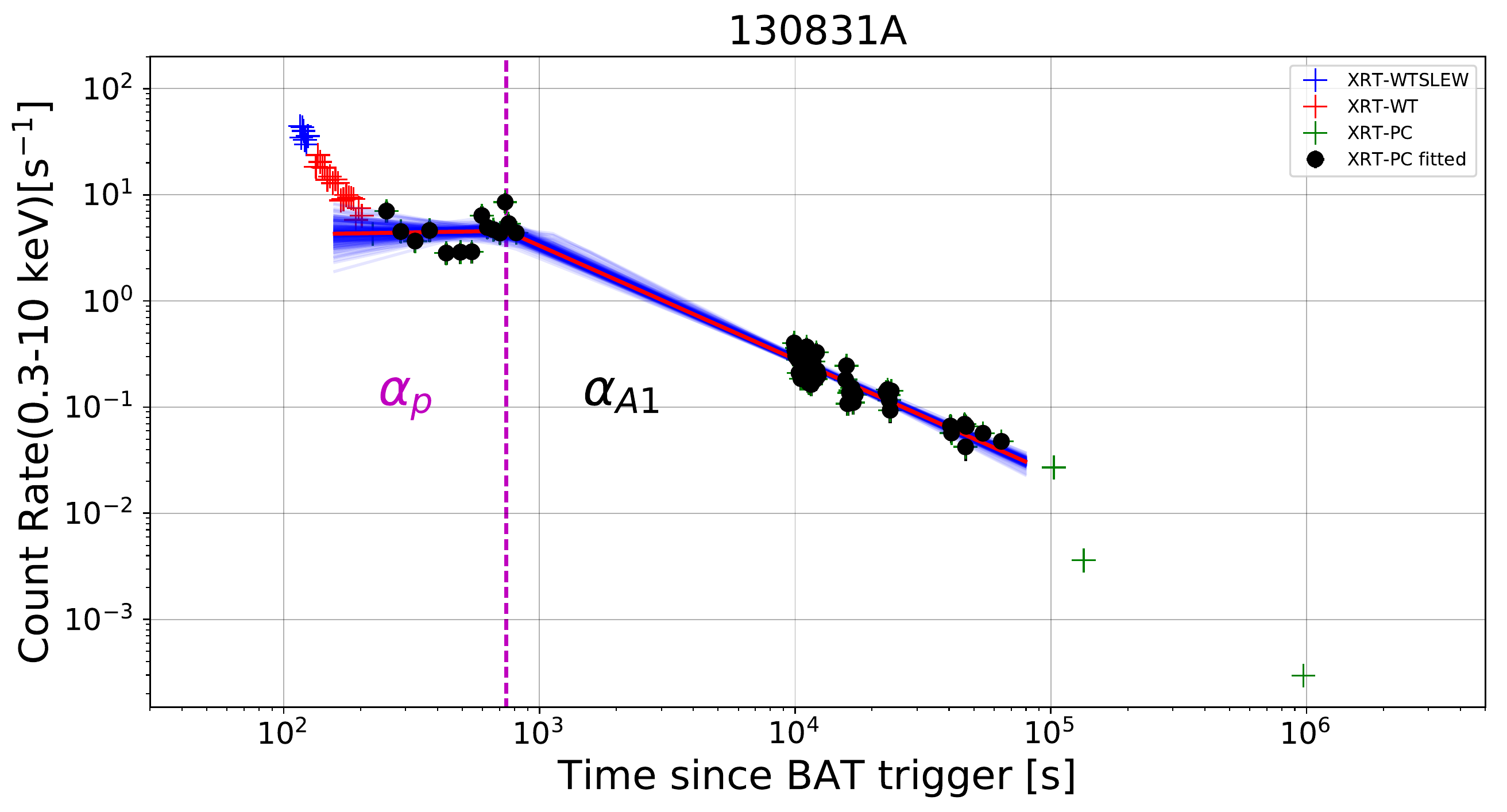}} \\
\Huge{b} & \raisebox{-.5\height}{\includegraphics[width=0.8\linewidth]{supp_figures/130831A-XRT-LC_with-fit.pdf}}
\end{tabular}
\caption{\textbf{\large (a)} \textbf{The X-ray LC of GRB 130831A.} The blue, red and green
crosses represent the XRT-WTSLEW, XRT-WT, XRT-PC mode data which are excluded from the fit. The black dots represent XRT-PC mode fitted data. The errors correspond to a significance of one sigma.
\textbf{\large (b)} \textbf{The optical LC of GRB 130831A.} The purple, yellow, red, green, blue, black, cyan colors together with squares, stars and points represent the data in GROND-r, GROND-i, GROND-z, GROND-J, GROND-H, Rc, B, V, Ic, UVOT-u, UVOT-b, UVOT-v and CR bands respectively. The errors correspond to a significance of one sigma. The data marked by blue shaded region excluded from the fit, see, Supplementary Method \LinkTo{sec:SuppMethod1b} for explanation. Although all the bands jointly fitted, for the demonstration purpose only UVOT-u band fit result displayed. In both panels \textbf{\large (a)} and \textbf{\large (b)}, the red line shows the mean of the posterior distribution and blue lines are 200 randomly selected samples from the MCMC sampling. The dashed vertical line (purple) represents the break times $T_{\rm a}$ (at the end of the plateau phase). The $\alpha_{\rm p}$, $\alpha_{\rm A1}$ are slopes during the plateau phase and self-similar phase respectively.}
\label{fig:fit_results_130831A_X-ray_opt}
\end{figure}
~

\begin{figure}[ht!]
\centering
\begin{tabular}{cc}
\Huge{a} &  \raisebox{-.5\height}{\includegraphics[width=0.8\linewidth]{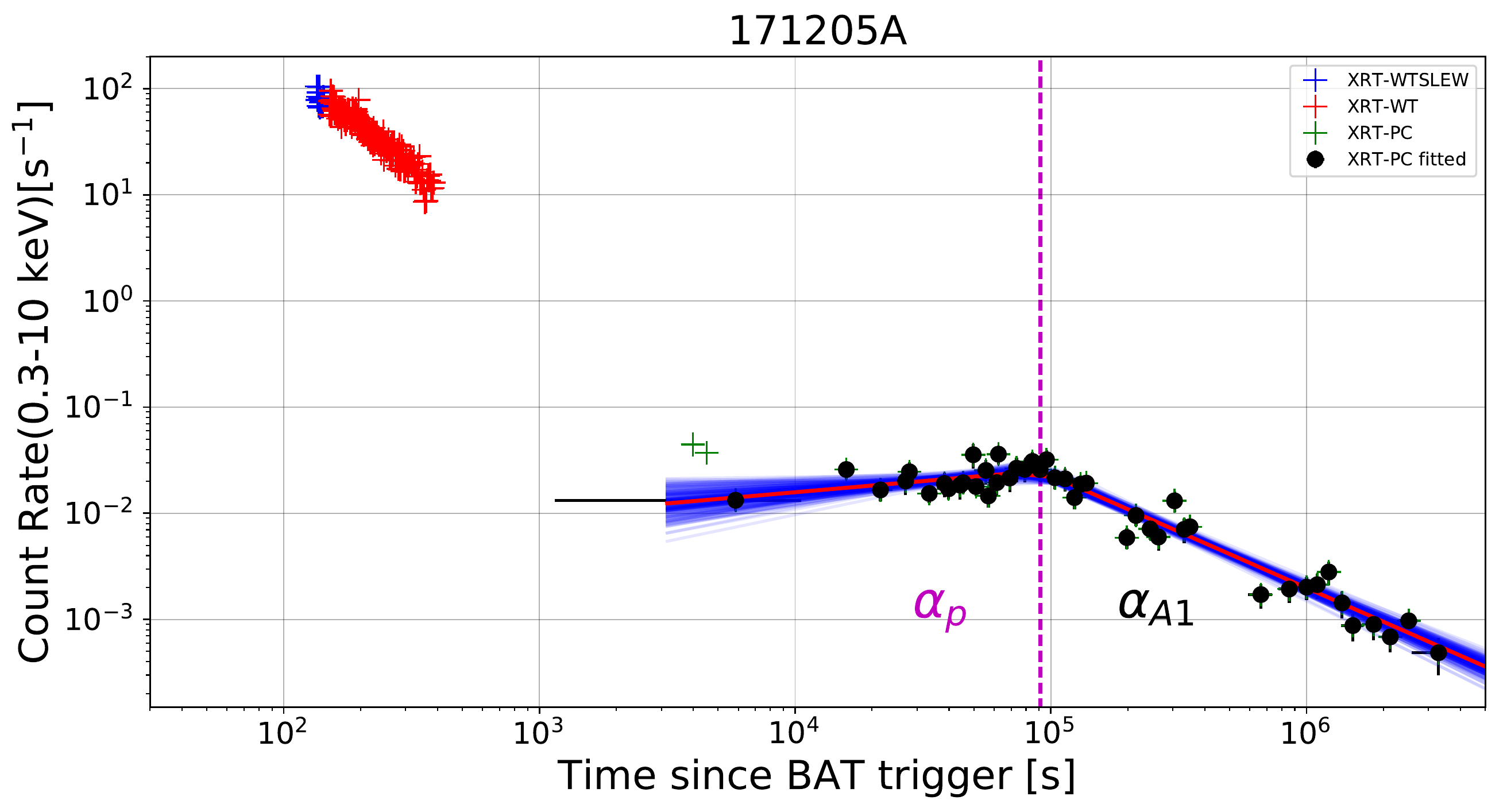}} \\
\Huge{b} & \raisebox{-.5\height}{\includegraphics[width=0.8\linewidth]{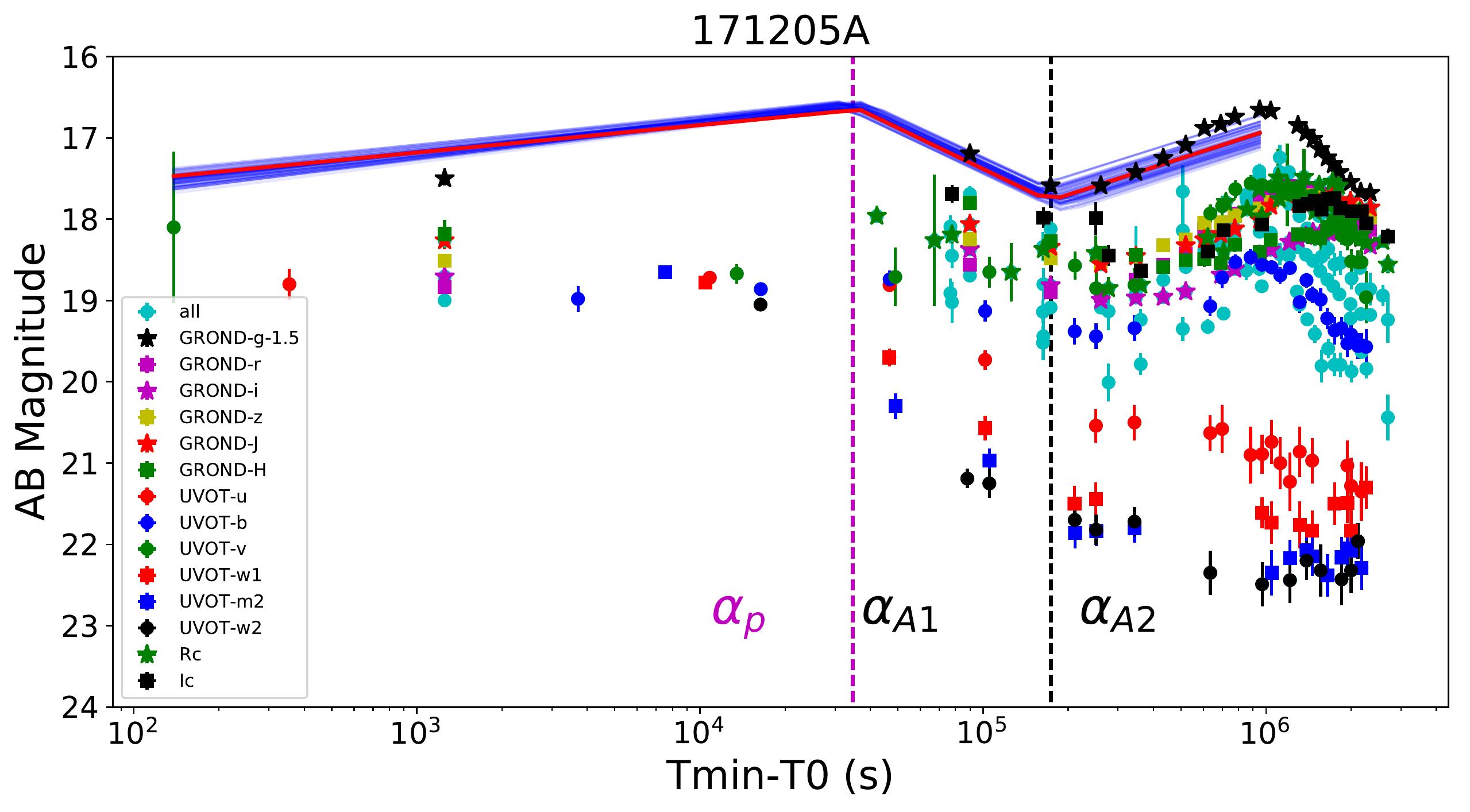}}
\end{tabular}
\caption{\textbf{\large (a)} \textbf{The X-ray LC of GRB 171205A.} The blue, red and green
crosses represent the XRT-WTSLEW, XRT-WT, XRT-PC mode data which are excluded from the fit. The black dots represent XRT-PC mode fitted data. The errors correspond to a significance of one sigma. The dashed vertical line (purple) represents the break times $T_{\rm a}$ (at the end of the plateau phase). The $\alpha_{\rm p}$, $\alpha_{\rm A1}$ are slopes during the plateau phase and self-similar phase respectively.
\textbf{\large (b)} \textbf{The optical LC of GRB 171205A.} The black, purple, yellow, red, green, blue colors together with points, stars, squares represent the data GROND-g, GROND-r, GROND-i, GROND-z, GROND-J, GROND-H, UVOT-u, UVOT-b, UVOT-v, UVOT-w1, UVOT-m2, UVOT-w2, Rc, Ic bands respectively. The errors correspond to a significance of one sigma. The cyan colors represent the other data that are not considered in the fitting process. For the details of the excluded data, see, Supplementary Method \LinkTo{sec:SuppMethod1b}. 
Although all the bands are jointly fitted, for the demonstration purpose, only the GROND-g band fit result is displayed and GROND-g band light curve distinguished by subtracting 1.5 from the AB magnitude. The dashed vertical lines (black and
purple) represent the break times $T_{\rm s}$ and $T_{\rm a}$ (at the end of the plateau
phase) respectively. The $\alpha_{\rm As}$, $\alpha_{\rm p}$, $\alpha_{\rm A1}$, are the slopes before the plateau phase, during the plateau phase and self-similar
phase respectively. In both panels \textbf{\large (a)} and \textbf{\large (b)}, the red line shows the mean of the posterior distribution and blue lines are 200 randomly selected samples from the MCMC sampling.}
\label{fig:fit_results_171205A_X-ray_opt}
\end{figure}
~

\begin{figure}[ht!]
\centering
\includegraphics[width=\linewidth]{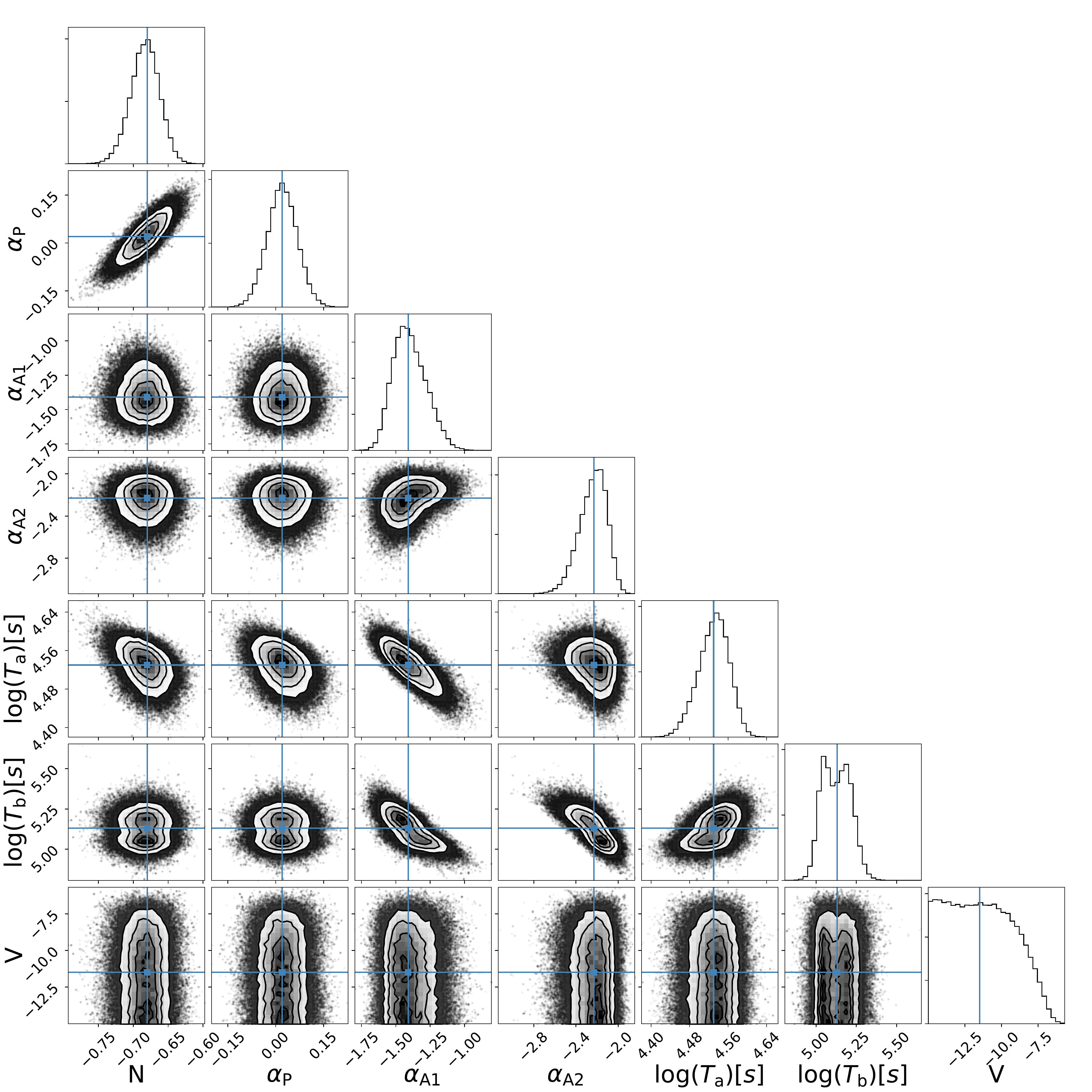}
\caption{\textbf{ Corner plot
showing the one-and two-dimensional posterior probability distributions for the fit
parameters  obtained from the X-ray LC of GRB 060614.} The blue lines present the mean
value of each parameter given in Table \ref{tab:fit_paramaters_060614}.
Contours gives from inward to outward 39.3\%, 68.3\% ,90.3\% confidence region.}
\label{fig:corner_plot_060614}
\end{figure}
~

\begin{figure}[ht!]
\centering
\begin{tabular}{cc}
\huge a & \huge b \\
\includegraphics[width=0.48\linewidth]{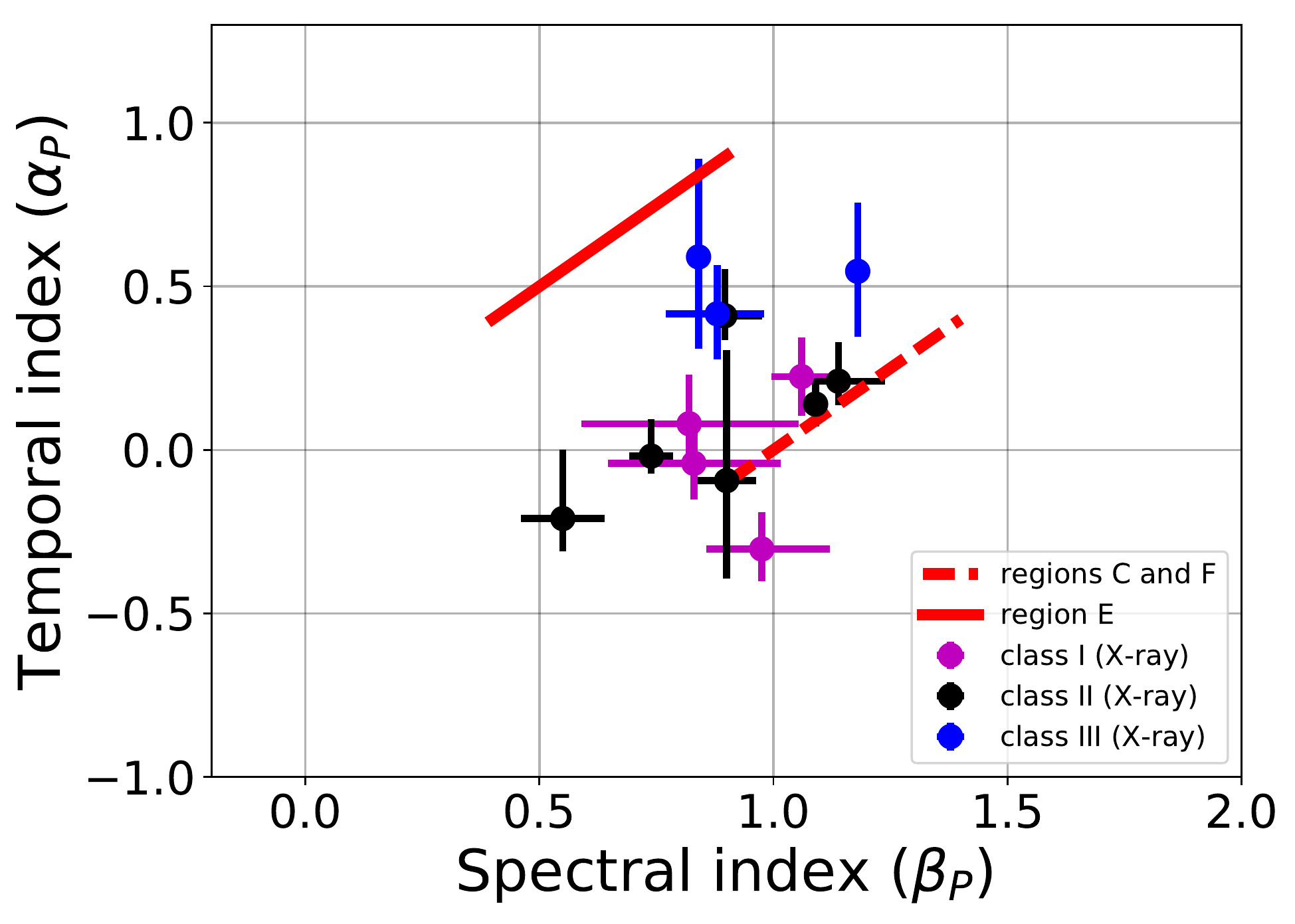} &
\includegraphics[width=0.48\linewidth]{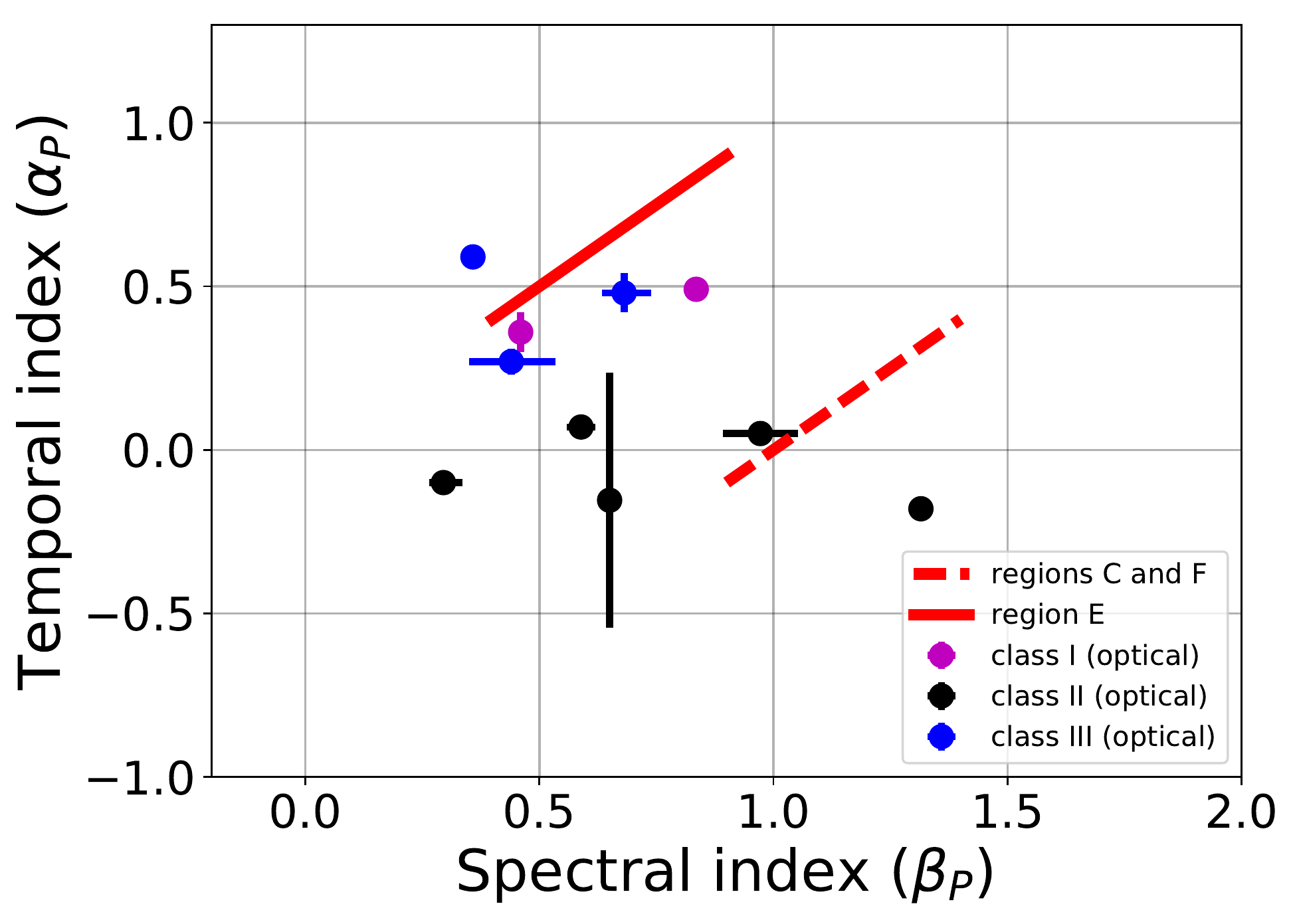}
\end{tabular}
\caption{\textbf{Temporal and spectral indices of the 13 GRBs in our sample during
the plateau phase \textbf{\large (a)} in the X-ray band and \textbf{\large (b)} in the optical band.} The purple, black, and blue data points correspond to the three classes (I, II and III respectively). The errors correspond to a significance of one sigma. The spectral indices in the optical
band of GRBs 100418A (in class II), 110213A and 130831A (both in class I) during
the plateau phase could not be retrieved from the
literature, and therefore these GRBs are excluded from panel
\textbf{\large (b)}. The red line shows the closure
relation obtained for region E in which $\nu_m < \nu_{\rm obs.}
< \nu_c$, and the dashed red line represents regions C and F in which $\nu_c <
\nu_m < \nu_{\rm obs.}$ and $\nu_m < \nu_c < \nu_{\rm obs.}$
respectively. These lines are computed by using an electron power-law
index $p = 1.8 - 2.8$. Despite the simplicity of the model, most data points
are in between the expected theoretical limits in both X-ray and optic bands. 
The source data necessary to reproduce this Figure are provided as a
Source Data file.}
\label{fig:CR-plateau}
\end{figure}
~

\begin{figure}[ht!]
\centering
\begin{tabular}{cc}
\huge a & \huge b \\
\includegraphics[width=0.48\linewidth]{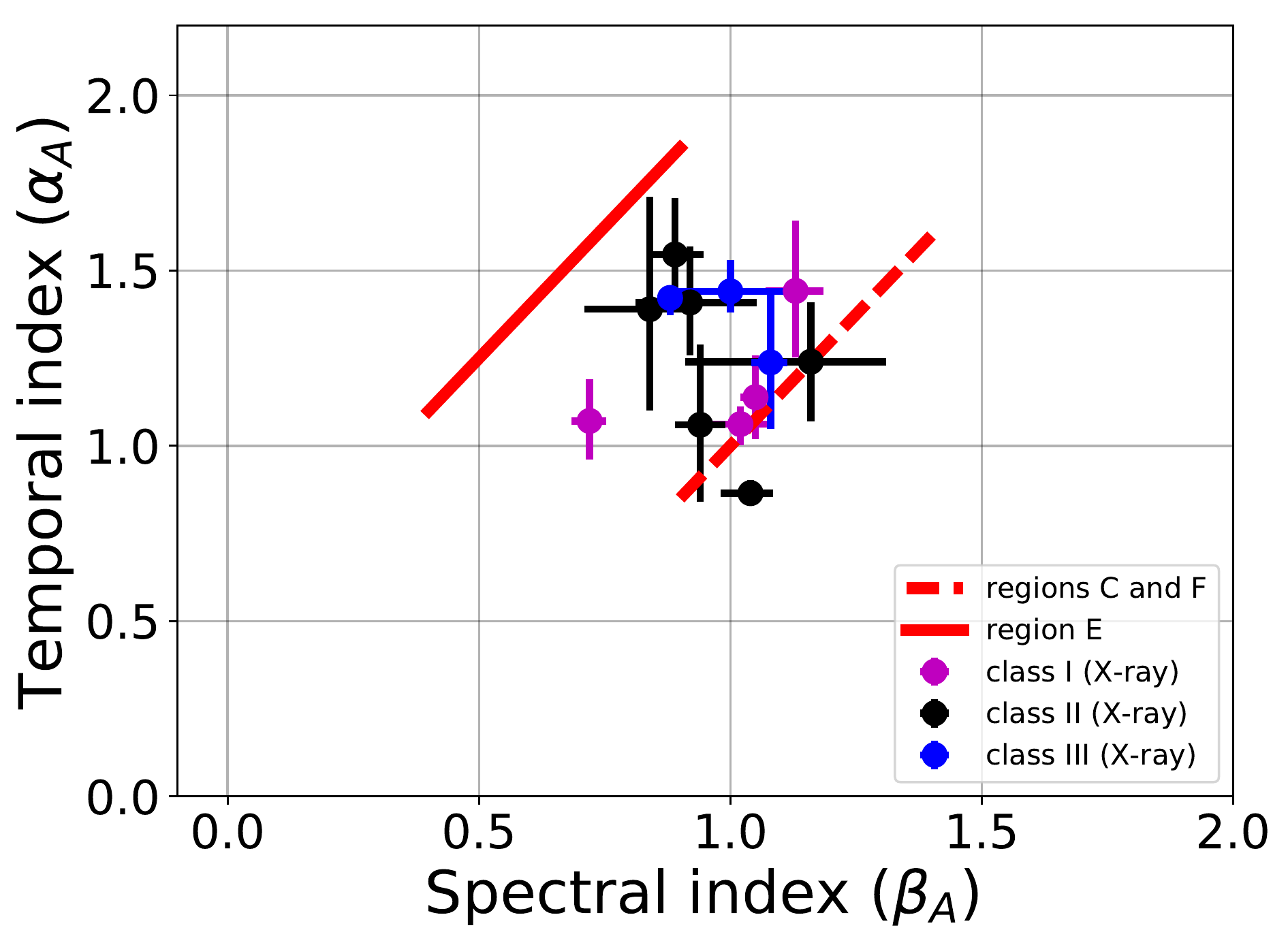} &
\includegraphics[width=0.48\linewidth]{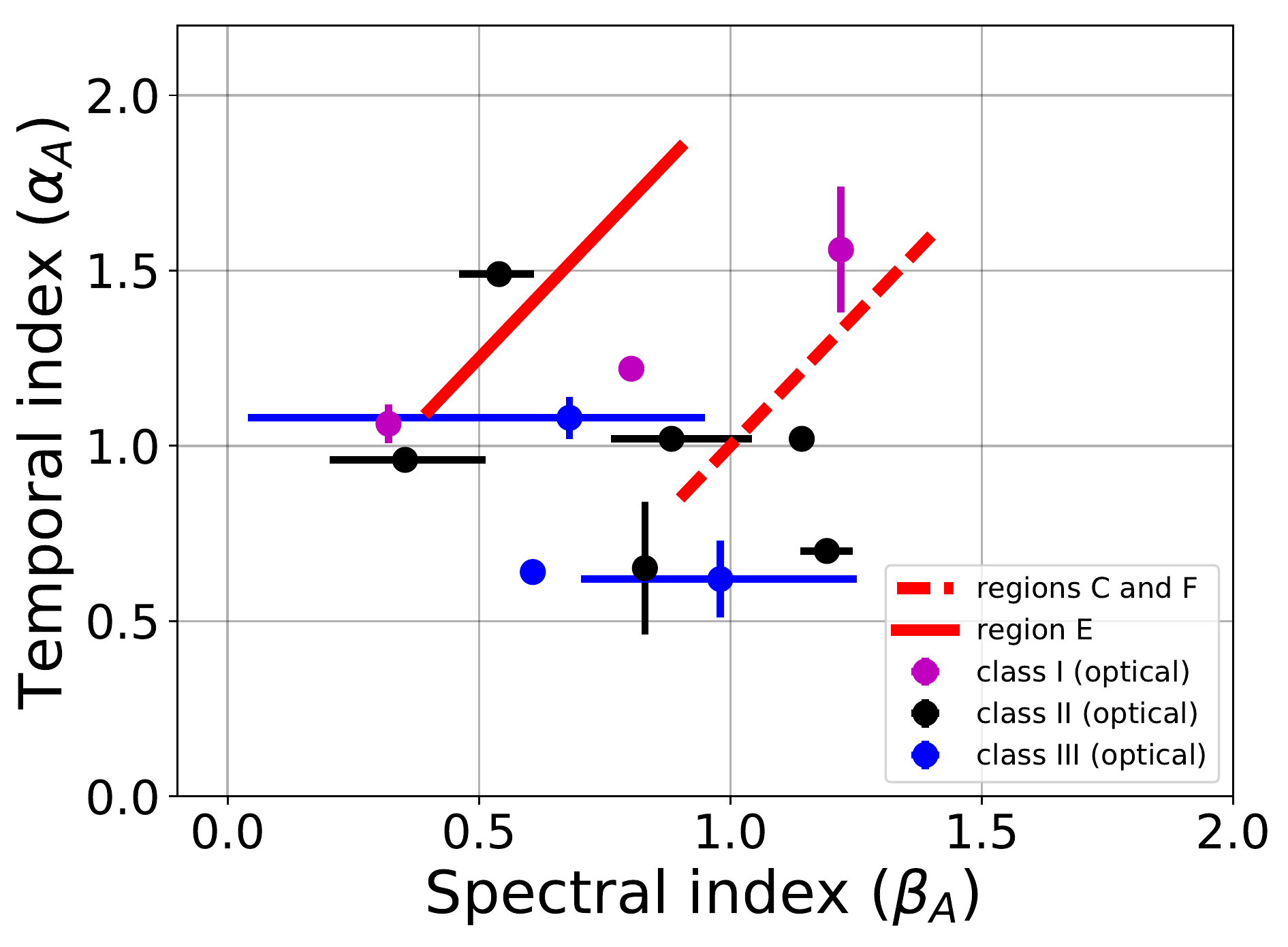}
\end{tabular}
\caption{\textbf{Temporal and spectral indices of the 13 GRBs during the self-similar phase \textbf{\large (a)} in  the X-ray band 
and \textbf{\large (b)} in the optical band.} The spectral index
in the optical band of 130831A (in class I) during the self-similar phase could not be retrieved from the literature,
and therefore this GRB is excluded from panel \textbf{\large (b)}.
The color coding is as in Supplementary Fig. \ref{fig:CR-plateau} for the data points and regions. In the X-ray band,
all the data points are within the expected theoretical limits of the closure
relations of the model during this phase as well, while in the optical band,
where the spectral indices are taken from the literature, the scatter is larger. The source data necessary to reproduce this Figure are provided as a
Source Data file.}
\label{fig:CR-self-similar}
\end{figure}
~

~

\begin{figure}[ht!]
\centering
\begin{tabular}{cc}
\huge a & \huge b \\
\includegraphics[width=0.48\linewidth]{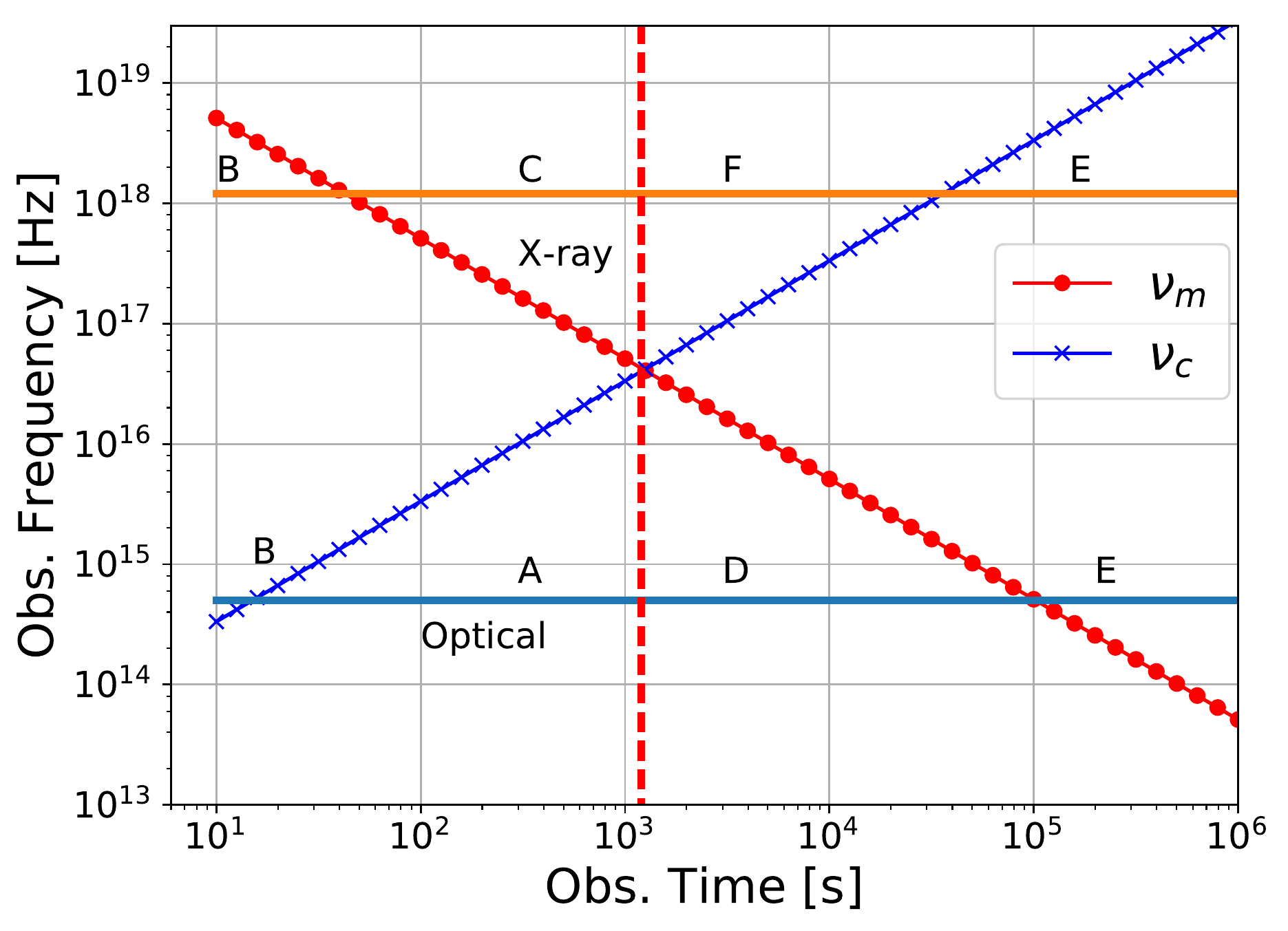} &
\includegraphics[width=0.48\linewidth]{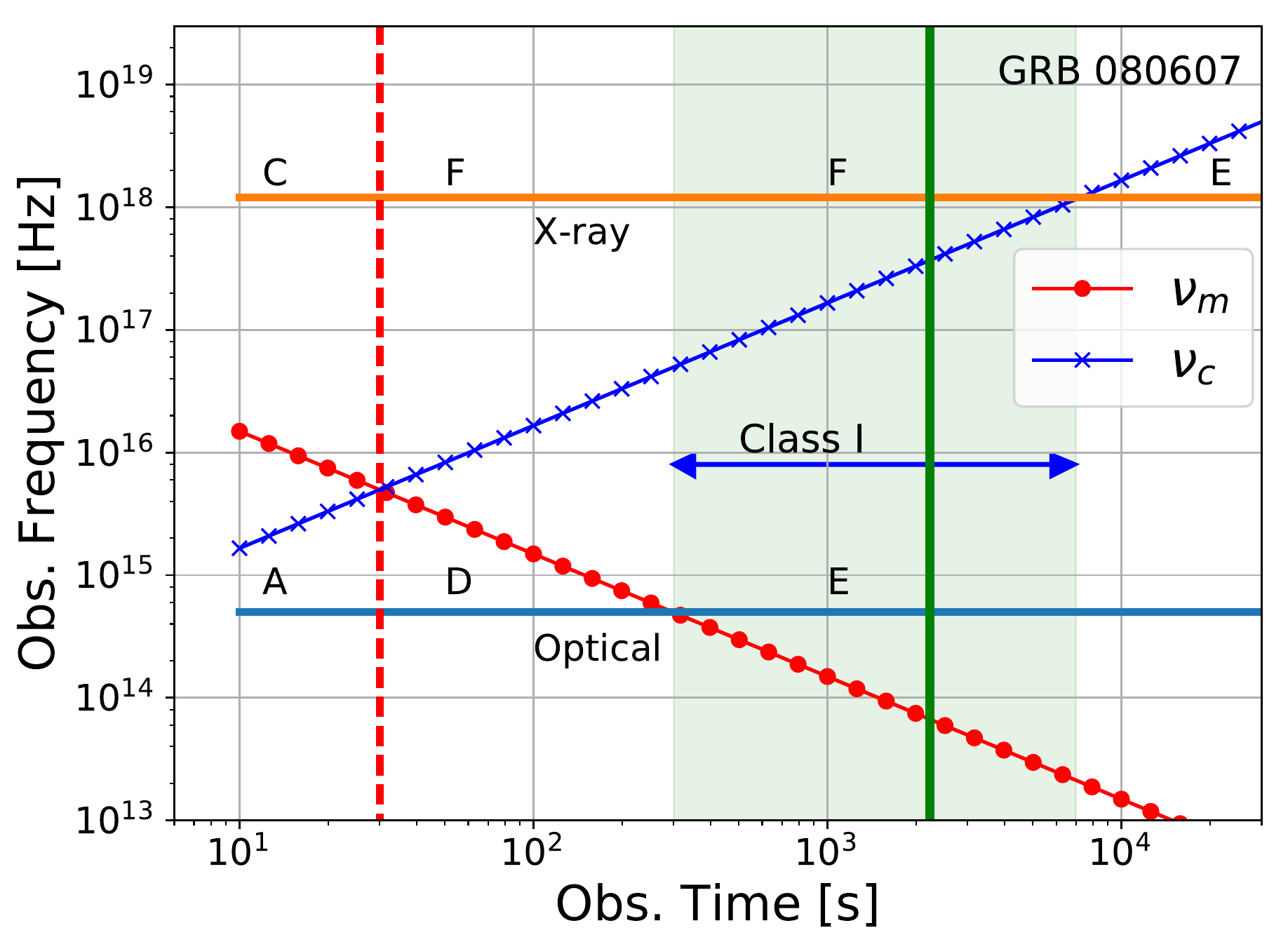} \\ 
\huge c & \huge d \\
\includegraphics[width=0.48\linewidth]{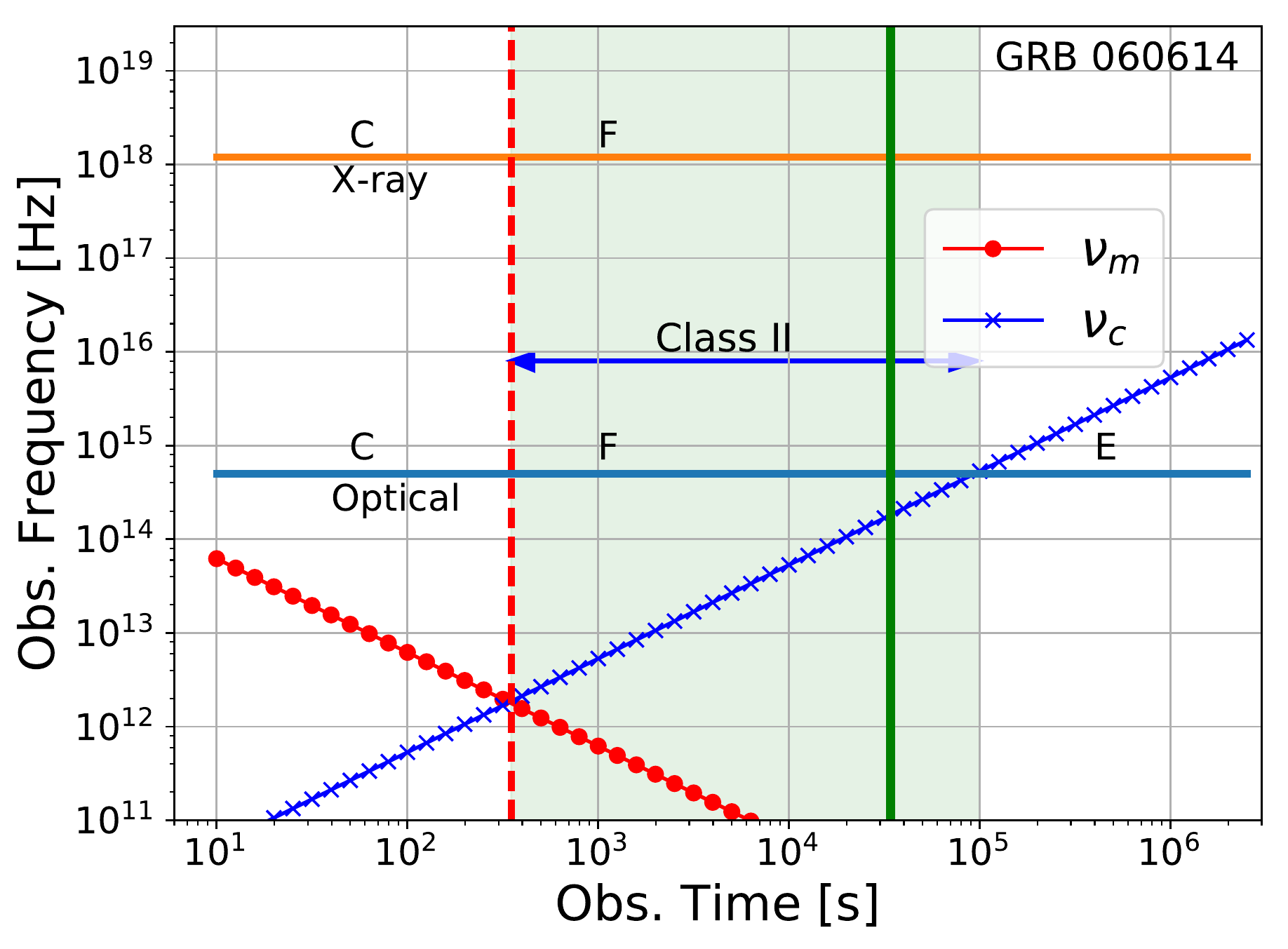} &
\includegraphics[width=0.48\linewidth]{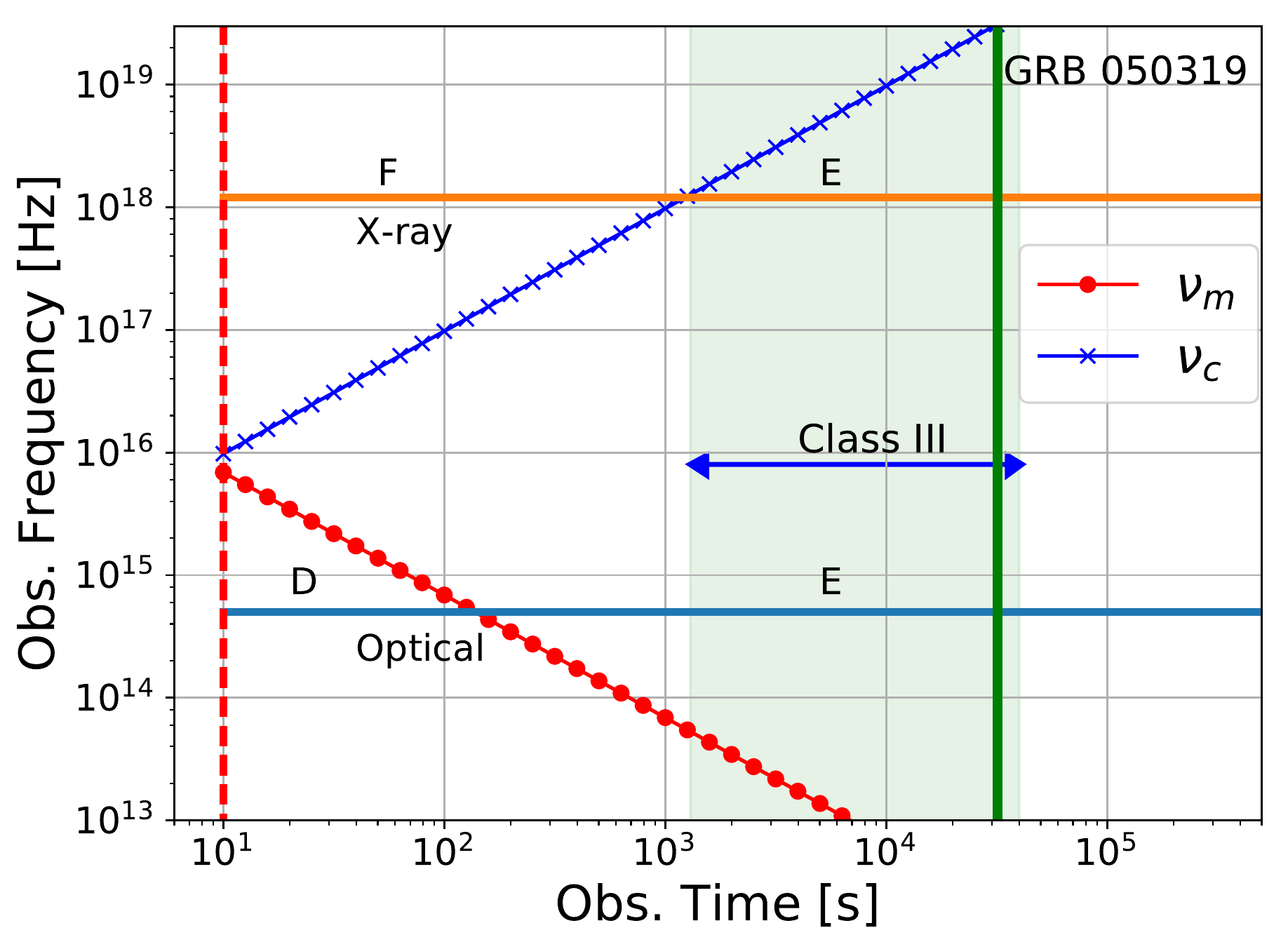}
\end{tabular}
\caption{\textbf{ Temporal evolution of the injection
frequency, $\nu_m$ (red points) and of the cooling frequency $\nu_c$
(blue crosses) in the coasting phase into a wind medium.} The letters (A, B,
C, D, E, F) represent the six possible spectral and temporal regions given
in Table \ref{tab:summary} and by Supplementary Equations
(\ref{eq:fast}) and (\ref{eq:slow}). The orange and
blue horizontal lines represent the X-ray and optical frequencies
respectively. The red dashed vertical line indicates
the crossing of the injection and of the cooling frequencies.
It is clear that a given observed frequency can shift from one region to another
only along one of the following paths: i) B-C-F-E or ii) B-A-D-E. Panel \textbf{\large (a)} is an
illustration while panels \textbf{\large (b)}, \textbf{\large (c)} and \textbf{\large (d)} are three examples showing
the occupied region (green
vertical span) of the light curves of GRBs 080607, 060614, 050319 obtained
from X-ray (in region F, F, E respectively) and optical data (in region E,
F, E). The color coding is as in
panel \textbf{\large (a)}, except that the green vertical line shows $T_{\rm a,X}$, the time marking the end of the
plateau phase. The characteristic frequencies $\nu_m$ and $\nu_c$ are computed using the
outflow parameters of GRBs 080607, 060614, 050319 presented in Table
\ref{tab:outflow_parameters} (main text). Clearly, during most of the plateau as well
as the early afterglow phases, GRBs 080607, 060614, 050319 are
classified as being in class I, II, III respectively.}
\label{fig:summary_regions_plateau}
\end{figure}

\end{document}